\pdfoutput = 1

\documentclass[12pt,a4paper]{article}

\usepackage{cite}
\usepackage{placeins}

\usepackage[utf8]{inputenc}

\usepackage{amsmath}
\usepackage[dvipsnames]{xcolor}
\usepackage{amssymb}
\usepackage{graphicx}
\usepackage[colorlinks=true, allcolors=Blue]{hyperref}
\usepackage{caption}
\usepackage{subcaption}
\usepackage{braket}
\usepackage{slashed}
\usepackage{float}
\usepackage{multirow}

\mathcode`l="8000
\begingroup
\makeatletter
\lccode`\~=`\l
\DeclareMathSymbol{\lsb@l}{\mathalpha}{letters}{`l}
\lowercase{\gdef~{\ifnum\the\mathgroup=\m@ne \ell \else \lsb@l \fi}}%
\endgroup

\begin{document}
\begin{titlepage}
\vspace*{-0.7truecm}
\begin{flushright}
Nikhef-2019-056 \\
MPP-2019-252
\end{flushright}

\vspace{1.6truecm}

\begin{center}
\boldmath
{\Large{\bf Testing Lepton Flavour Universality with
 (Semi)-Leptonic $\boldsymbol{D_{(s)}}$ Decays }
}
\unboldmath
\end{center}

\vspace{0.8truecm}

\begin{center}
{\bf Robert Fleischer\,${}^{a,b}$, Ruben Jaarsma\,${}^{a}$ and Gabri\"el Koole\,${}^{a,c}$}

\vspace{0.5truecm}

${}^a${\sl Nikhef, Science Park 105, NL-1098 XG Amsterdam, Netherlands}

${}^b${\sl  Faculty of Science, Vrije Universiteit Amsterdam,\\
NL-1081 HV Amsterdam, Netherlands}

${}^c${\sl  Max Planck Institute for Physics, \\
F\"ohringer Ring 6, 80805 M\"unchen, Germany}

\end{center}

\vspace*{1.7cm}

\begin{abstract}
\noindent
Data in $B$-meson decays indicate violations of lepton flavour universality, thereby raising the question about such phenomena in
the charm sector. We perform a model-independent analysis of NP contributions in (semi)-leptonic decays of $D_{(s)}$ 
mesons which originate from $c \to d \bar{l} \nu_l$  and $c \to s \bar{l} \nu_l$  charge-current interactions. Starting from the most
general low-energy effective Hamiltonian containing four-fermion operators and the corresponding short-distance coefficients,
we explore the impact of new (pseudo)-scalar, vector and tensor operators and constrain their effects through the interplay with 
current data. We pay special attention to the elements $|V_{cd}|$ and $|V_{cs}|$ of the Cabibbo--Kobayashi--Maskawa 
matrix and extract them from the $D_{(s)}$ decays in the presence of possible NP decay contributions, comparing them with determinations utilizing unitarity. We find a picture in agreement with the Standard Model within the current uncertainties. Using the results from our analysis, we make also predictions for leptonic $D_{(s)}^+ \to e^+ \nu_e$ modes which could be hugely enhanced 
with respect to their tiny Standard Model branching ratios. It will be interesting to apply our strategy at the future high-precision frontier.
\end{abstract}

\vspace*{2.1truecm}

\vfill

\noindent
December 2019

\end{titlepage}

\thispagestyle{empty}
\vbox{}
\newpage

\setcounter{page}{1}


\section{Introduction}

Forty-five years after the discovery of the charm quark, flavour physics has developed into a broad line of research, allowing us to probe the Standard Model (SM) with unprecedented sensitivity to new interactions and particles at energy scales far beyond the TeV regime which is directly accessible at the Large Hadron Collider (LHC) today \cite{Buras:2014zga}. 

In these explorations, decays of $B$ mesons usually play the key role. Measurements by the BaBar, Belle and LHCb 
collaborations of processes originating from $b \to c l^- \bar{\nu}_{l}$ and $ b \to s l^+ l^-$ quark-level transitions indicate
deviations from the SM, where the following observables are in the focus ($l = e, \mu$):
\begin{equation}\label{RD-RK}
\mathcal{R}_{D^{(*)}} \equiv \frac{\mathcal{B}(B \to D^{(*)} \tau^- \nu_{\tau})}{\mathcal{B}(B \to D^{(*)} l^- \nu_{l})}, \hspace{10mm} R_{K^{(*)}} \equiv \frac{\mathcal{B}(B \to K^{(*)} \mu^+ \mu^-)}{\mathcal{B}(B \to K^{(*)} e^+ e^-)}.
\end{equation}
The experimental values of these ratios show tensions with respect to the corresponding SM 
predictions at the $(2\mbox{--}3)\sigma$ level (for recent reviews, see, e.g., Refs.~\cite{Li:2018lxi, Bifani:2018zmi}). 
In particular, the data raise the exciting question of a possible violation of a central feature of the SM: lepton flavour universality (LFU).

On the theory side, various models of New Physics (NP), i.e.\ physics lying beyond the SM, have been proposed that could 
explain the $B$-decay anomalies, allowing in particular also for violations of LFU. Important specific scenarios are given by
leptoquark models \cite{Becirevic:2017jtw, Calibbi:2017qbu, BC, Crivellin:2018yvo, Cornella:2019hct, Popov:2019tyc}, $Z'$ models \cite{Bian:2017xzg, Allanach:2018odd, Calibbi:2019lvs} or Two-Higgs-Doublet models \cite{Niehoff:2015bfa, Iguro:2017ysu, Crivellin:2019dun}, implying usually a rich phenomenology of patterns and correlations among various observables. 

In view of the potential violation of LFU in $B$-meson decays, it is interesting to search for such phenomena and possible signals
of physics beyond the SM also in the charm sector. In fact, in NP scenarios allowing us to describe the $B$ decay anomalies, 
effects may also arise in weak decay processes of $D$ mesons. In Ref.~\cite{Hiller}, rare decays of the kind $D\to\pi\ell\ell$ and $D_s\to K\ell\ell$ have recently been analyzed. These modes arise form flavour-changing neutral current (FCNC) interactions and 
are the counterparts of the rare $B$ decays entering the $R_{K^{(*)}}$ ratios in Eq.~(\ref{RD-RK}). Such processes are usually 
considered as particularly powerful NP probes as they are not allowed in the SM at the tree level but originate from quantum 
fluctuations at the loop level. For an analysis of rare $s \to d$ kaon processes, see Ref.~\cite{Crivellin:2016vjc}. However, decays caused by charged-current interactions at the SM tree level may also be affected by NP effects, as indicated by the $\mathcal{R}_{D^{(*)}}$ observables in Eq.~(\ref{RD-RK}). This opens up the door to investigate such NP effects in the corresponding decays of charmed mesons as well \cite{Fajfer:2015ixa}.   

In this paper, we shall probe LFU violating effects through (semi)-leptonic $D_{(s)}$-meson decays, applying the strategy proposed in 
Refs.~\cite{Banelli:2018fnx,BFJTX-SciPost} for (semi)-leptonic decays of $B_{(s)}$ mesons. 
Precise lattice QCD calculations and experimental information on leptonic and semileptonic decays of $D_{(s)}$ mesons, arising
 from $c \to d \bar{l} \nu_l$ or $c \to s \bar{l} \nu_l$ quark-level transitions, allow us to test LFU in the charm sector. In particular, 
 we constrain short-distance coefficients describing physics beyond the SM through a comparison of theoretical calculations with 
experimental data. Furthermore, we will extract the Cabibbo--Kobayashi--Maskawa (CKM) matrix elements $|V_{cd}|$ and $|V_{cs}|$ 
from weak charm decays, also in the presence of NP contributions, and will make predictions for leptonic $D_{(s)}^+ \to e^+ \nu_e$
decays. These modes could be hugely enhanced through new pseudoscalar contributions, in fact close to the current experimental upper bounds on the corresponding branching ratios. 

The outline of this paper is as follows: in Section \ref{sec:thfr}, we introduce the most general basis of local operators describing (semi)-leptonic $D_{(s)}$ decays, and discuss the resulting low-energy effective Hamiltonian. Furthermore, we exploit the unitarity of the CKM matrix to determine $|V_{cd(s)}|$ without any use of information following from $D_{(s)}$ decay data. In Section~\ref{sec:lept}, we utilize current experimental information on leptonic $D_{(s)}$ decays to constrain the short-distance coefficients for NP contributions. Subsequently, in Section~\ref{sec:semilep} we perform a similar analysis for semileptonic $D_{(s)}$ decays. In Section~\ref{sec:VcdVcsNP}, we use some of the obtained constraints to determine $|V_{cd}|$ and $|V_{cs}|$ in the presence of pseudoscalar NP interactions. In Section~\ref{sec:predic}, we discuss predictions for leptonic $D_{(s)}^+ \to e^+ \nu_e$ decays. Finally, we present our conclusions and outlook in Section~\ref{sec:concl}.


\section{Theoretical Framework}\label{sec:thfr}

\subsection{Low-Energy Effective Hamiltonian}

The charged-current interaction processes underlying weak decays of $D_{(s)}$ mesons can be described by local four-fermion operators with their associated short-distance Wilson coefficient functions. 
Considering all possible Lorentz structures for decays originating
from $c \rightarrow d \bar{l}\nu_l$ or $c \rightarrow s \bar{l}\nu_l$ transitions (with $ l = e, \mu, \tau$) and assuming neutrinos to 
be left-handed, we obtain the following operator basis (with $q=d, s$) \cite{Tanaka:2012nw, Sakaki:2013bfa}:
\begin{equation}\label{Eq_Operator_Basis}
\begin{alignedat}{4}
 	&\mathcal{O}^l_{V_L} 	&&= (\bar{q}_L \gamma_{\mu}  c_L) (\bar{\nu}_{l L} \gamma^{\mu}  l_L), \hspace{1cm} &&\mathcal{O}^l_{V_R} &&= (\bar{q}_R \gamma_{\mu}  c_R) (\bar{\nu}_{l L} \gamma^{\mu}  l_L),\\
	&\mathcal{O}^l_{S_1} 	&&= (\bar{q}_L c_R) (\bar{\nu}_{l L}   l_R), &&\mathcal{O}^l_{S_2} &&= (\bar{q}_R c_L) (\bar{\nu}_{l L}   l_R),\\  
 	&\mathcal{O}^l_{T} 	&&= (\bar{q}_L \sigma^{\mu \nu} c_R) (\bar{\nu}_{l L} \sigma_{\mu \nu}  l_R).
\end{alignedat}
\end{equation}
The antisymmetric tensor is defined as $\sigma^{\mu \nu} \equiv \frac{i}{2}[\gamma^{\mu},\gamma^{\nu}]$. Using the appropriate Fierz identity, one can show that the tensor operator with opposite quark chiralities vanishes. In our analysis, it will be convenient to switch to an operator basis that contains a single operator describing the scalar interactions and, similarly, a single pseudoscalar operator. To this end, we define the scalar and pseudoscalar operators in the following way:
\begin{alignat}{2}
    &\mathcal{O}^l_{S} \equiv \frac{1}{2} \big( \mathcal{O}^l_{S_1} + \mathcal{O}^l_{S_2} \big) = \frac{1}{2}(\bar{q}  c)(\bar{\nu}_{l L}   l_R), \hspace{8mm}&&\mathcal{O}^l_{P} \equiv \frac{1}{2} \big( \mathcal{O}^l_{S_1} - \mathcal{O}^l_{S_2} \big) = \frac{1}{2}(\bar{q} \gamma_5 c)(\bar{\nu}_{l L}   l_R).
\end{alignat}
The most general effective Hamiltonian containing all possible local operators of the lowest dimension for $c \rightarrow q \bar{l} \nu_l$ transitions can therefore be written as
\begin{equation}\label{Eq_Eff_Hamiltonian}
    \mathcal{H}_{\text{eff}} = \frac{4 G_F}{\sqrt{2}} V_{cq} \Big[ (1 + C^l_{V_L}) \mathcal{O}^l_{V_L} + C^l_{V_R}\mathcal{O}^l_{V_R} + C^l_{S} \mathcal{O}^l_S + C^l_{P} \mathcal{O}^l_{P} + C^l_{T}\mathcal{O}^l_{T} \Big], 
\end{equation}
where the subscripts $V_L$, $V_R$, $S$, $P$ and $T$ denote the left-handed vector, right-handed vector, scalar, pseudoscalar and tensor contributions, respectively. In the SM, only the left-handed vector operator is present with an overall 
Wilson coefficient equal to one. In our analysis, we shall assume real Wilson coefficients for simplicity, i.e.\ that the NP effects do not involve new sources of CP violation. For a discussion of such effects in the analogous $B$ decays, we refer the reader to Ref.~\cite{Banelli:2018fnx,BFJTX-SciPost}.


\subsection{\boldmath $|V_{cd}|$ and $|V_{cs}|$ from Unitarity}\label{Sec_Vcd_Vcs_Unitarity}

The CKM matrix elements $V_{cd}$ and $V_{cs}$ are usually determined directly from leptonic and semileptonic $D_{(s)}$ decays and assuming the SM (such an extraction from experimental $D$ decay rates using lattice QCD form factors
was performed in Ref.~\cite{lattice-Vcx}). In this work, we are investigating these decays in the presence of NP contributions, 
hence we need an independent determination of $V_{cd}$ and $V_{cs}$. To this end, we adopt the Wolfenstein parametrization \cite{Wolfenstein:1983yz} of the CKM matrix, exploiting its unitarity. Here, $V_{cd}$ and $V_{cs}$ are related to the Wolfenstein parameters $\{\lambda, A, \rho, \eta \}$. Including corrections up to $\mathcal{O}(\lambda^5)$ yields the following expressions \cite{Buras:1994ec, Buras:2002sd}:
\begin{equation}\label{Eq_Vcd_Wolfenstein}
V_{cd} = - \lambda + \frac{1}{2} A^2 \lambda^5[1-2(\rho + i \eta)] + \mathcal{O}(\lambda^7),
\end{equation}
\begin{equation}
V_{cs} = 1- \frac{1}{2} \lambda^2 - \frac{1}{8} \lambda^4 (1 + 4A^2) + \mathcal{O}(\lambda^6) .
\end{equation}
The Wolfenstein parameters entering here can be determined without any information from $D_{(s)}$ decays, which is a very advantageous feature of the charm system. The parameters $\lambda$ and $A$ are related to the CKM elements $V_{us}$ and $V_{cb}$, respectively. The absolute value of $V_{us}$ is determined from the experimental information on kaon decays and assuming the SM. The current average of the results from semileptonic $K^0_S$, $K^0_L$ and $K^{\pm}$ decays, combined with $K \to \mu \nu (\gamma)$ decays, is given as follows \cite{Tanabashi:2018oca}: 
\begin{equation}
|V_{us}| = 0.2243 \pm 0.0005,
\end{equation}
where the SM has been assumed. For $|V_{cb}|$, the current world average from exclusive and inclusive semileptonic decays of $B$ mesons to charm takes the following value \cite{Tanabashi:2018oca}:
\begin{equation}\label{Vcb}
|V_{cb}| = A\lambda^2+{\cal O}(\lambda^8)=(42.2 \pm 0.8) \times 10^{-3} ;
\end{equation}
measurements of $|V_{cb}|$ obtained from $\mathcal{B}(B \to D^{(*)} \tau \bar{\nu})$ are not included. 
To determine the remaining Wolfenstein parameters, we further exploit the unitarity of the CKM matrix. The side $R_b$ of the unitarity triangle (UT) of the CKM matrix together with the UT angle $\gamma$ allows us to determine $\rho$ and $\eta$. For the determination of $R_b$, we use the following
current world average for $|V_{ub}|$ obtained from inclusive and exclusive semileptonic $B$ decays assuming the SM \cite{Tanabashi:2018oca}:
\begin{equation}
|V_{ub}| = (3.94 \pm 0.36) \times 10^{-3} ,
\end{equation}
combined with the result in Eq.\ (\ref{Vcb}). There exist tensions at the 3\,$\sigma$ level between the inclusive and exclusive determinations of $|V_{cb}|$ and $|V_{ub}|$\cite{Bouchard:2019all}. However, for our analysis, it has an essentially negligible impact as these CKM parameters enter only through
strongly suppressed higher-order corrections: $|V_{ub}|$ appears only in the corrections in Eq.\ (\ref{Eq_Vcd_Wolfenstein}), which differ from the leading term by four orders of $\lambda$. The angle $\gamma$ is usually determined from the tree-dominated $B \to DK$ decays which yield $\gamma = (73.5^{+4.2}_{-5.1})^{\circ}$ \cite{Tanabashi:2018oca}. However, as possible NP could slightly affect its value, we allow $\gamma$ to be within $[60^{\circ}, 80^{\circ}]$. Varying $\gamma$ within this range has a very minor impact. Finally, we obtain the following values for $|V_{cd}|$ and $|V_{cs}|$ from the unitarity of the CKM matrix:
\begin{equation}\label{Vcd-SM}
|V_{cd}| = 0.2242 \pm 0.0005 ,
\end{equation}
\begin{equation}\label{Vcs-SM}
|V_{cs}| =   0.9736 \pm 0.0001.
\end{equation}
The important feature of these results is that they are independent of possible NP contributions to (semi)-leptonic charged-current charm transitions, which are usually exploited to determine these CKM matrix elements from experimental data. We shall use them as reference values for our analysis discussed below.


\section{Leptonic Decays}\label{sec:lept}

Leptonic $D^+_{(s)}\rightarrow l^+ \nu_l$ decays are the simplest and cleanest weak decay class of charmed mesons. All the hadronic dynamics is captured by a single parameter: the $D_{(s)}$-meson decay constant $f_{D^+_{(s)}}$. Leptonic decays of $D$ and $D_s$ mesons contain the flavour-changing quark transitions $c \rightarrow d$ and $c  \rightarrow s$, respectively. Accurate non-perturbative calculations of the decay constants combined with precise experimental results provide excellent opportunities to perform tests of lepton flavour universality. 

In the SM, the branching fraction for leptonic $D^+_{(s)}$ decays is given as
\begin{equation}\label{BFleptonicDdecaySM}
    \mathcal{B}(D^+_{(s)}\rightarrow l^+ \nu_l)\big|_{\text{SM}} = \frac{G_F^2}{8 \pi} |V_{cq}|^2 f_{D^+_{(s)}}^2 M_{D^+_{(s)}} m^2_l \Big(1 -  \frac{m^2_l}{M^2_{D^+_{(s)}}}\Big)^2 \tau_{D^+_{(s)}} ,
\end{equation}
where $G_F$ is Fermi's constant, $\tau_{D^+_{(s)}}$ is the lifetime of the $D_{(s)}^+$ meson, and $M^+_{D_{(s)}}$ and $m_l$ are the masses of the $D^+_{(s)}$ meson and the lepton ($l=e,\mu,\tau$), respectively. The decay constants $f_{D^+}$ and $f_{D^+_{s}}$ are determined from  lattice QCD calculations. For our analysis, we use the values determined by the FLAG working group \cite{Aoki:2019cca}: 
\begin{equation}
f_{D^+} = (209.0 \pm 2.4) \text{ MeV}, 
\end{equation}
\begin{equation}
f_{D^+_s} = (248.0 \pm 1.6)  \text{ MeV}. 
\end{equation}
The resulting leptonic branching fractions in the SM and their experimental values are given in Table \ref{SM_BFsLep_summary}. Here, we use the values for $|V_{cd}|$ and $|V_{cs}|$ obtained from unitarity, given in Eqs.~(\ref{Vcd-SM}) and (\ref{Vcs-SM}), respectively. For the $D_{(s)} \to e^+ \nu_e$ decays, only experimental upper bounds are available. This is due to the extremely strong helicity suppression in these processes, which is reflected by the proportionality of the branching fraction to $m_l^2$. However, potential contributions from pseudoscalar NP interactions could lift the helicity suppression in these decays, thereby making them excellent probes for NP.  For previous studies of NP effects in these decays, see Refs.~\cite{Akeroyd:2003jb,Akeroyd:2009tn,Akeroyd:2002pi}.

\begin{table}[t]
\begin{center}
\renewcommand{\arraystretch}{1.2}
\begin{tabular}{ cccc } 
 \hline
 Decay & SM & Experiment  \\ 
 \hline
 \hline
 $\mathcal{B}(D^+ \rightarrow e^+ \nu_{e}) $ 		& $(9.16 \pm 0.22) \times 10^{-9}  $ & $<  8.8 \times 10^{-6}$ & \cite{Tanabashi:2018oca} \\ 
 $\mathcal{B}(D^+ \rightarrow \mu^+ \nu_{\mu})$ 	& $(3.89 \pm 0.09) \times 10^{-4}$ & $(3.74 \pm 0.17) \times 10^{-4} $& \cite{Tanabashi:2018oca} \\ 
 $\mathcal{B}(D^+ \rightarrow \tau^+ \nu_{\tau})$	& $(1.04 \pm 0.03) \times 10^{-3}$ & $(1.20 \pm 0.27)\times 10^{-3}$ & \cite{Ablikim:2019rpl}  \\
 \hline
 $\mathcal{B}(D_s^+ \rightarrow e^+ \nu_{e}) $		& $(1.24 \pm 0.02) \times 10^{-7}$ &$<  8.3 \times 10^{-5}  $& \cite{Tanabashi:2018oca}\\ 
 $\mathcal{B}(D_s^+ \rightarrow \mu^+ \nu_{\mu}) $	& $(5.28 \pm 0.08) \times 10^{-3}$& $(5.50 \pm 0.23) \times 10^{-3} $&  \cite{Tanabashi:2018oca} \\ 
 $\mathcal{B}(D_s^+ \rightarrow \tau^+ \nu_{\tau}) $	& $(5.15 \pm 0.08) \times 10^{-2}$& $(5.48 \pm 0.23)\times 10^{-2} $&  \cite{Tanabashi:2018oca}\\ 
 \hline
\end{tabular}
\end{center}
\caption{Branching ratios of leptonic $D_{s}^+$ decays calculated in the SM and comparison with the currently available 
experimental values.}\label{SM_BFsLep_summary}
\end{table}

\subsection{Constraints on Pseudoscalar Coefficients}

As the scalar contributions coming from $\mathcal{O}_S$ vanish due to parity conservation, we start our NP analysis by considering contributions from new pseudoscalar particles. Using the low-energy effective 
Hamiltonian in Eq.\ (\ref{Eq_Eff_Hamiltonian}), we complement the SM branching fraction with a pseudoscalar contribution and obtain the following expression:
\begin{equation}\label{Eq_Lep_BF_PseuNP}
    \mathcal{B}(D_{(s)}^+ \rightarrow l^+ \nu_l)=\mathcal{B}(D_{(s)}^+ \rightarrow l^+ \nu_l)\big|_{\text{SM}} \Bigg|1 + C^{l}_{P} \frac{M^2_{D^+_{(s)}}}{m_l(m_c+m_q)} \Bigg|^2 ,
\end{equation}
where $C^{l}_{P}$ is the short-distance coefficient for the pseudoscalar NP contribution, and $m_c$ and $m_q$ are the masses of the charm and down (strange) quarks. In these branching fractions, the decay constants are the source of the largest theoretical uncertainties. Moreover, they contain the CKM elements $|V_{cd}|$ or $|V_{cs}|$, which we would finally like to determine from these
decays, also in the presence of NP contributions. With this in mind, we consider the following ratio of two leptonic decays:
\begin{equation}\label{Eq_Ratio_lep_BFs_PseuNP}
    R^{l_1}_{l_2} \equiv \frac{\mathcal{B}(D_{(s)}^+ \rightarrow l_1^+ \nu_{l_1})}{\mathcal{B}(D_{(s)}^+ \rightarrow l_2^+ \nu_{l_2})}   = \frac{\alpha^{l_1} \big|1 + \beta^{l_1} C^{l_1}_P \big|^2 }{\alpha^{l_2}\big|1 + \beta^{l_2} C^{l_2}_P \big|^2},
\end{equation}
where 
\begin{equation}
\alpha^{l_{1(2)}} = m^2_{l_{1(2)}} \Big(1- \frac{m^2_{l_{1(2)}}}{M^2_{D^+_{(s)}}}\Big)^2 \quad\mbox{and}\quad 
\beta^{l_{1(2)}} = \frac{M^2_{D^+_{(s)}}}{[m_{l_{1(2)}}(m_c + m_q)]}. 
\end{equation}
This observable is theoretically clean as the decay constants and also the CKM matrix elements cancel. 

\begin{figure}[t!]
\centering
\begin{subfigure}{.45\linewidth}
  \centering
  \includegraphics[width=.8\linewidth]{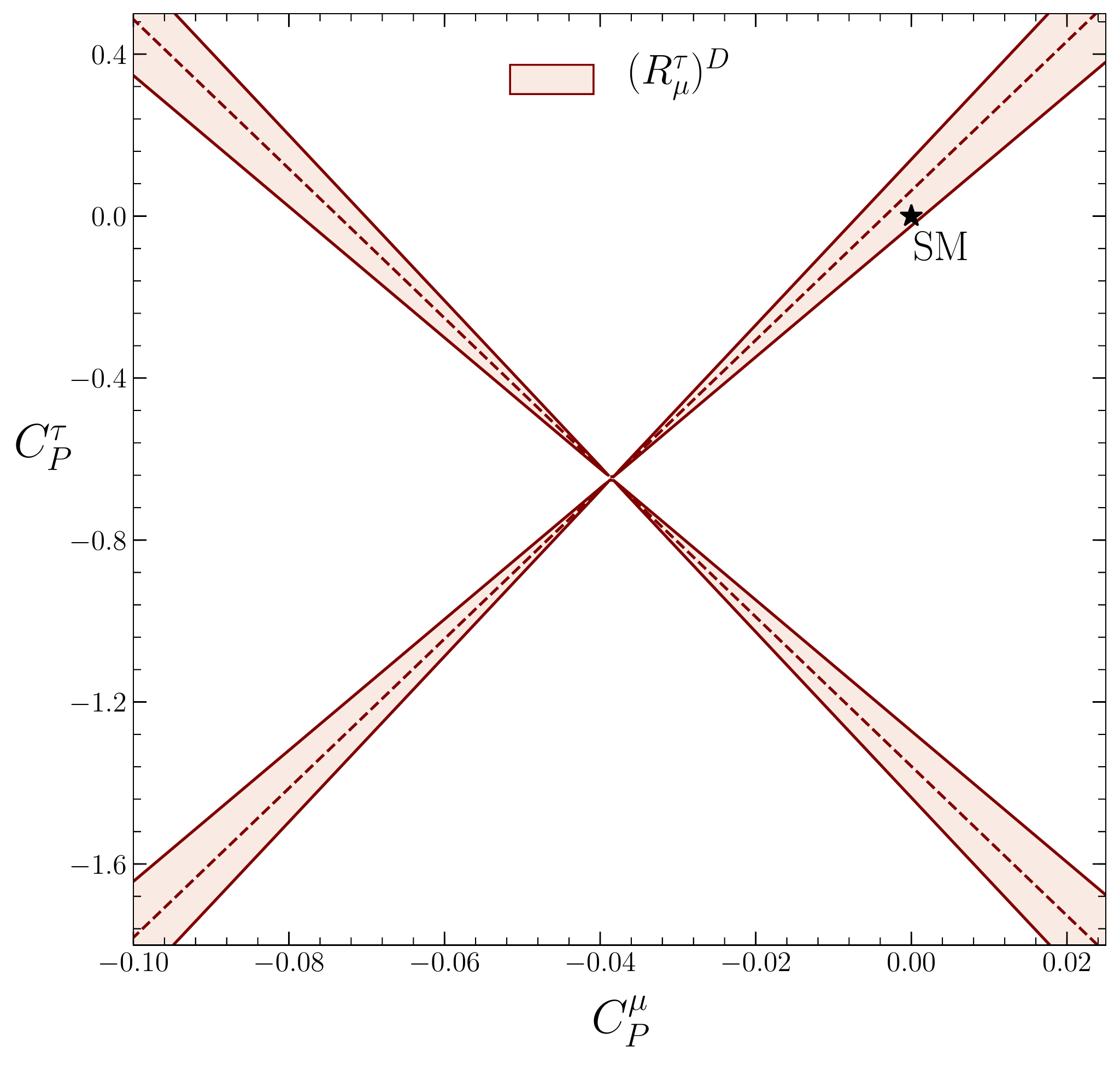}
\end{subfigure}
\begin{subfigure}{.45\linewidth}
  \centering
  \includegraphics[width=.8\linewidth]{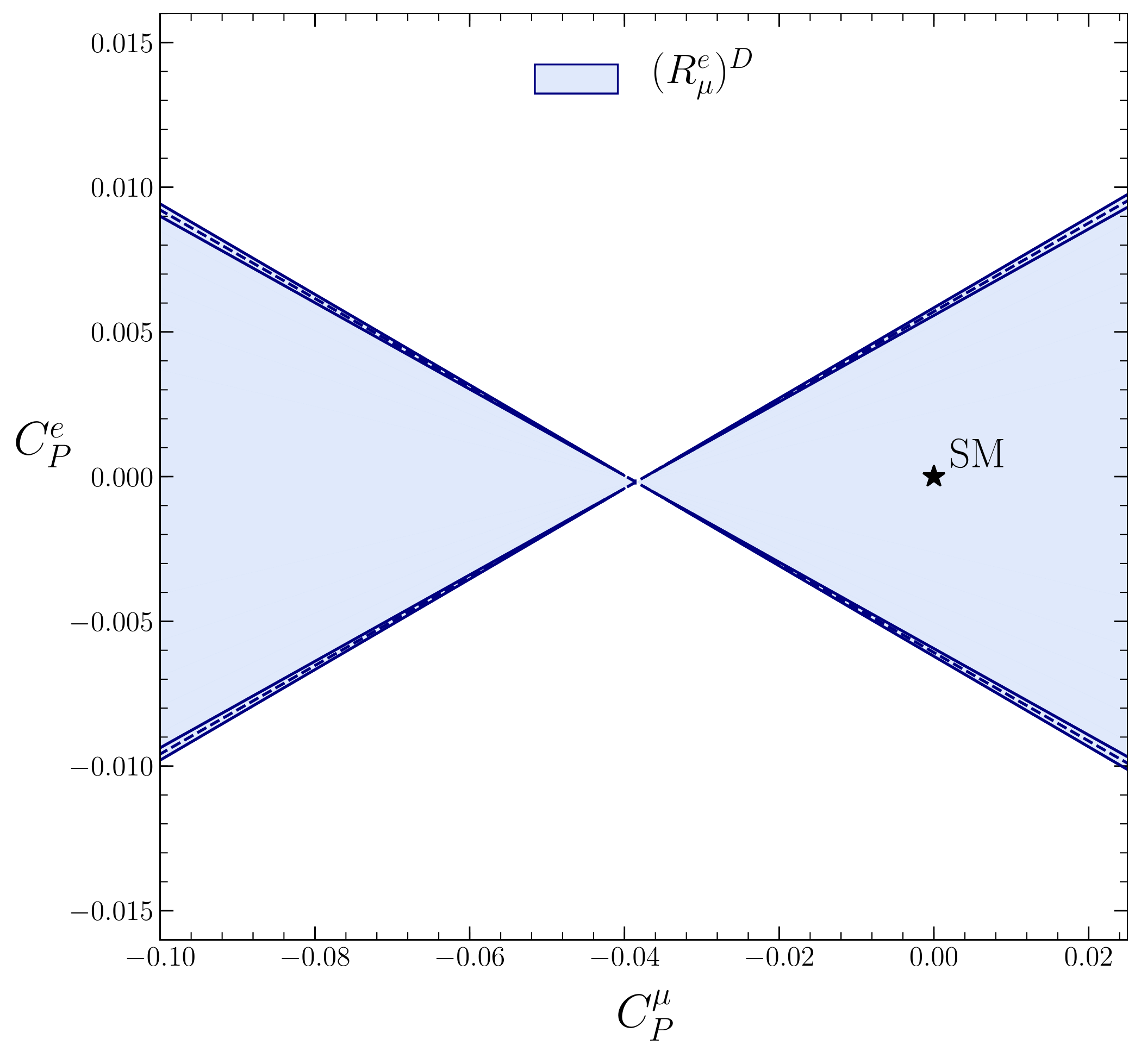}
\end{subfigure}
\caption{The allowed regions in the $C^{\mu}_P$--$C^{\tau}_P$ (left) and $C^{\mu}_P$--$C^{e}_P$ (right) planes 
following from the ratios $(R^{\tau}_{\mu})^{D}$ and $(R^{e}_{\mu})^{D}$, respectively. The black stars indicate the SM predictions.}
\label{Fig_Lep_Constr_PseuNP_D}
\end{figure}

Let us first consider leptonic decays of $D$ mesons.
Using the experimental information in Table~\ref{SM_BFsLep_summary}, we obtain the following value for the ratio between two leptonic $D$ decays with tau leptons and muons in the final state:
\begin{equation}\label{Eq_R_tau_mu_lep_exp} 
(R^{\tau}_{ \mu})^D = 3.21 \pm 0.73 ,
\end{equation}
where we have utilized the recent first observation of the decay $D^+ \to \tau^+ \nu_{\tau}$ by the BESIII collaboration \cite{Ablikim:2019rpl}. By comparing the experimental value with the corresponding theoretical expression, we determine the allowed regions in the $C^{\mu}_P$--$C^{\tau}_P$ plane. The result is presented in Fig.\ \ref{Fig_Lep_Constr_PseuNP_D} (left), where the uncertainties coming from the masses are neglected due to their smallness. The SM prediction ($C^{\mu}_P=C^{\tau}_P=0$), indicated by the black star, is in agreement with the obtained constraints at the 1$\sigma$ level. 
For electrons and muons in the final state, we obtain
\begin{equation}\label{Eq_R_e_mu_lep_exp} 
(R^{e}_{ \mu})^D < (2.35 \pm 0.11) \times 10^{-2}  .
\end{equation}

\noindent Comparison with the corresponding theoretical expression allows us to calculate the allowed regions in the $C^{\mu}_P$--$C^{e}_P$ plane. For the branching fraction $\mathcal{B}(D^+ \to e^+ \nu_e)$, only an experimental upper bound is available, resulting in the large wedge-shaped allowed regions in the panel on the right-hand side of Fig.\ \ref{Fig_Lep_Constr_PseuNP_D}. The SM prediction ($C^{\mu}_P=C^{e}_P=0$) lies within the obtained 1\,$\sigma$ contour. A future measurement of $\mathcal{B}(D^+ \to e^+ \nu_e)$ would allow us to determine more stringent constraints in the $C^{\mu}_P$--$C^{e}_P$ plane from the $(R^{e}_{\mu})^D$ ratio, 
in analogy to the constraints following from $(R^{\tau}_{\mu})^D$.

\begin{figure}[]
\centering
\begin{subfigure}{.45\linewidth}
  \centering
  \includegraphics[width=.8\linewidth]{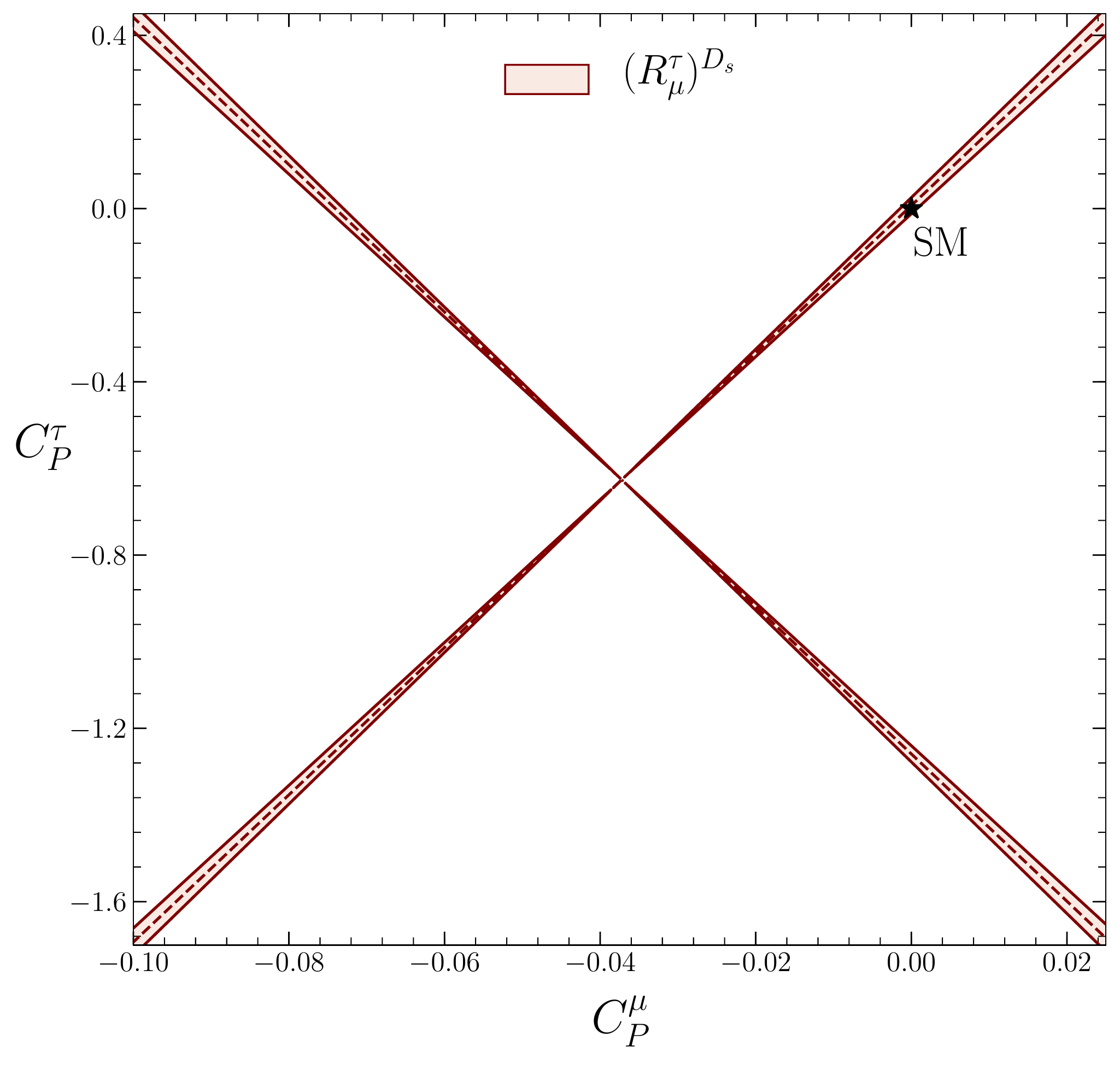}
\end{subfigure}
\begin{subfigure}{.45\linewidth}
  \centering
  \includegraphics[width=.8\linewidth]{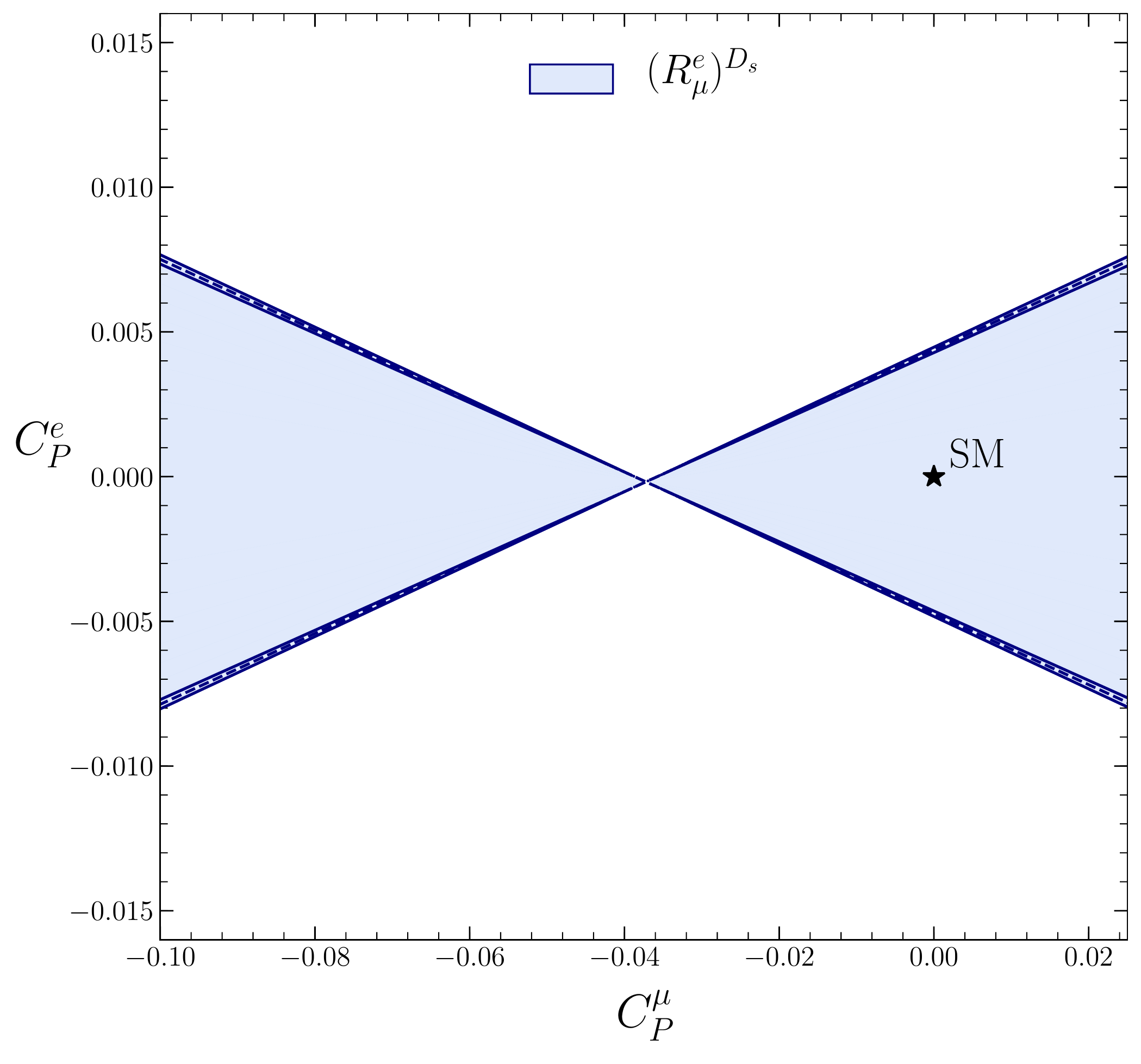}
\end{subfigure}
\caption{The allowed regions in the $C^{\mu}_P$--$C^{\tau}_P$ (left) and $C^{\mu}_P$--$C^{e}_P$ (right) planes following from 
the ratios $(R^{\tau}_{\mu})^{D_s}$ and $(R^{e}_{\mu})^{D_s}$, respectively. The black stars indicate the SM predictions.}
\label{Fig_Lep_Constr_PseuNP_Ds}
\end{figure}

For $D_s$ decays, we perform a similar analysis in the presence of pseudoscalar NP. From the experimental information on leptonic $D_s$ decays in Table \ref{SM_BFsLep_summary}, we obtain the following value for $(R^{\tau}_{\mu})^{D_s}$:
\begin{equation}\label{Eq_R_tau_mu_lep_Ds}  
(R^{\tau}_{\mu})^{D_s} = 9.96 \pm 0.59 .
\end{equation}
Using this result, we constrain the corresponding short-distance coefficients. The resulting allowed regions in the $C^{\mu}_P$--$C^{\tau}_P$ plane are shown in the panel on the left-hand side in Fig.~\ref{Fig_Lep_Constr_PseuNP_Ds}. For $(R^{e}_{\mu})^{D_s}$, we obtain the experimental value
\begin{equation}\label{Eq_R_e_mu_lep_Ds} 
(R^{e}_{ \mu})^{D_s} < (1.51 \pm 0.06) \times 10^{-2} .
\end{equation}
By comparing our theoretical expression with the experimental information, we obtain the allowed regions in the $C^{\mu}_P$--$C^{e}_P$ plane shown in the panel on the right-hand side of Fig. \ref{Fig_Lep_Constr_PseuNP_Ds}. We observe that the obtained 1$\sigma$ contours for both $(R^{\tau}_{\mu})^{D_s}$ and $(R^{e}_{\mu})^{D_s}$ contain the SM predictions. We would like to note that also for $\mathcal{B}(D^+_s \to e^+ \nu_e)$, only an experimental upper bound is available. A future measurement of this branching fraction would allow us to put more stringent constraints on the relevant NP coefficients. We shall return to $D^+_{(s)} \to e^+ \nu_e$ decays in Section~\ref{sec:predic}.

\subsection{Constraints on Vector Coefficients}

\begin{figure}[h!]
\centering
\begin{subfigure}{.45\linewidth}
  \centering
  \includegraphics[width=.8\linewidth]{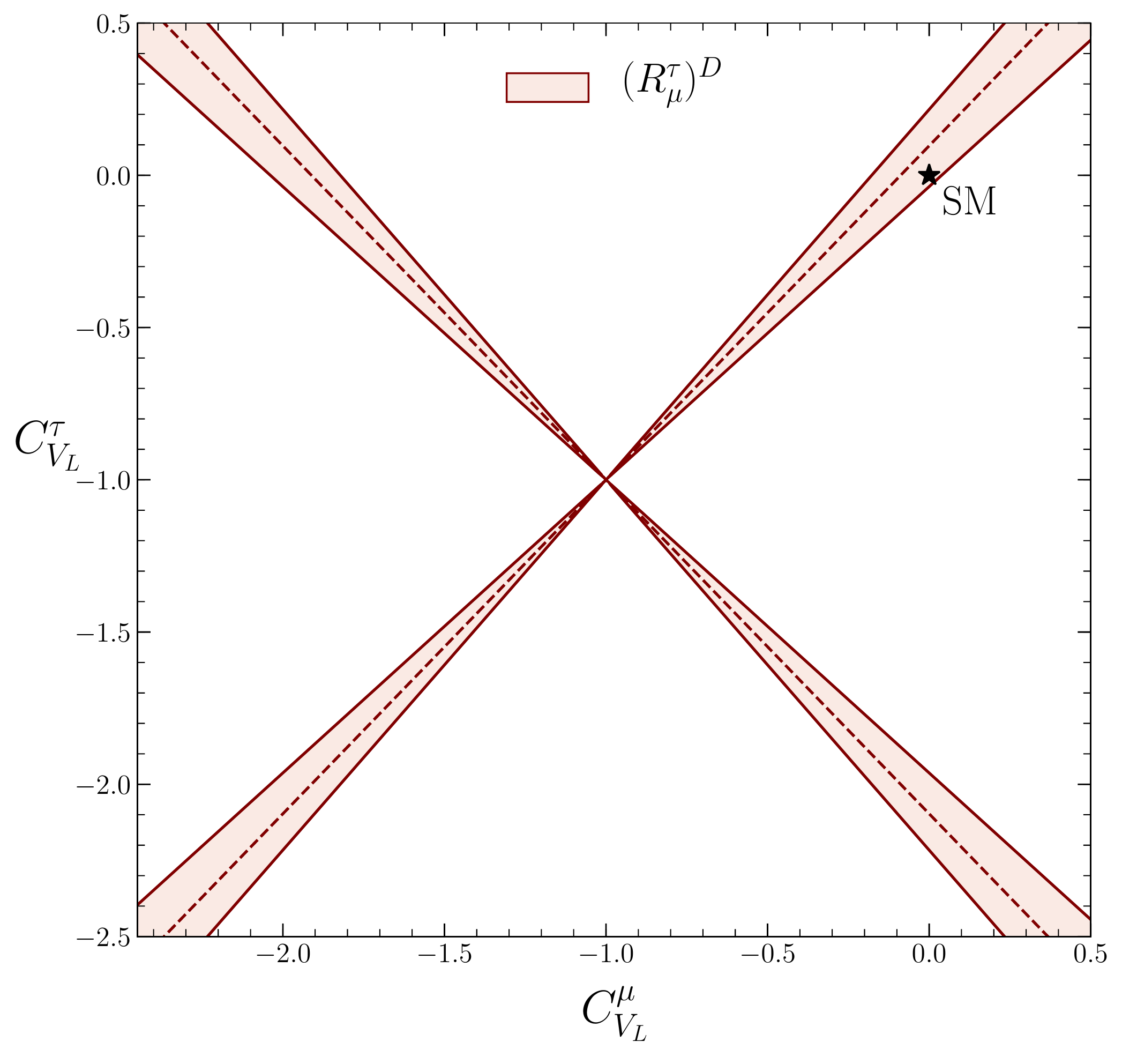}
\end{subfigure}%
\begin{subfigure}{.45\linewidth}
  \centering
  \includegraphics[width=.8\linewidth]{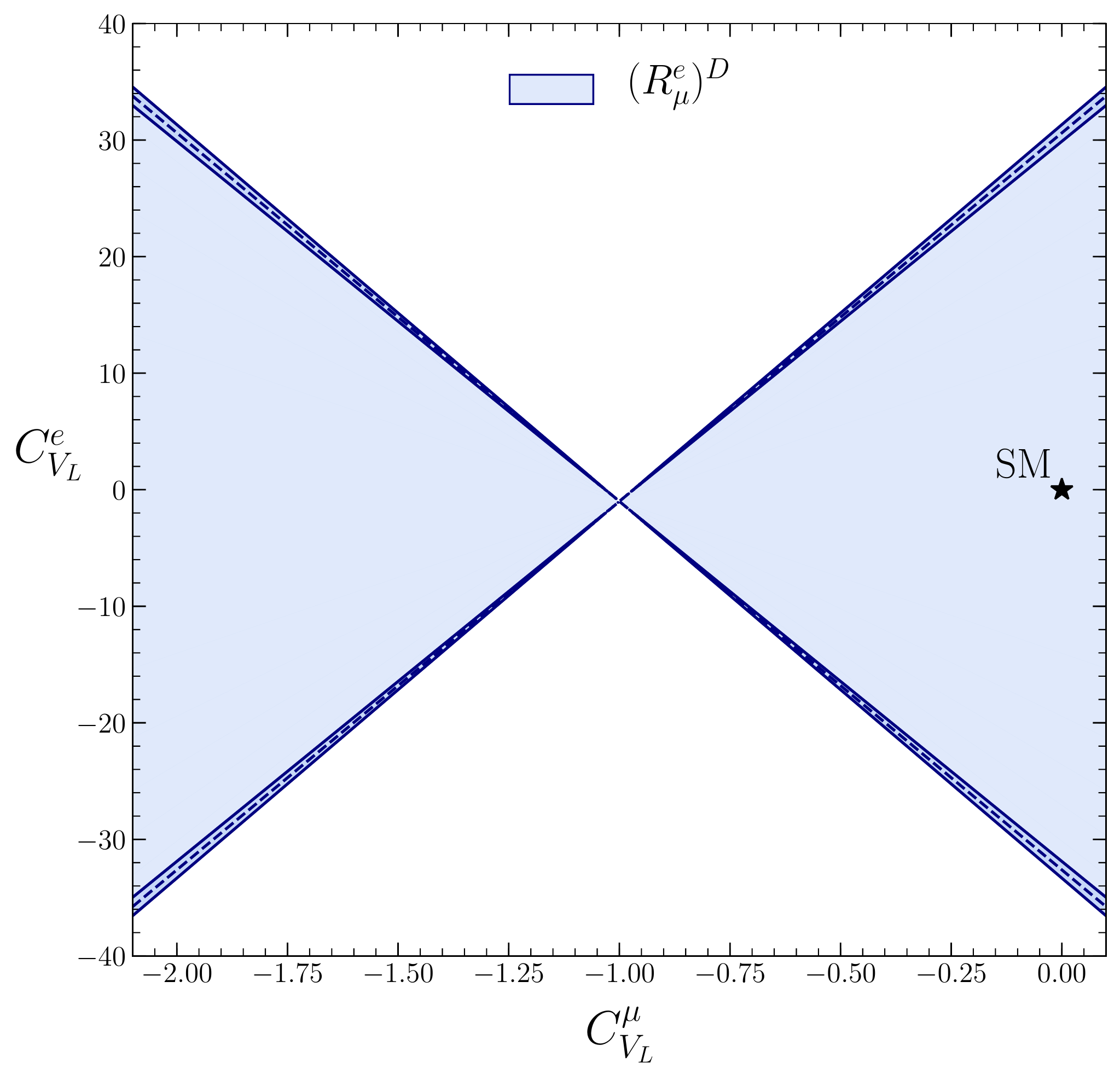}
\end{subfigure}
\begin{subfigure}{.45\linewidth}
  \centering
  \includegraphics[width=.8\linewidth]{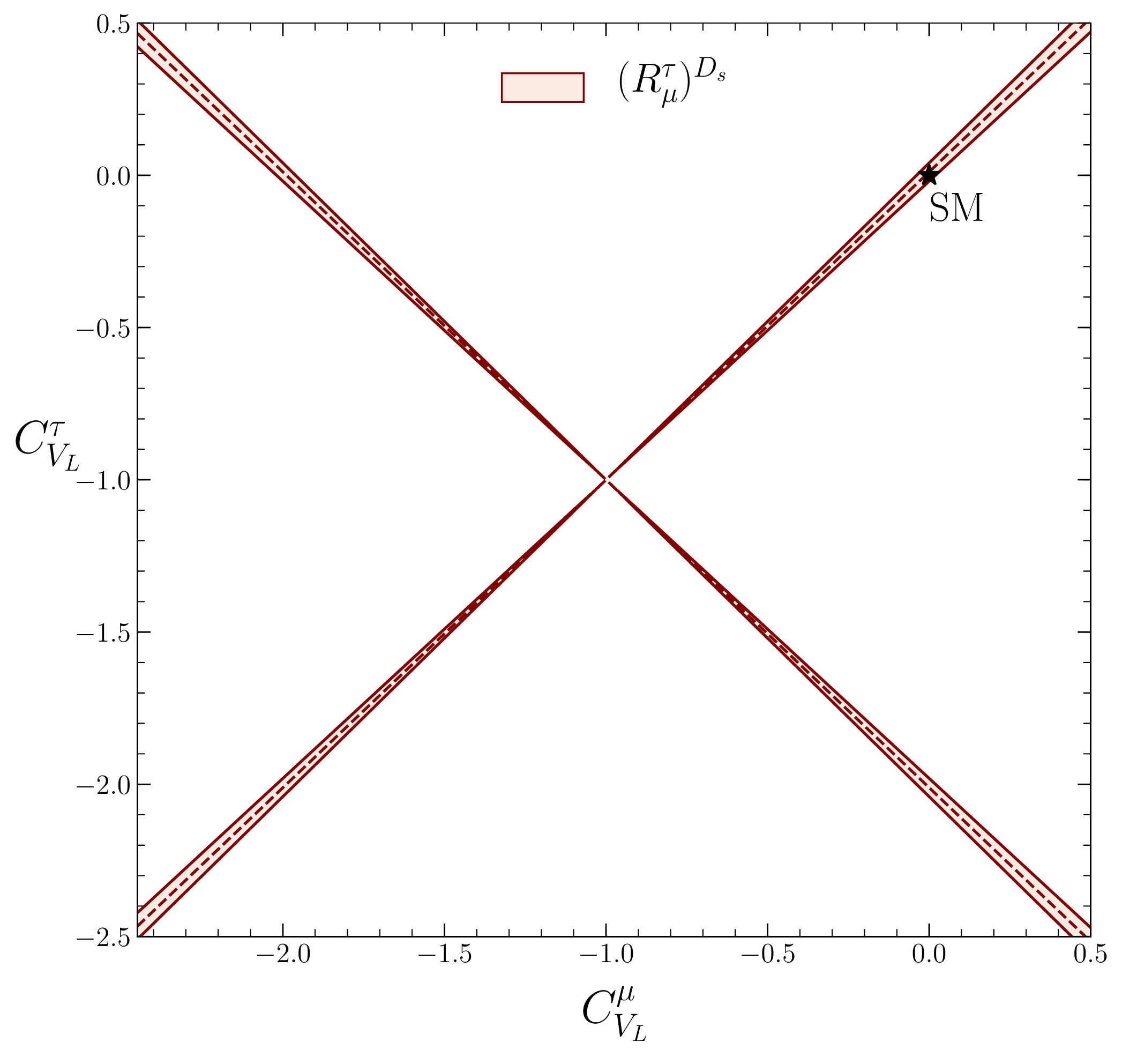}
\end{subfigure}%
\begin{subfigure}{.45\linewidth}
  \centering
  \includegraphics[width=.8\linewidth]{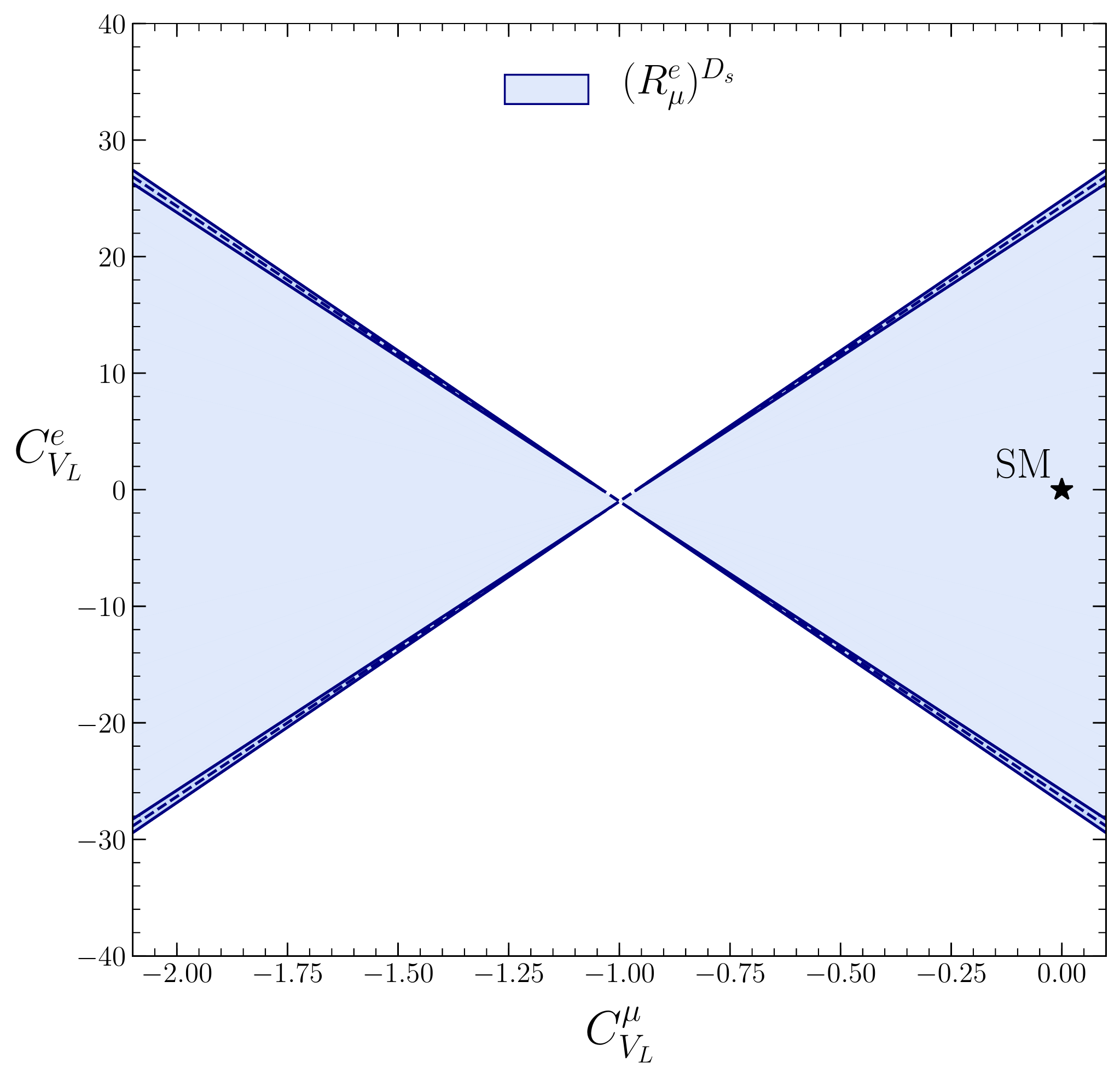}
\end{subfigure}
\caption{Allowed regions in the $C^{\mu}_{V_L}$--$C^{\tau}_{V_L}$ plane using the ratios $(R^{\tau}_{\mu})^{D}$ (top left) and $(R^{\tau}_{\mu})^{D_s}$ (bottom left), and in the $C^{\mu}_{V_L}$--$C^{e}_{V_L}$ plane using the ratios $(R^{e}_{\mu})^{D}$ (top right) and $(R^{e}_{\mu})^{D_s}$ (bottom right). The black stars indicate the SM predictions.}
\label{Fig_Lep_Constr_LHvecNP_D_Ds}
\end{figure}

Besides pseudoscalar NP contributions, the SM branching fraction may also be complemented by contributions from additional vector interactions. Including possible left-handed (LH) vector interactions leads to the following branching fraction:
\begin{equation}\label{Eq_BF_Lep_LHvecNP}
    \mathcal{B}(D_{(s)}^+ \rightarrow l^+ \nu_l)=\mathcal{B}(D_{(s)}^+ \rightarrow l^+ \nu_l)\big|_{\text{SM}} \Big|1 + C^{l}_{V_L} \Big|^2 ,
\end{equation}
where $C^{l}_{V_L}$ is the short-distance coefficient for the LH vector interaction. Just as for the pseudoscalar coefficients, we need a theoretically clean observable for the extraction of constraints on the vector coefficients. In analogy to Eq.\ (\ref{Eq_Ratio_lep_BFs_PseuNP}), we define the ratio of two leptonic branching fractions with different leptons in the final state:
\begin{equation}\label{Eq_Ratio_Lep_BFs_LHvec}
    R^{l_1}_{l_2} \equiv \frac{\mathcal{B}(D_{(s)}^+ \rightarrow l_1^+ \nu_{l_1})}{\mathcal{B}(D_{(s)}^+ \rightarrow l_2^+ \nu_{l_2})}   = \frac{\alpha^{l_1} \big|1 + C^{l_1}_{V_L} \big|^2 }{\alpha^{l_2}\big|1 + C^{l_2}_{V_L} \big|^2}.
\end{equation}
We constrain the LH vector coefficients by comparing the theoretical expressions to the experimental information in Eqs.\ (\ref{Eq_R_tau_mu_lep_exp}--\ref{Eq_R_e_mu_lep_Ds}). The obtained allowed regions in the $C^{\mu}_{V_L}$--$C^{\tau}_{V_L}$ and $C^{\mu}_{V_L}$--$C^{e}_{V_L}$ planes are given in Fig.\ \ref{Fig_Lep_Constr_LHvecNP_D_Ds}.

\begin{figure}[t!]
\centering
\begin{subfigure}{.45\linewidth}
  \centering
  \includegraphics[width=.8\linewidth]{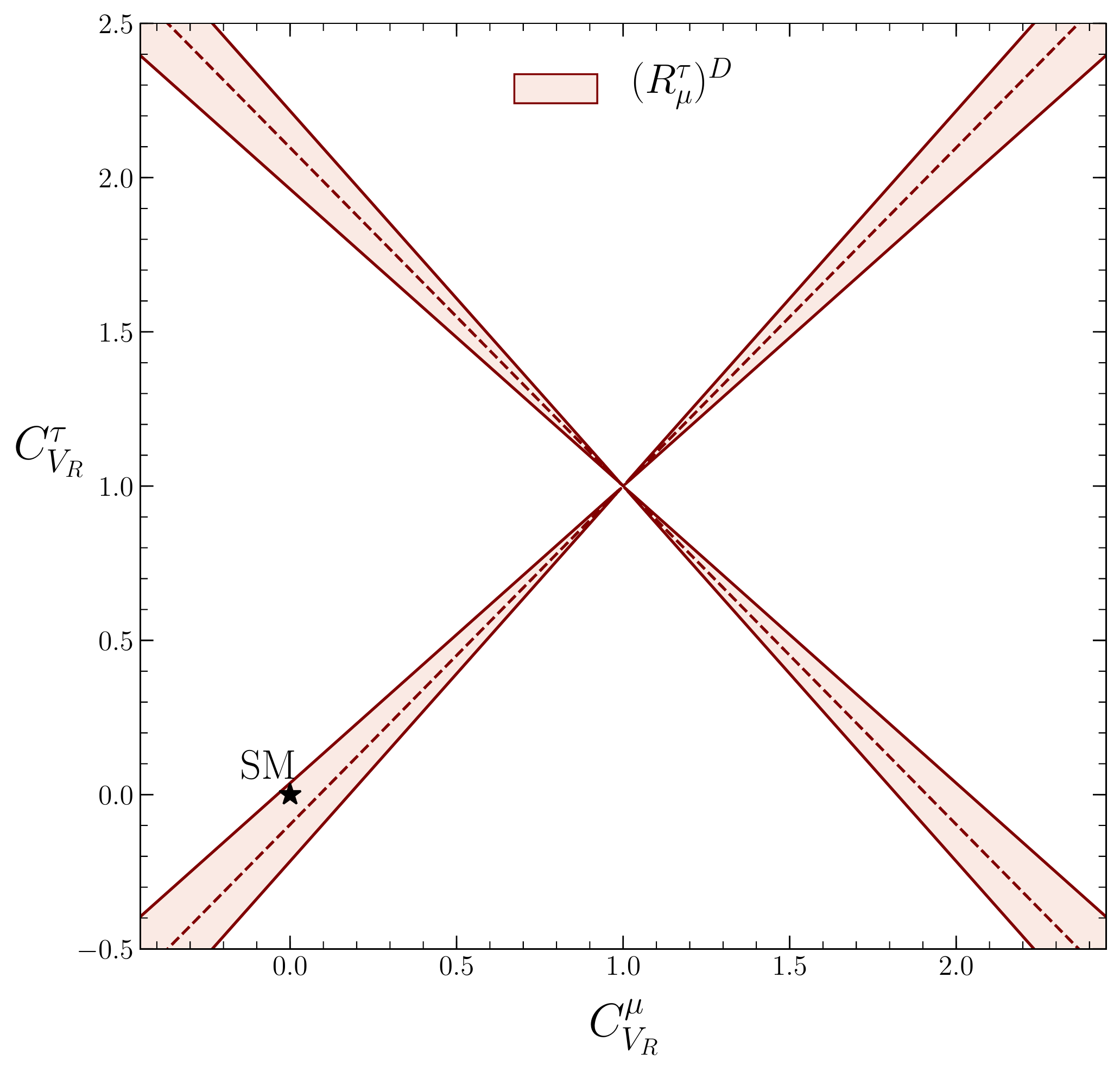}
\end{subfigure}%
\begin{subfigure}{.45\linewidth}
  \centering
  \includegraphics[width=.8\linewidth]{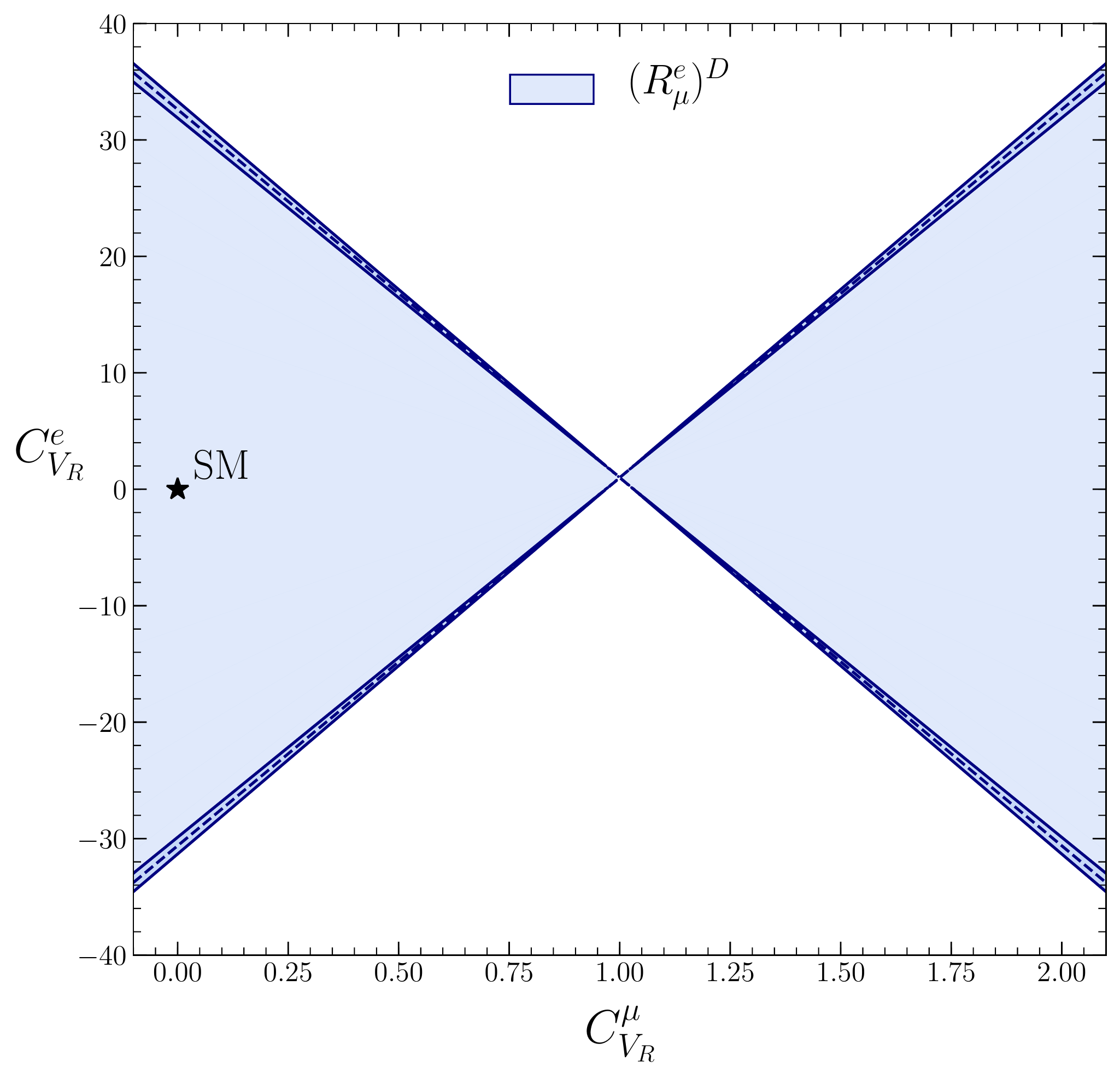}
\end{subfigure}
\begin{subfigure}{.45\linewidth}
  \centering
  \includegraphics[width=.8\linewidth]{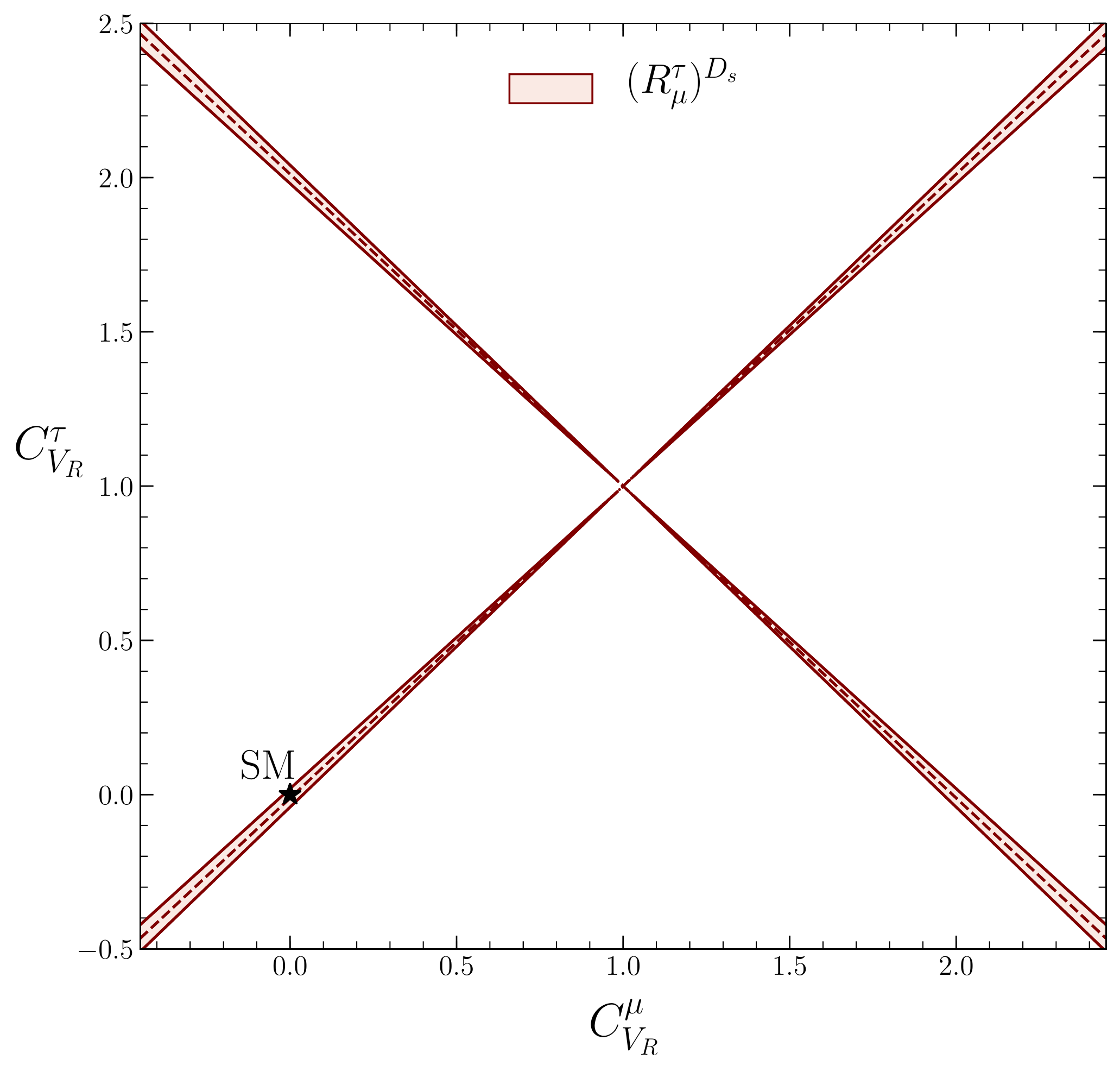}
\end{subfigure}%
\begin{subfigure}{.45\linewidth}
  \centering
  \includegraphics[width=.8\linewidth]{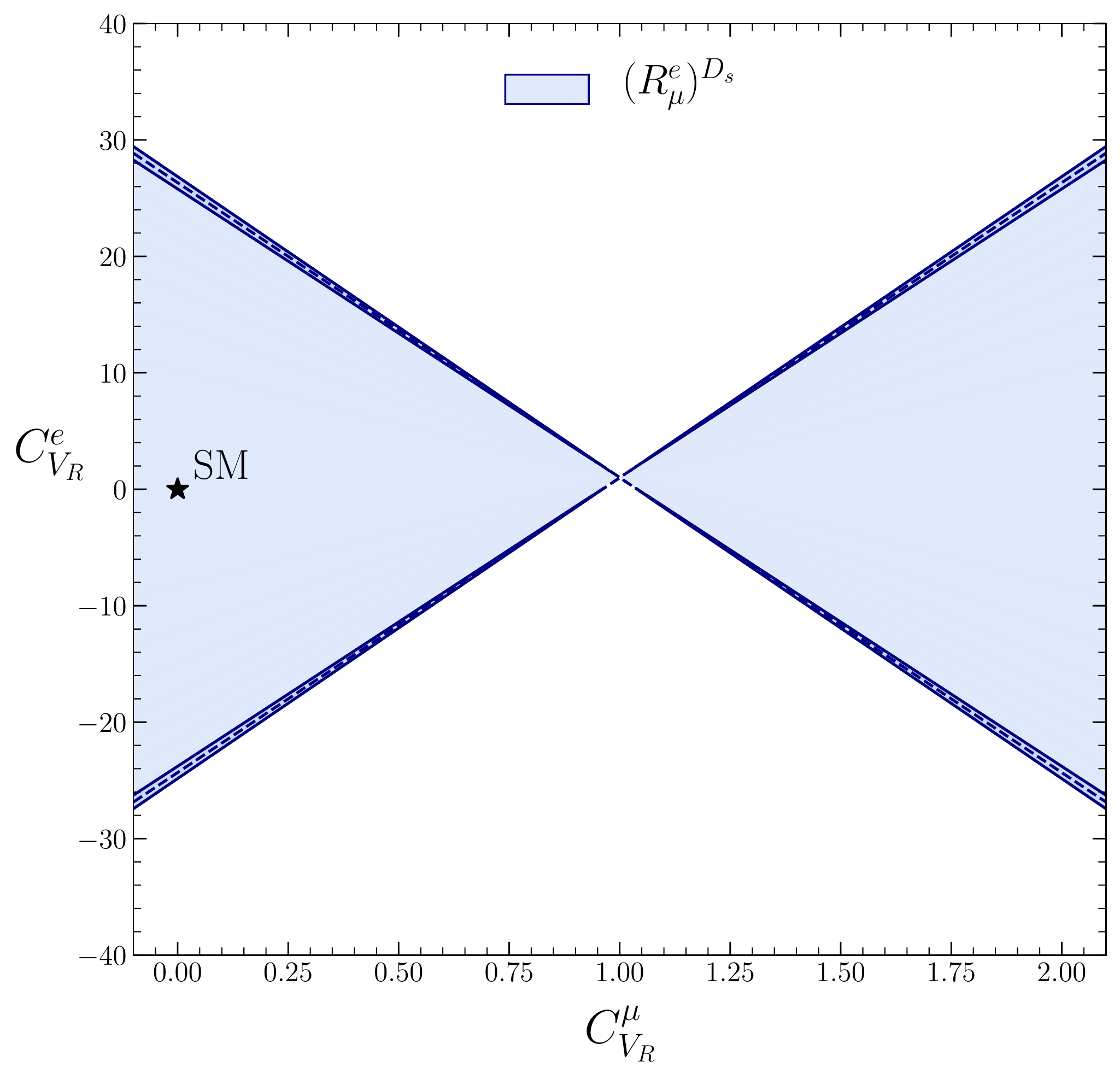}
\end{subfigure}
\caption{Allowed regions in the $C^{\mu}_{V_R}$--$C^{\tau}_{V_R}$ plane using the ratios $(R^{\tau}_{\mu})^{D}$ (top left) and $(R^{\tau}_{\mu})^{D_s}$ (bottom left), and in the $C^{\mu}_{V_R}$--$C^{e}_{V_R}$ plane using the ratios $(R^{e}_{\mu})^{D}$ (top right) and $(R^{e}_{\mu})^{D_s}$ (bottom right). The black stars indicate the SM predictions.}
\label{Fig_Lep_Constr_RHvecNP_D_Ds}
\end{figure}

We proceed our analysis by considering right-handed (RH) vector NP interactions. The branching fraction including RH vector contributions takes the following form:
\begin{equation}\label{D_to_lv_BF_NP_RHvec}
    \mathcal{B}(D_{(s)}^+ \rightarrow l^+ \nu_l)=\mathcal{B}(D_{(s)}^+ \rightarrow l^+ \nu_l)\big|_{\text{SM}} \Big|1 - C^{l}_{V_R} \Big|^2 ,
\end{equation}
where $C^{l}_{V_R}$ is the short-distance coefficient for the RH vector interaction. 
\newpage
\noindent Just as for the LH vector coefficients, we define the ratio of two leptonic branching fractions with different leptons in the final state:
\begin{equation}\label{Eq_Ratio_Lep_BFs_RHvec}
    R^{l_1}_{l_2} \equiv \frac{\mathcal{B}(D_{(s)}^+ \rightarrow l_1^+ \nu_{l_1})}{\mathcal{B}(D_{(s)}^+ \rightarrow l_2^+ \nu_{l_2})}   = \frac{\alpha^{l_1} \big|1 - C^{l_1}_{V_R} \big|^2 }{\alpha^{l_2}\big|1 - C^{l_2}_{V_R} \big|^2}.
\end{equation}
Using the experimental constraints, we obtain the allowed regions in the $C^{\mu}_{V_R}$--$C^{\tau}_{V_R}$ and 
$C^{\mu}_{V_R}$--$C^{e}_{V_R}$ planes shown in Fig.\ \ref{Fig_Lep_Constr_RHvecNP_D_Ds}.

\section{Semileptonic  Decays}\label{sec:semilep}
Semileptonic $D_{(s)}$ decays provide further powerful ways to constrain the short-distance NP coefficients. Since they are driven by the same $c \to d \bar{l} \nu_l$ and $c \to s  \bar{l} \nu_l$ quark transitions, these processes are effectively described by the same local operators. Consequently,  constraints on the short-distance coefficients coming from semileptonic decays can be utilized to complement those from leptonic decays. 
%
%
\subsection{\boldmath $D \to P \bar{l} \nu_l$ Decays}
First, we consider the semileptonic $D$ decays of the form $D \to P \bar{l} \nu_l$, where $P$ denotes a pseudoscalar meson. The differential branching fraction is given in the SM as follows \cite{Sakaki:2013bfa}:
\begin{equation}
\begin{split}
    \frac{d \mathcal{B}(D \rightarrow P \bar{l} \nu_l)}{d q^2 } \Bigg|_{\text{SM}} &= \frac{G_F^2 \tau_{D} |V_{cq}|^2}{24 \pi^3 M_{D}^2}  \Bigg[  \frac{(H^{P}_{V,0})^2}{4} \Big(1 + \frac{m_l^2}{2q^2} \Big) + \frac{3}{8} \frac{m_l^2}{q^2} (H^{P}_{V,s})^2   \Bigg]  \frac{(q^2 - m_l^2)^2}{q^2} |\vec{p}_{P}|,
\end{split}
\end{equation}
where $q^2$ is the four-momentum of the lepton-neutrino pair, satisfying the relation
\begin{equation}
m_l^2 \leq q^2 \leq (M_{D} - M_P)^2 .
\end{equation}
In order to calculate the  amplitudes $H^{P}_{V,0}$ and $H^{P}_{V,s}$, hadronic form factors are needed, requiring non-perturbative methods. In the literature, various form-factor parametrizations were proposed (for a more detailed discussion, see Appendix \ref{AppFFs}). In Table \ref{Tab_SM_BFsPseudo_Summary}, the SM branching fractions for semileptonic $D \to \pi$ and $D \to K$ decays with electrons or muons in the final states are given, along with their experimental counterparts. For the SM predictions, we used the values of  $|V_{cd}|$ and $|V_{cs}|$ in Eqs.~(\ref{Vcd-SM}) and (\ref{Vcs-SM}), respectively, obtained from the unitarity of the CKM 
matrix. Furthermore, we applied the lattice results given in Refs.\  \cite{Aubin:2004ej} and \cite{Lubicz:2017syv}, adopting double-pole and $z$-series parametrizations, respectively. As the latter approach results in smaller uncertainties for the branching fractions, we will 
use it in our analysis outlined below. For a studies using quark model caclulations, we refer the reader to Refs.~\cite{QuarkModel1,QuarkModel2}.

\begin{table}[t]
\renewcommand{\arraystretch}{1.2}
\begin{center}
\begin{tabular}{ cccc } 
 \hline
 Decay & Double-pole & $z$-series & Experiment  \\ 
 \hline
 \hline
  $\mathcal{B}(D^0 \rightarrow \pi^- e^+ \nu_{e})$ & $(3.21 \pm 0.68) \times 10^{-3}$ &$(2.64 \pm 0.31) \times 10^{-3}$& $(2.91 \pm 0.04)\times 10^{-3} $ \\ 
  $\mathcal{B}(D^0 \rightarrow K^- e^+ \nu_{e})$& $(3.64 \pm 0.76) \times 10^{-2}$ &$(3.49 \pm 0.29) \times 10^{-2}$& $(3.542 \pm 0.035)\times 10^{-2} \hspace{.6mm} $  \\ 
   $\mathcal{B}(D^+ \rightarrow \pi^0 e^+ \nu_{e})$ & $(4.17 \pm 0.88) \times 10^{-3}$ & $(3.42 \pm 0.41) \times 10^{-3}$ & $(3.72 \pm 0.17)\times 10^{-3} \hspace{4mm} $ \\ 
  $\mathcal{B}(D^+ \rightarrow \overline{K}^0 e^+ \nu_{e})$& $(9.31 \pm 1.95) \times 10^{-2}$ &$(8.92 \pm 0.75) \times 10^{-2}$& $(8.73 \pm 0.10)\times 10^{-2} \hspace{4mm} $ \\ 
 \hline
 $\mathcal{B}(D^0 \rightarrow \pi^- \mu^+ \nu_{\mu})$ & $(3.17  \pm 0.67) \times 10^{-3}$ &$(2.60 \pm 0.31) \times 10^{-3}$& $(2.67 \pm 0.12)\times 10^{-3} \hspace{4mm} $ \\ 
  $\mathcal{B}(D^0 \rightarrow K^- \mu^+ \nu_{\mu})$& $(3.56 \pm 0.74) \times 10^{-2}$ &$(3.40 \pm 0.29) \times 10^{-2}$& $(3.41 \pm 0.04)\times 10^{-2} \hspace{4mm} $  \\ 
   $\mathcal{B}(D^+ \rightarrow \pi^0 \mu^+ \nu_{\mu})$ & $(4.12 \pm 0.87) \times 10^{-3}$ & $(3.38 \pm 0.40) \times 10^{-3}$ & $(3.50 \pm 0.15)\times 10^{-3} \hspace{4mm} $ \\ 
  $\mathcal{B}(D^+ \rightarrow \overline{K}^0 \mu^+ \nu_{\mu})$& $(9.11  \pm 1.90) \times 10^{-2}$ &$(8.70 \pm 0.73) \times 10^{-2}$& $(8.76 \pm 0.19)\times 10^{-2} \hspace{4mm} $ \\ 
  \hline
\end{tabular}
\end{center}
\caption{Branching fractions for semileptonic $D$ 
decays calculated in the SM using double-pole \cite{Aubin:2004ej} and 
$z$-series \cite{Lubicz:2017syv} parametrizations, and comparison with the current experimental results given 
in Ref.\ \cite{Tanabashi:2018oca}.}\label{Tab_SM_BFsPseudo_Summary}
\end{table}

\subsubsection{Constraints on (Pseudo)-Scalar Coefficients}

\begin{figure}[h!]
\centering
\begin{subfigure}{.45\linewidth}
  \centering
  \includegraphics[width=.8\linewidth]{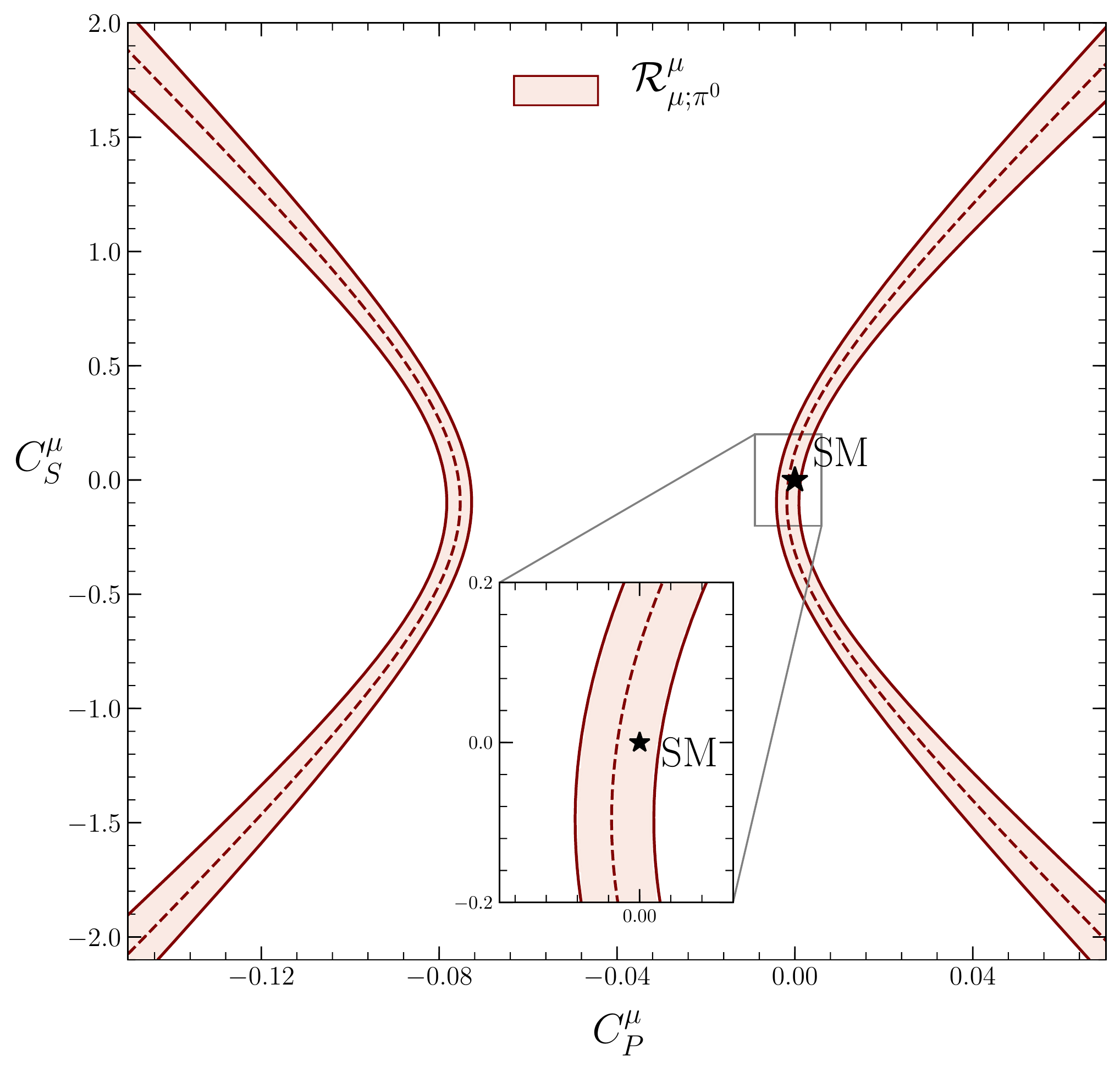}
\end{subfigure}
\begin{subfigure}{.45\linewidth}
  \centering
  \includegraphics[width=.8\linewidth]{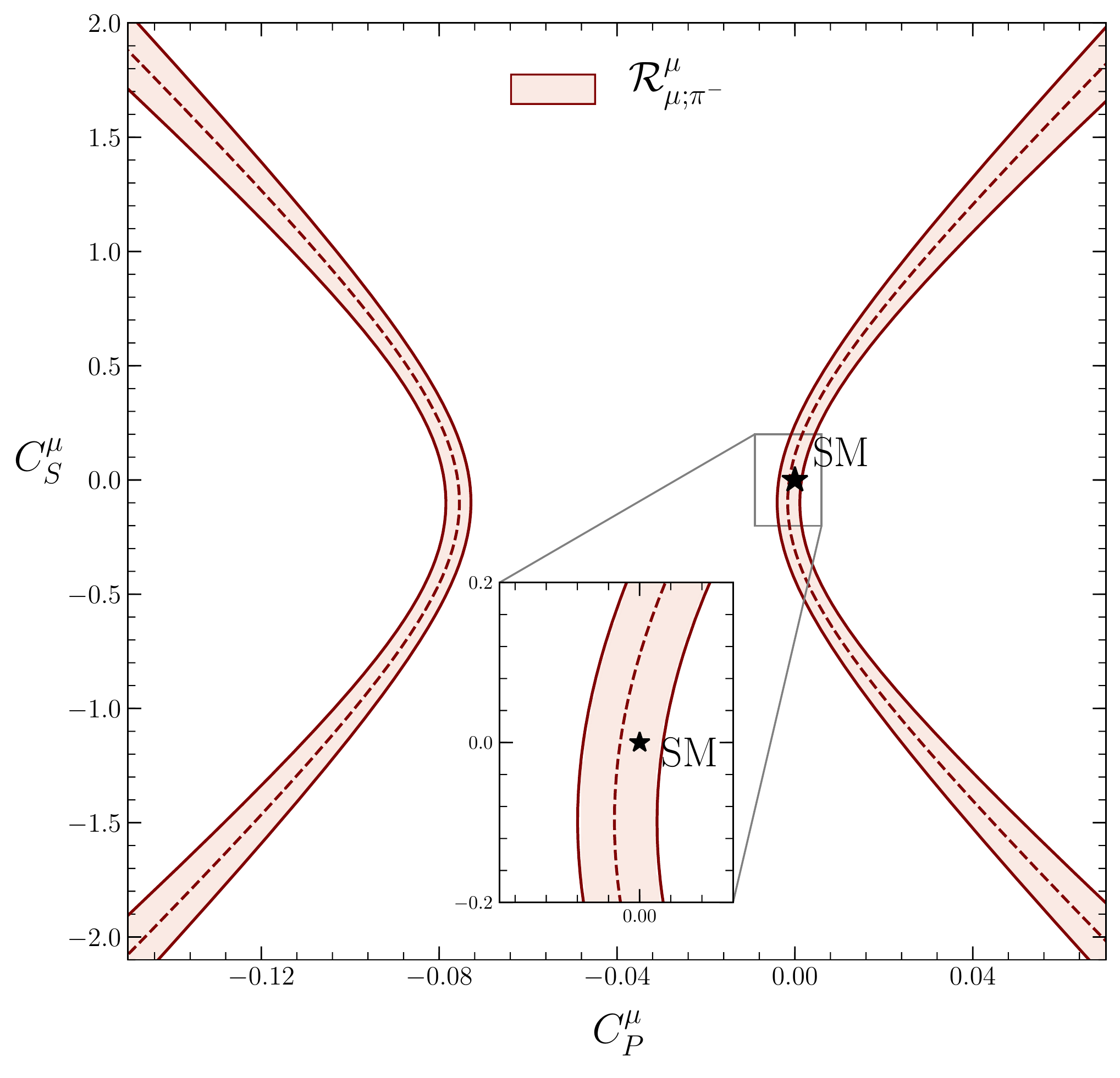}
\end{subfigure}
\caption{Allowed regions in the $C^{\mu}_P$--$C^{\mu}_S$ plane using the ratios $\mathcal{R}^{\mu}_{\mu ; \pi^0}$ (left) and $\mathcal{R}^{\mu}_{\mu ; \pi^-}$ (right).}
\label{Fig_LepSemilep_Constr_ScalPseuNP_Pi}
\end{figure}

\begin{figure}[H]
\centering
\begin{subfigure}{.45\linewidth}
  \centering
  \includegraphics[width=.8\linewidth]{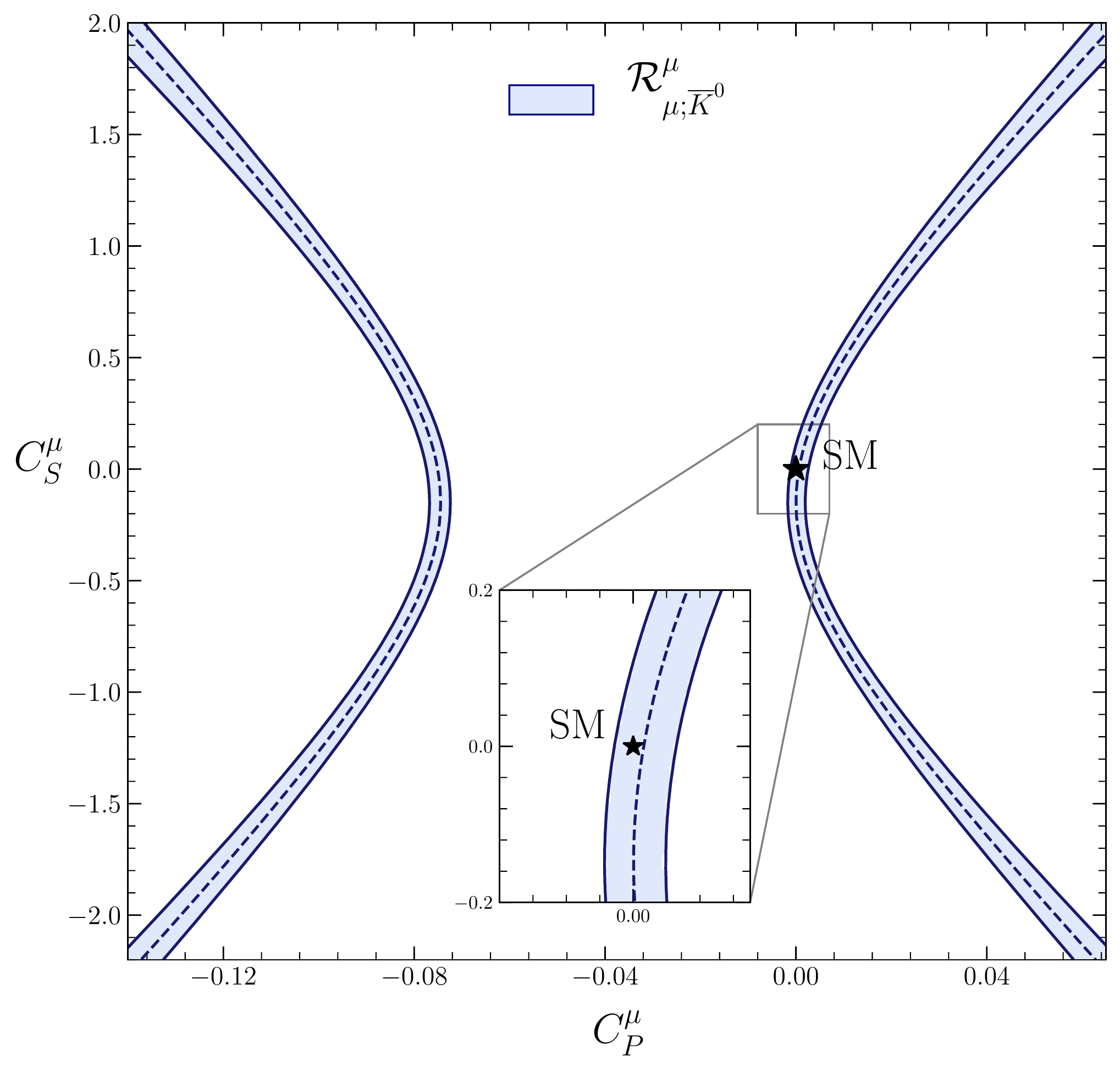}
\end{subfigure}
\begin{subfigure}{.45\linewidth}
  \centering
  \includegraphics[width=.8\linewidth]{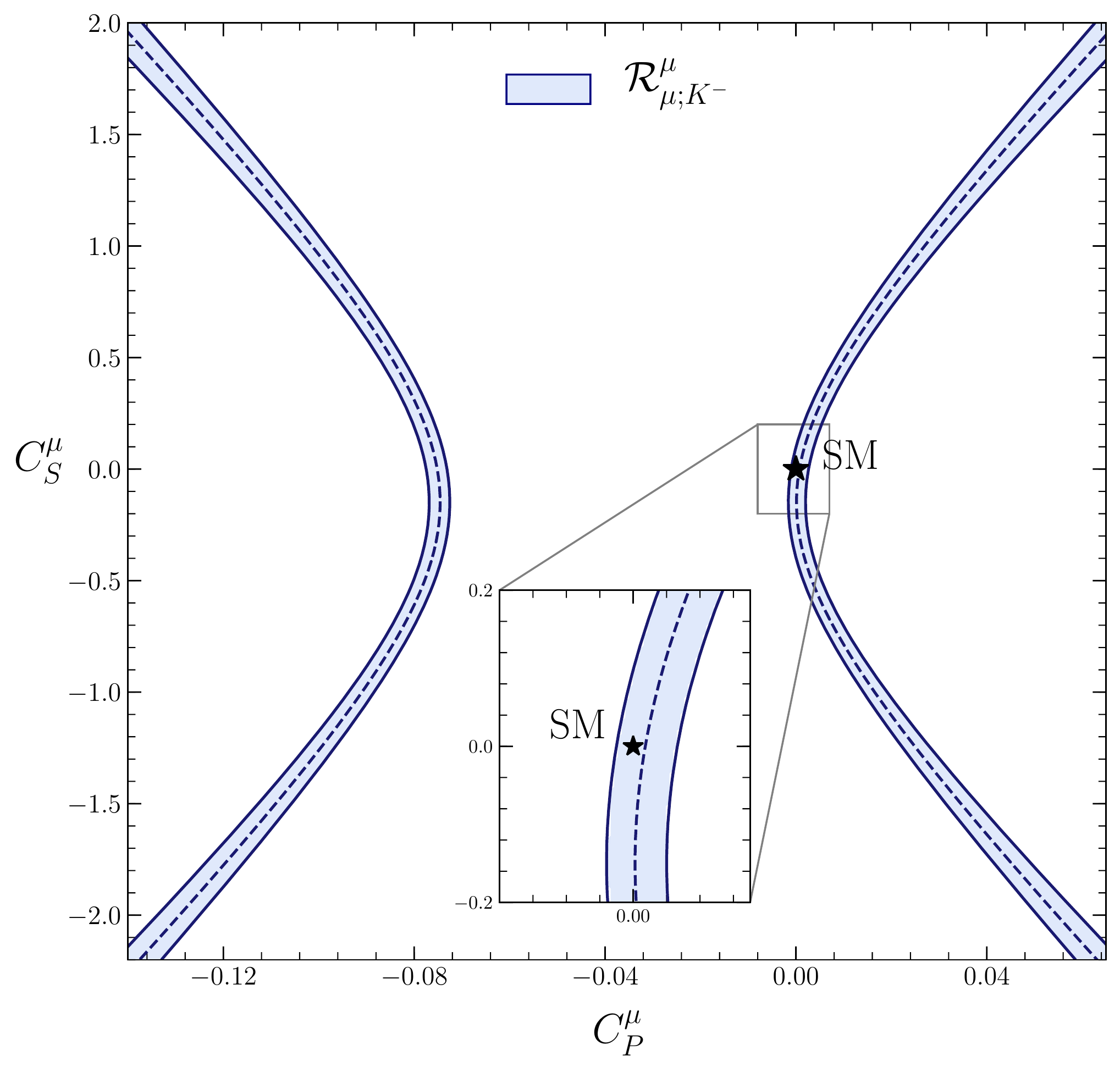}
\end{subfigure}
\caption{Allowed regions in the $C^{\mu}_P$--$C^{\mu}_S$ plane using the ratios $\mathcal{R}^{\mu}_{\mu ; \overline{K}^0}$ (left) and $\mathcal{R}^{\mu}_{\mu ; K^-}$ (right).}
\label{Fig_LepSemilep_Constr_ScalPseuNP_K}
\end{figure}

Allowing for scalar NP interactions, the differential branching fraction for semileptonic $D$ decays into a pseudoscalar meson and a lepton--neutrino pair takes the following form:
\begin{equation}\label{Eq_Semilep_P_BF_ScalNP}
\begin{split}
    \frac{d \mathcal{B}(D \rightarrow P \bar{l} \nu_l)}{d q^2 } &= \frac{G_F^2 \tau_D |V_{cq}|^2}{24 \pi^3 M_D^2}  \Bigg\{\Bigg[  \frac{(H^{P}_{V,0})^2}{4} \Big(1 + \frac{m_l^2}{2q^2} \Big) + \frac{3}{8} \frac{m_l^2}{q^2} (H^{P}_{V,t})^2   \Bigg]  \\
    &\hspace{15mm} + \frac{3}{8} |C_S^l |^2 (H^{P}_{S})^2 + \frac{3}{4} \mathcal{R}e(C_S^{l*} ) \frac{m_l}{\sqrt{q^2}} H^{P}_{S}H^{P}_{V,t} \Bigg\}  \frac{(q^2 - m_l^2)^2}{q^2} |\vec{p}_{P}| ,
\end{split}
\end{equation}
where the term in the square brackets is the SM part. The other terms, containing the NP contributions, are sensitive to the scalar  coefficient $C_S^l$. In order to obtain an observable that is independent of $V_{cd}$ or $V_{cs}$, we consider the ratio between a leptonic $D_{(s)}$ decay and a semileptonic $D$ decay containing the same quark transition and lepton flavour in the final state. Semileptonic $D$ decays with tau leptons are kinematically forbidden, and for leptonic decays to electrons only an experimental upper bound is available. Therefore, we restrict ourselves to decays with muons in the final state and define the following ratios:
\begin{equation}\label{Eq_Ratio_Lep_Semilep_Pseu_pseuNP_Pi}
\mathcal{R}^{\mu}_{\mu;\pi^0} \equiv \frac{\mathcal{B}(D^+ \to \mu^+ \nu_{\mu})}{ \mathcal{B}(D^+ \to \pi^0 \mu^+ \nu_{\mu} )} , \hspace{10mm} \mathcal{R}^{\mu}_{\mu;\pi^-} \equiv \frac{\mathcal{B}(D^+ \to \mu^+ \nu_{\mu})}{ \mathcal{B}(D^0 \to \pi^- \mu^+ \nu_{\mu} )} , 
\end{equation}
and
\begin{equation}\label{Eq_Ratio_Lep_Semilep_Pseu_pseuNP_K}
\mathcal{R}^{\mu}_{\mu;\overline{K}^0} \equiv \frac{\mathcal{B}(D_s^+ \to \mu^+ \nu_{\mu})}{ \mathcal{B}(D^+ \to \overline{K}^0 \mu^+ \nu_{\mu} )} ,  \hspace{10mm} \mathcal{R}^{\mu}_{\mu;K^-} \equiv \frac{\mathcal{B}(D_s^+ \to \mu^+ \nu_{\mu})}{ \mathcal{B}(D^0 \to K^- \mu^+ \nu_{\mu} )} .
\end{equation}
These observables are sensitive to the scalar and pseudoscalar coefficients $C^l_S$ and $C^l_P$. For the $D^+ \to \pi^0 \mu^+ \nu_{\mu}$ decay, an isospin factor of 1/2 has to be taken into account since $\pi^0 = (u\bar{u} - d\bar{d})/\sqrt{2}$. From the measured branching fractions given in Table \ref{Tab_SM_BFsPseudo_Summary}, we obtain the following experimental values:
\begin{equation}\label{Eq_Ratio_Lep_Semilep_Pseu_pseuNP_Pi_Exp}
\mathcal{R}^{\mu}_{\mu;\pi^0} = (1.07 \pm 0.07) \times 10^{-1} , \hspace{10mm} \mathcal{R}^{\mu}_{\mu;\pi^-}  = (1.40 \pm 0.09) \times 10^{-1} , 
\end{equation}
and
\begin{equation}\label{Eq_Ratio_Lep_Semilep_Pseu_pseuNP_K_Exp}
\mathcal{R}^{\mu}_{\mu;\overline{K}^0} = (6.28 \pm 0.30) \times 10^{-2} ,  \hspace{10mm} \mathcal{R}^{\mu}_{\mu;K^-}= (1.61 \pm 0.07) \times 10^{-1} .
\end{equation}
By comparing the theoretical expressions for the ratios in Eqs.\ (\ref{Eq_Ratio_Lep_Semilep_Pseu_pseuNP_Pi}) and (\ref{Eq_Ratio_Lep_Semilep_Pseu_pseuNP_K}) with the corresponding experimental values, we constrain the (pseudo)-scalar coefficients. From $\mathcal{R}^{\mu}_{\mu;\pi^0}$ and $\mathcal{R}^{\mu}_{\mu;\pi^-}$, we obtain the allowed regions in the $C_P^{\mu}$--$C_S^{\mu}$ plane shown in Fig. \ref{Fig_LepSemilep_Constr_ScalPseuNP_Pi}. The results are in agreement with the SM prediction $C_P^{\mu}$ = $C_S^{\mu}$ = 0 at the 1$\sigma$ level. This is also the case for the allowed regions obtained from $\mathcal{R}^{\mu}_{\mu;\overline{K}^0}$ and $\mathcal{R}^{\mu}_{\mu;K^-}$, shown in Fig. \ref{Fig_LepSemilep_Constr_ScalPseuNP_K}. 


\subsubsection{Constraints on Vector Coefficients}

Next, we consider contributions from LH  and RH vector NP interactions. The differential branching fraction can be written as follows:
\begin{equation}\label{D_to_P_lv_BF_NP_vec}
\begin{split}
    \frac{d \mathcal{B}(D \rightarrow P \bar{l} \nu_l)}{d q^2 } &= \frac{d \mathcal{B}(D \rightarrow P \bar{l} \nu_l)}{d q^2 } \Bigg|_{\text{SM}}   \Big| 1 + C^l_{V_{L(R)}}\Big|^2 .
\end{split}
\end{equation}
In principle, we could define ratios in analogy to Eqs.\ (\ref{Eq_Ratio_Lep_Semilep_Pseu_pseuNP_Pi}) and (\ref{Eq_Ratio_Lep_Semilep_Pseu_pseuNP_K}) given in the previous section, but include vector contributions instead of (pseudo)-scalar ones. However, Eqs.\ (\ref{Eq_BF_Lep_LHvecNP}) and  (\ref{D_to_P_lv_BF_NP_vec}) indicate that the LH vector NP contributions would cancel. On the other hand, in the case of RH vector contributions, they would not cancel due to the relative sign difference, but the structure of the formulae does prohibit the extraction of stringent constraints. 

\begin{figure}[h!]
\centering
\begin{subfigure}{.45\linewidth}
  \centering
  \includegraphics[width=.8\linewidth]{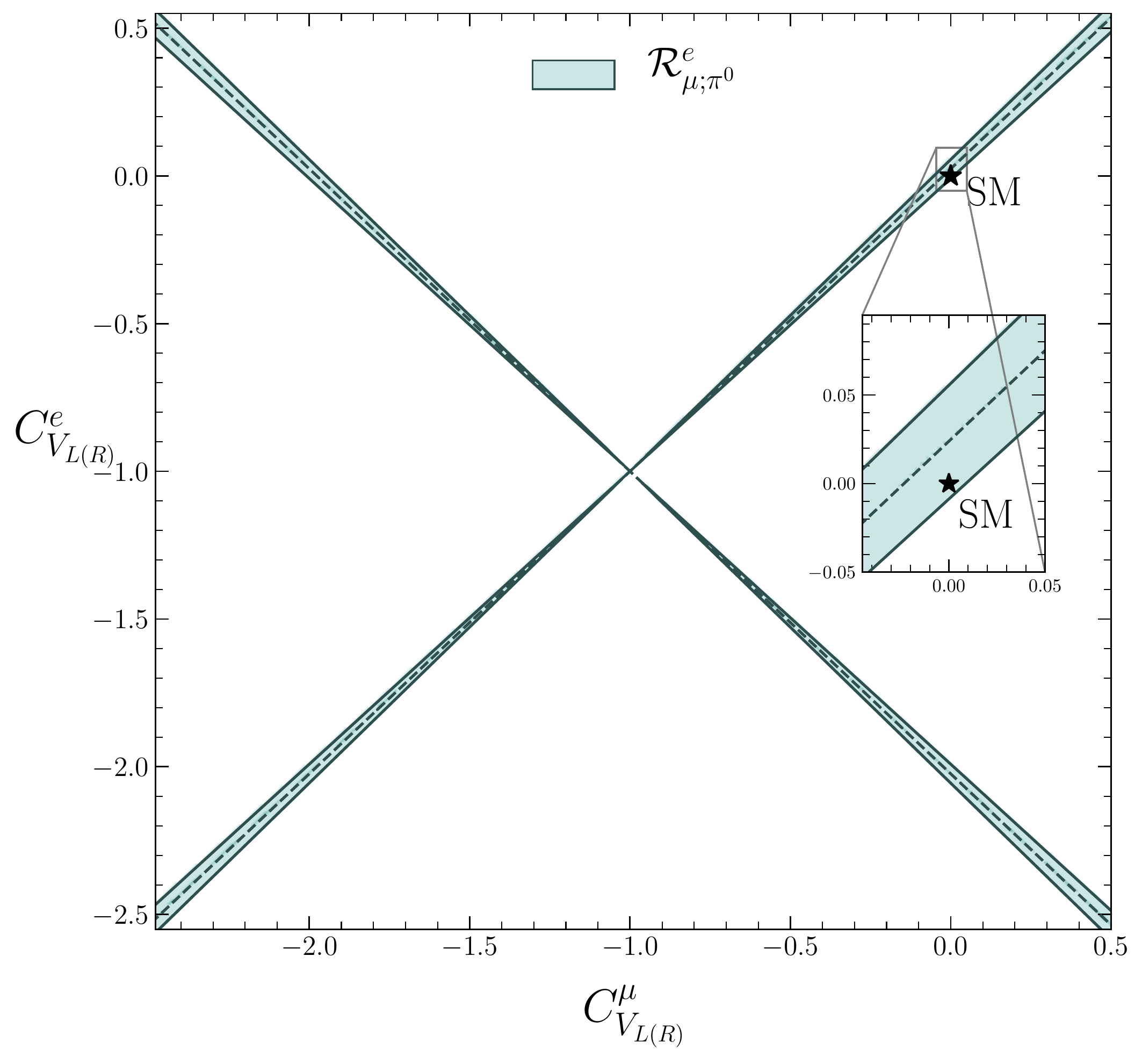}
\end{subfigure}
\begin{subfigure}{.45\linewidth}
  \centering
  \includegraphics[width=.8\linewidth]{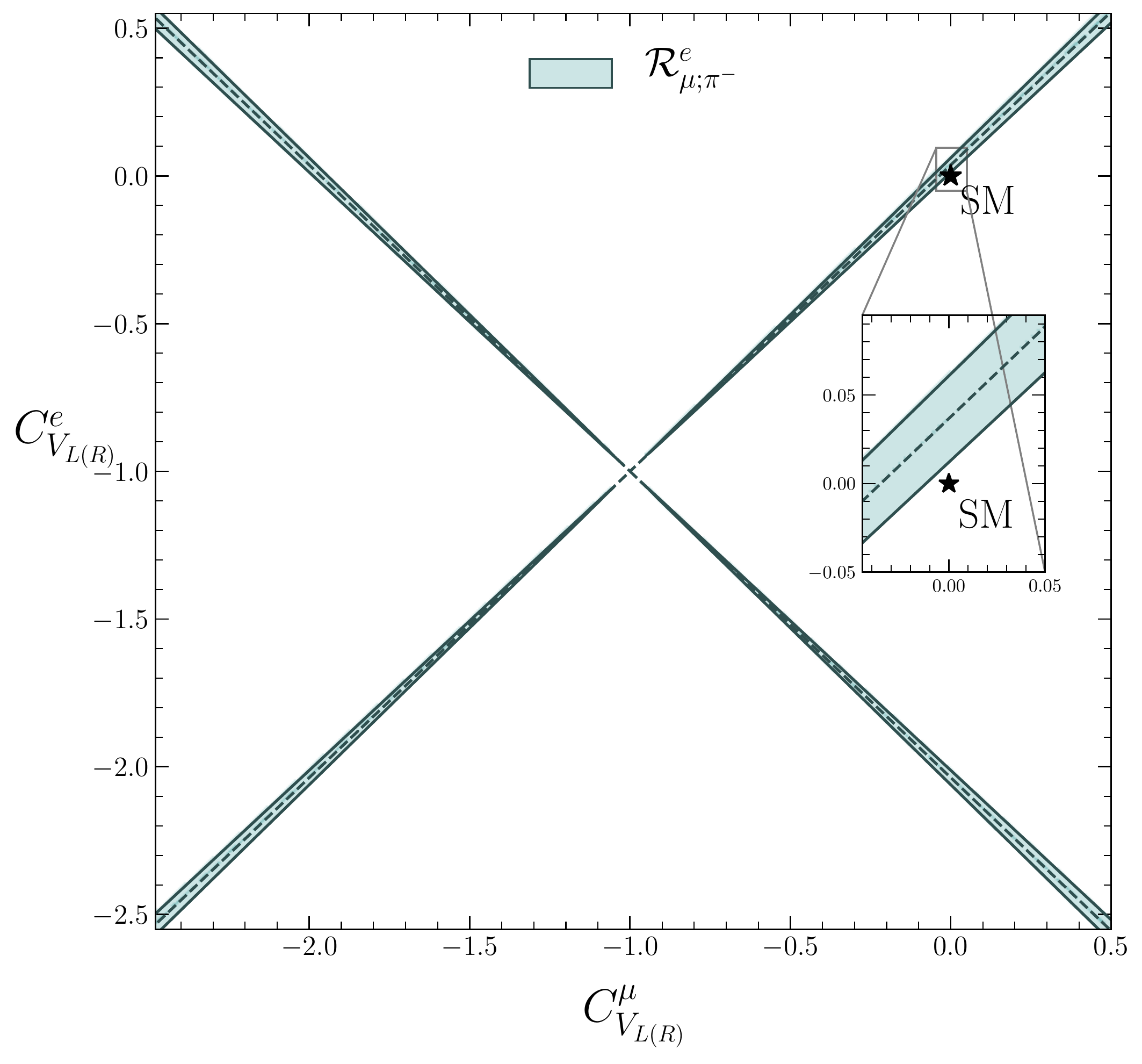}
\end{subfigure}
\caption{Allowed regions in the $C^{\mu}_{V_{L(R)}}$--$C^{e}_{V_{L(R)}}$ plane using the ratios $\mathcal{R}^{\mu}_{\mu ; \pi^0}$ (left) and $\mathcal{R}^{\mu}_{\mu ; \pi^-}$ (right).}
\label{Fig_SemilepPseu_Constr_VecNP_Pi}
\end{figure}

\begin{figure}[]
\centering
\begin{subfigure}{.45\linewidth}
  \centering
  \includegraphics[width=.8\linewidth]{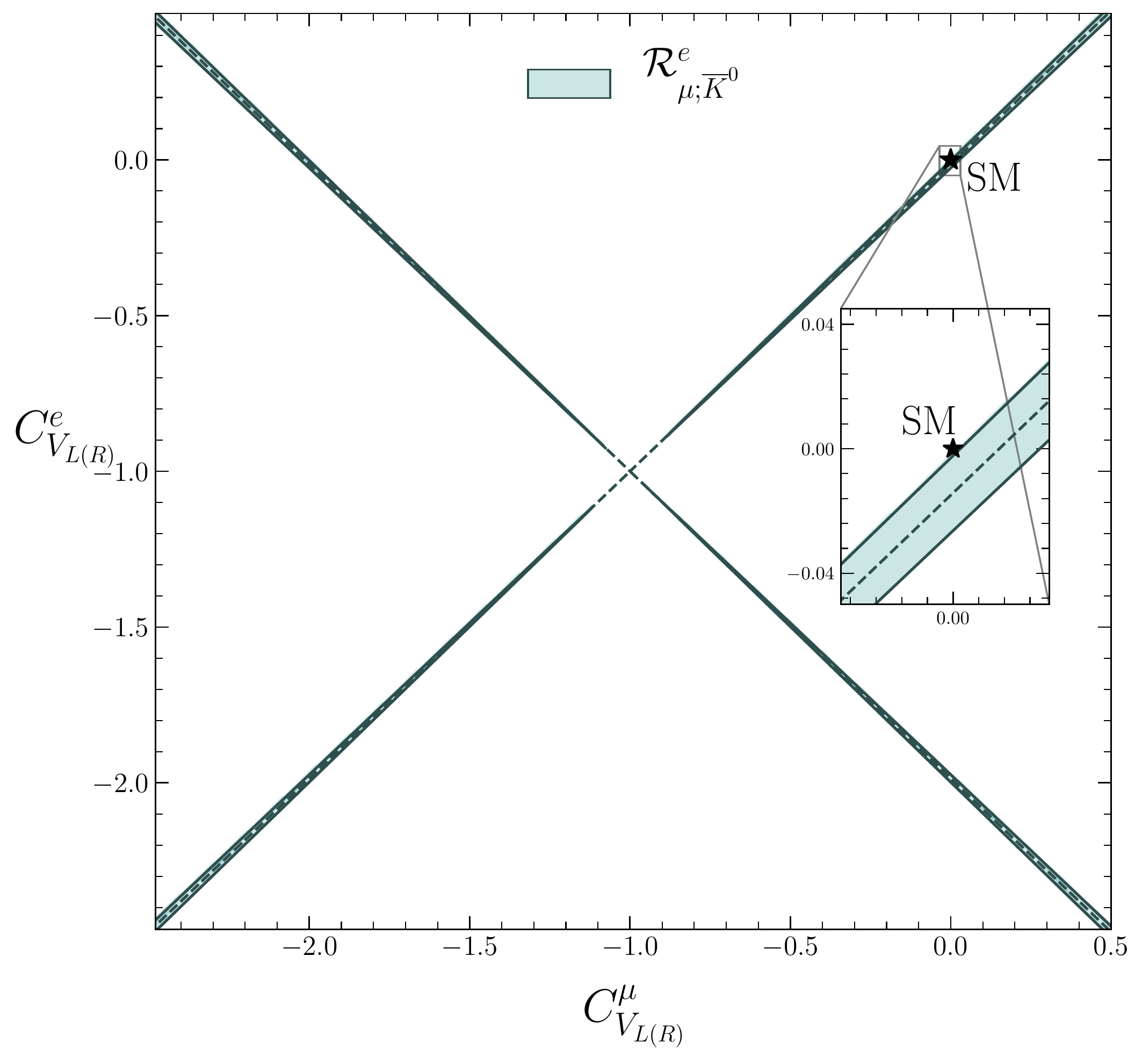}
\end{subfigure}
\begin{subfigure}{.45\linewidth}
  \centering
  \includegraphics[width=.8\linewidth]{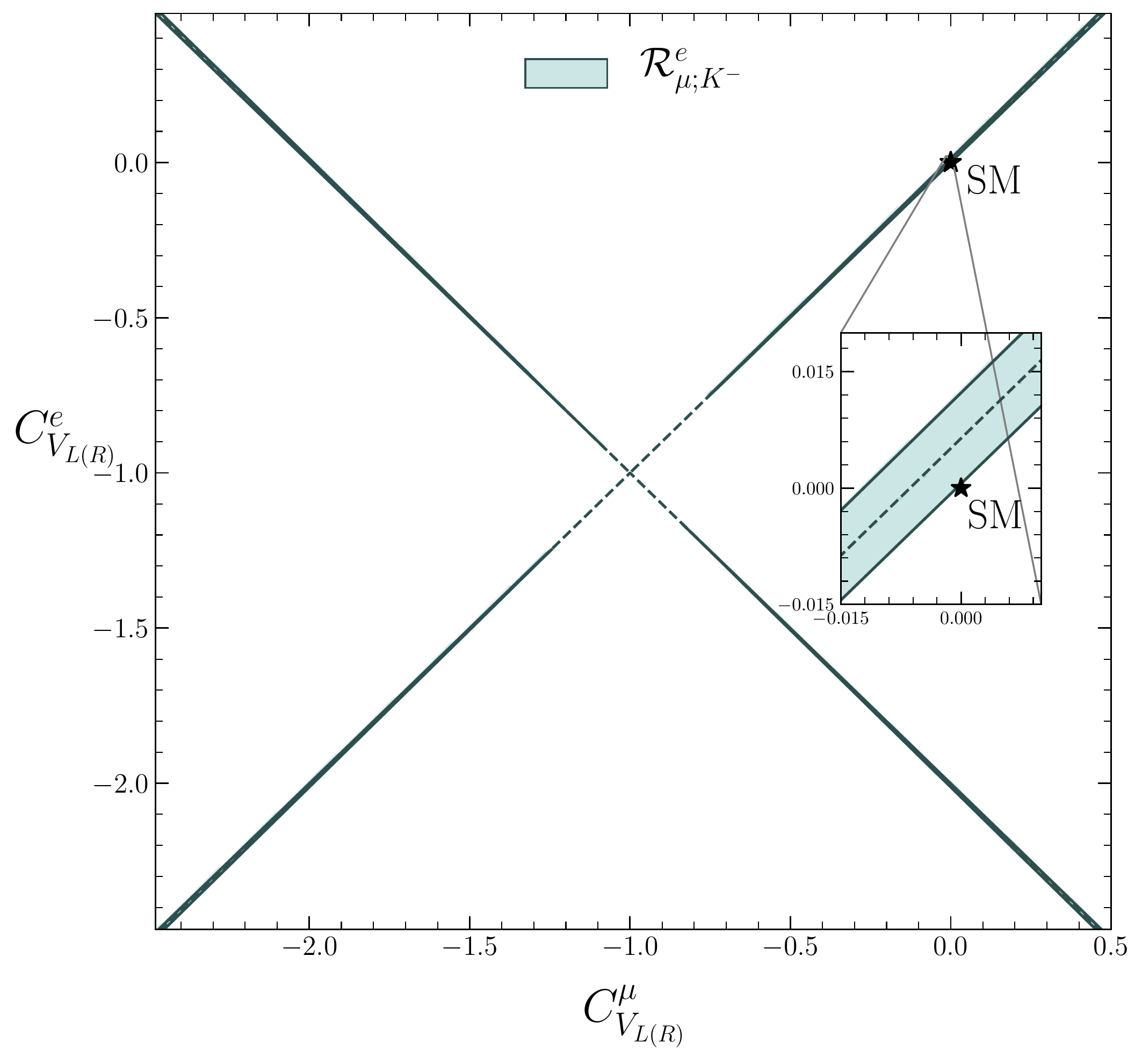}
\end{subfigure}
\caption{Allowed regions in the $C^{\mu}_{V_{L(R)}}$--$C^{e}_{V_{L(R)}}$ plane using the ratios $\mathcal{R}^{\mu}_{\mu ; \overline{K}^0}$ (left) and $\mathcal{R}^{\mu}_{\mu ; K^-}$ (right).}
\label{Fig_SemilepPseu_Constr_VecNP_K}
\end{figure}

We can, however, investigate possible vector NP interactions through ratios between two semileptonic decays with different flavours of leptons. To this end, we define the following observables:
\begin{equation}\label{Eq_Remu_Pseu_VecNP_Pi}
\mathcal{R}^{e}_{\mu;\pi^0} \equiv \frac{\mathcal{B}(D^+ \to \pi^0 e^+ \nu_{e} )}{ \mathcal{B}(D^+ \to \pi^0 \mu^+ \nu_{\mu} )} , \hspace{10mm} \mathcal{R}^{e}_{\mu;\pi^-} \equiv \frac{\mathcal{B}(D^0 \to \pi^- e^+ \nu_{e} )}{ \mathcal{B}(D^0 \to \pi^- \mu^+ \nu_{\mu} )} , 
\end{equation}
and
\begin{equation}\label{Eq_Remu_Pseu_VecNP_K}
\mathcal{R}^{e}_{\mu;\overline{K}^0} \equiv \frac{\mathcal{B}(D^+ \to \overline{K}^0 e^+ \nu_{e} )}{ \mathcal{B}(D^+ \to \overline{K}^0 \mu^+ \nu_{\mu} )} ,  \hspace{10mm} \mathcal{R}^{e}_{\mu;K^-} \equiv \frac{\mathcal{B}(D^0 \to K^- e^+ \nu_{e} )}{ \mathcal{B}(D^0 \to K^- \mu^+ \nu_{\mu} )} .
\end{equation}
These observables are independent of CKM matrix elements and sensitive to the coefficients $C^{l}_{V_{L(R)}}$, with $l =e ,\mu$. Hence, they provide interesting opportunities to test lepton flavour universality in $D$ decays. From the measured branching fractions in Table \ref{Tab_SM_BFsPseudo_Summary}, we obtain the following experimental values:
\begin{equation}\label{Eq_Remu_Pseu_VecNP_Pi_Exp}
\mathcal{R}^{e}_{\mu;\pi^0} = 1.06 \pm 0.07 , \hspace{10mm} \mathcal{R}^{e}_{\mu;\pi^-}  = 1.09 \pm 0.05 , 
\end{equation}
and
\begin{equation}\label{Eq_Remu_Pseu_VecNP_K_Exp}
\mathcal{R}^{e}_{\mu;\overline{K}^0} = (9.97 \pm 0.24)\times 10^{-1},  \hspace{10mm} \mathcal{R}^{e}_{\mu;K^-}= 1.04 \pm 0.01 .
\end{equation}
The structure of the theoretical expressions for these observables, following from Eq.\ (\ref{D_to_P_lv_BF_NP_vec}), is identical for LH and RH vector NP interactions. Therefore, the allowed regions in the $C^{\mu}_{V_{L}}$--$C^{e}_{V_{L}}$ and 
$C^{\mu}_{V_{R}}$--$C^{e}_{V_{R}}$ planes, obtained by comparing these expression with the experimental values, are the same. This is can be seen in Figs.\ \ref{Fig_SemilepPseu_Constr_VecNP_Pi} and \ref{Fig_SemilepPseu_Constr_VecNP_K}. Furthermore, as leptonic and semileptonic decays are described by the same operators, we may compare these constraints with the ones obtained from leptonic decays in Fig.\ \ref{Fig_Lep_Constr_LHvecNP_D_Ds} (right) and Fig.\ \ref{Fig_Lep_Constr_RHvecNP_D_Ds} (right). The constraints presented there are a significant improvement, resulting from the fact that semileptonic decays do not suffer from helicity suppression. It is interesting to note that the SM predictions, $C^{\mu}_{V_{L(R)}} =  C^{e}_{V_{L(R)}} = 0$, lie just outside the 1$\sigma$ contours in three of the four plots. 

\subsubsection{Constraints on Tensor Coefficients}
Finally, let us probe the tensor operator in the operator basis given in Eq.~(\ref{Eq_Operator_Basis}) through its impact on  the semileptonic decays. The differential branching fraction including such tensor interactions is given as
\begin{equation}\label{Eq_D_to_P_lv_BF_tensorNP}
\begin{split}
    \frac{d \mathcal{B}(D \rightarrow P \bar{l} \nu_l)}{d q^2 } &= \frac{G_F^2 \tau_D |V_{cq}|^2}{24 \pi^3 M_D^2}  \Bigg\{\Bigg[  \frac{(H^{P}_{V,0})^2}{4} \Big(1 + \frac{m_l^2}{2q^2} \Big) + \frac{3}{8} \frac{m_l^2}{q^2} (H^{P}_{V,t})^2   \Bigg]  \\
    &\hspace{5mm} + 2 |C_T^l |^2 \Big(1 + \frac{2m_l^2}{q^2} \Big) (H^{P}_{T})^2 -3 \mathcal{R}e(C_T^{l*} ) \frac{m_l}{\sqrt{q^2}} H^{P}_{T}H^{P}_{V,0} \Bigg\}  \frac{(q^2 - m_l^2)^2}{q^2} |\vec{p}_{P}| .
\end{split}
\end{equation}
For the amplitude $H^P_T$, we use the lattice results obtained in Ref.\ \cite{Lubicz:2018rfs} for the corresponding form factors (for details, see Appendix \ref{AppFFs}). This leads to  expressions for the semileptonic branching fractions, dependent on the tensor NP coefficients $C^l_T$. As the tensor operator in Eq.\ (\ref{Eq_Operator_Basis}) is antisymmetric in $\mu$ and $\nu$, there are no tensor contributions to leptonic decays. Consequently, we cannot probe tensor NP through ratios between leptonic and semileptonic decays. Hence, we take the ratios in Eqs.\ (\ref{Eq_Remu_Pseu_VecNP_Pi}) and  (\ref{Eq_Remu_Pseu_VecNP_K}), but allow for tensor contributions instead of vector contributions. These ratios are then sensitive to the tensor coefficients $C^{e}_T$ and $C^{\mu}_T$. 

\begin{figure}[]
\centering
\begin{subfigure}{.45\linewidth}
  \centering
  \includegraphics[width=.8\linewidth]{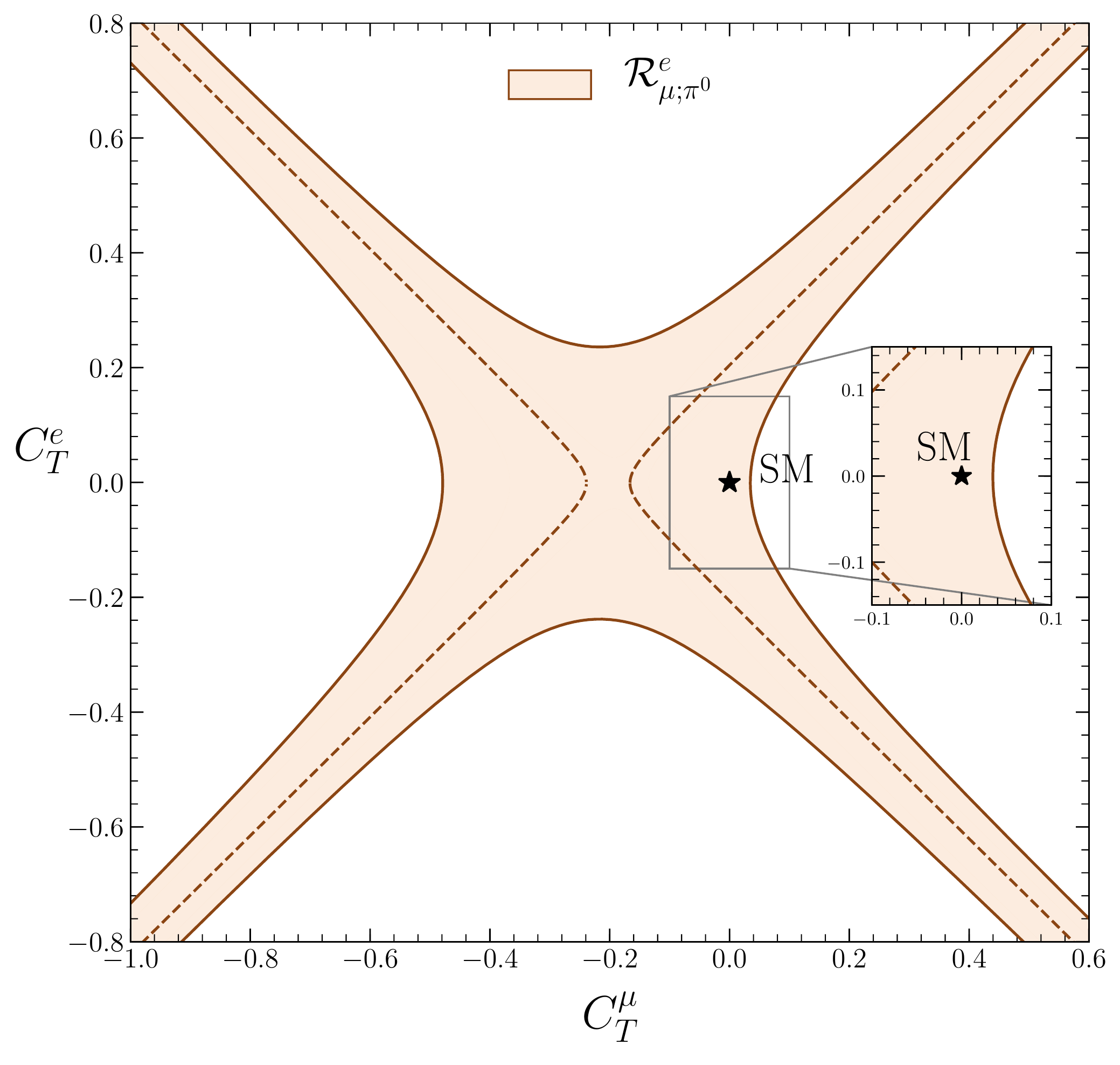}
\end{subfigure}
\begin{subfigure}{.45\linewidth}
  \centering
  \includegraphics[width=.8\linewidth]{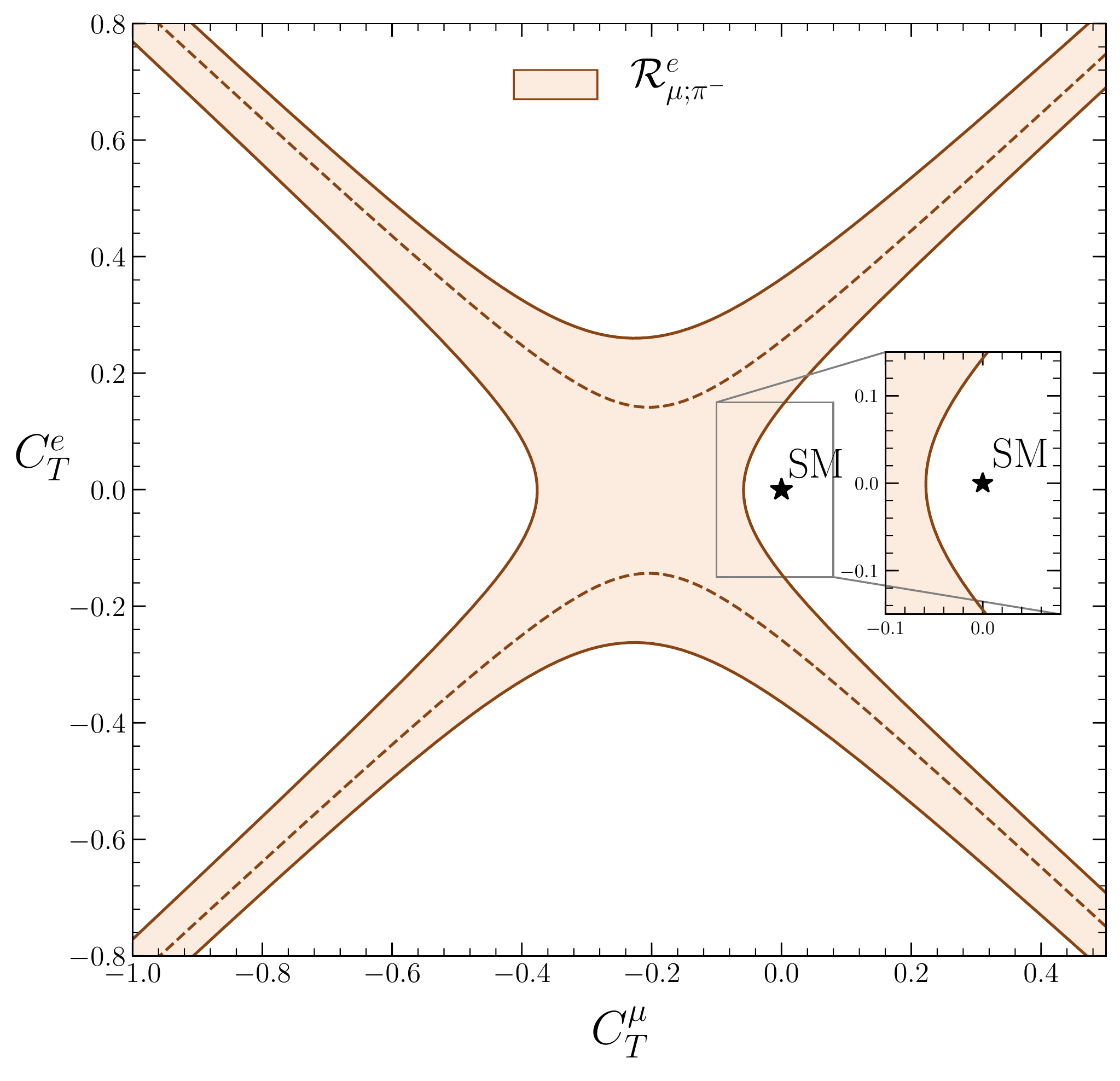}
  \end{subfigure}
\caption{Allowed regions in the $C^{\mu}_{T}- C^{e}_{T}$ plane using the ratios $\mathcal{R}^{e}_{\mu;\pi^0}$ (left) and $\mathcal{R}^{e}_{\mu ; \pi^-}$ (right).}
\label{Fig_Tensor_Pseu_Pi_Constr}
\end{figure}

\begin{figure}[]
\centering
\begin{subfigure}{.45\linewidth}
  \centering
  \includegraphics[width=.8\linewidth]{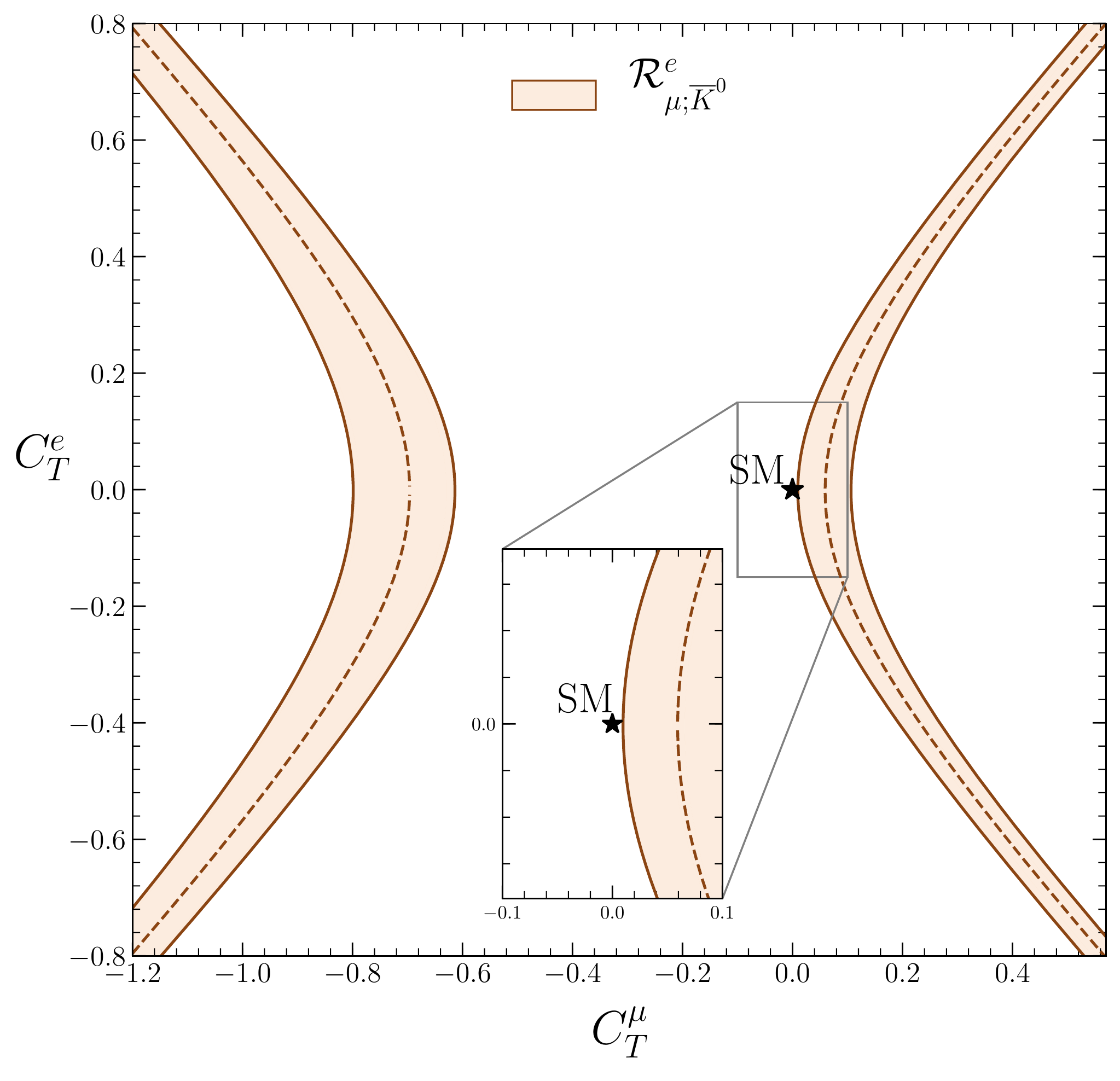}
\end{subfigure}%
\begin{subfigure}{.45\linewidth}
  \centering
  \includegraphics[width=.8\linewidth]{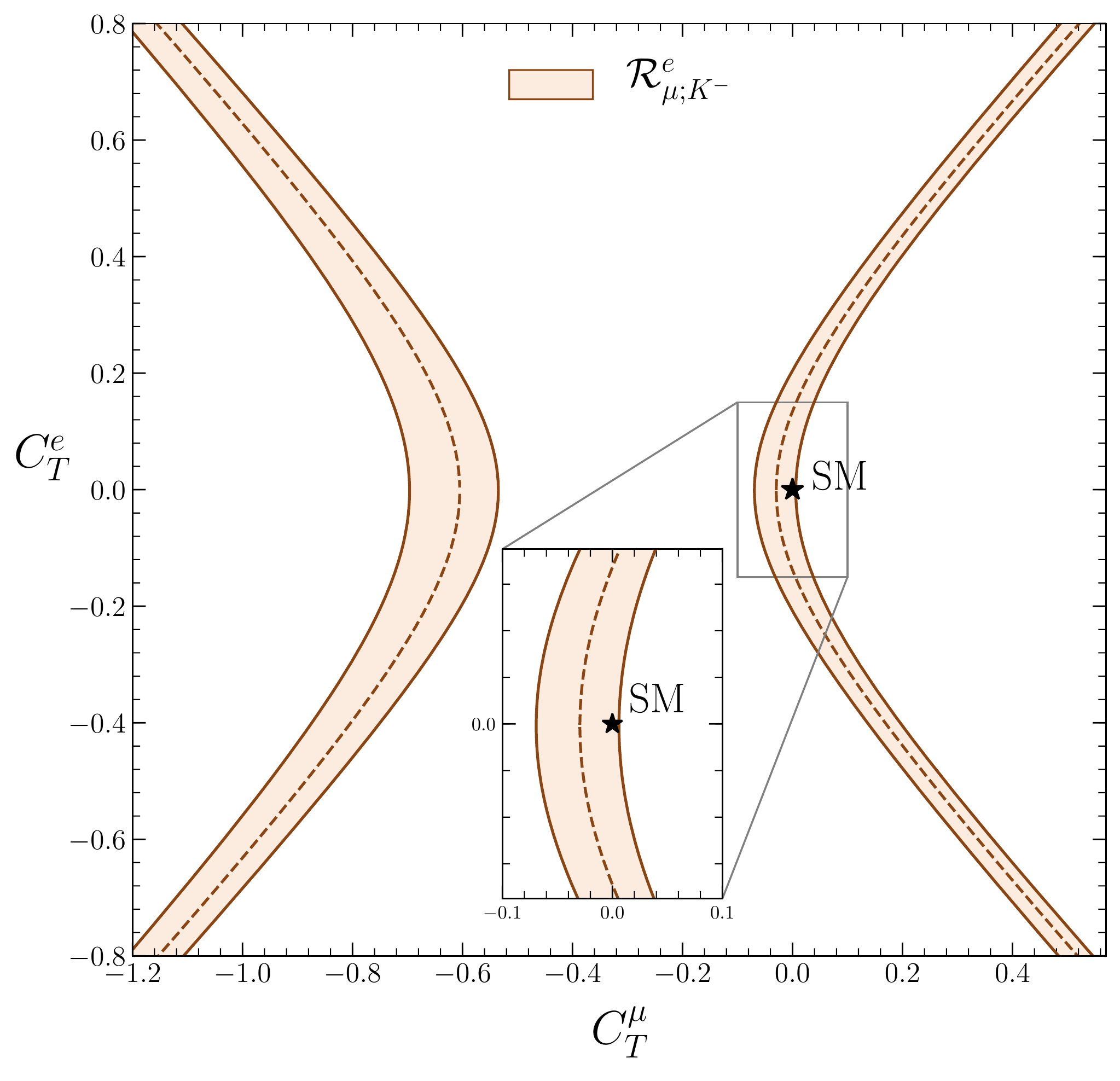}
  \end{subfigure}
\caption{Allowed regions in the $C^{\mu}_{T}- C^{e}_{T}$ plane using the ratios $\mathcal{R}^{e}_{\mu;\overline{K}^0}$ (left) and $\mathcal{R}^{e}_{\mu ; K^-}$ (right).}
\label{Fig_Tensor_Pseu_K_Constr}
\end{figure}

By comparing our theoretical expressions with the experimental information in Eqs.\ (\ref{Eq_Remu_Pseu_VecNP_Pi_Exp}) and 
(\ref{Eq_Remu_Pseu_VecNP_K_Exp}), we determine the allowed regions in the $C^{\mu}_{T}$--$C^{e}_{T}$ plane. The constraints from $\mathcal{R}^{e}_{\mu;\pi^0} $ and $\mathcal{R}^{e}_{\mu;\pi^-} $ are shown in the panels on the left- and right-hand sides of
Fig.~\ref{Fig_Tensor_Pseu_Pi_Constr}, respectively. In the latter case, it is interesting to see that the SM prediction corresponding to $C^{\mu}_{T}=C^{e}_{T} = 0$ is excluded at the 1\,$\sigma$ level. In Fig.\ \ref{Fig_Tensor_Pseu_K_Constr}, the allowed regions in the $C^{\mu}_{T}$--$C^{e}_{T}$ plane are shown, following from the constraints on $\mathcal{R}^{e}_{\mu;\overline{K}^0} $ (left) and $\mathcal{R}^{e}_{\mu;K^-} $ (right). We observe that in the case of $\mathcal{R}^{e}_{\mu;\overline{K}^0} $, the SM prediction lies just outside the 1\,$\sigma$ contours.

\subsection{\boldmath $D \to V \bar{l} \nu_l$ Decays} 

We continue our analysis of semileptonic $D_{(s)}$ decays by considering decays of the form $D \to V \bar{l} \nu_l$, where $V$ denotes a vector meson. Measurements of the branching fractions for $D \to \rho$ and $D \to K^*$ decays allow us to further constrain the short-distance NP coefficients. In particular, $D \to V$ decays are sensitive to the pseudoscalar NP coefficient $C_P^l$, thereby offering an interesting complement to the constraints following from the leptonic decays in Section~\ref{sec:lept}.

The  SM expression for the differential branching fraction for $D \rightarrow V \bar{l} \nu_l$ decays takes the form \cite{Sakaki:2013bfa}
\begin{equation}
\begin{split}
\frac{d \mathcal{B}(D \rightarrow V \bar{l} \nu_l)}{dq^2} &= \frac{G_F^2 \tau_{D} |V_{cq}|^2}{24 \pi^3 M^2_{D}} \Bigg\{  \frac{1}{4} \Big( 1 + \frac{m_l^2}{2q^2} \Big) \Big[ (H^{V}_{V,+})^2 + (H^{V}_{V,-})^2 + (H^{V}_{V,0})^2 \Big] \\
& \hspace{25mm} + \frac{3}{8}\frac{m_l^2}{q^2} (H^{V}_{V,t})^2   \Bigg\} \frac{(q^2 - m_l^2)^2}{q^2}|\vec{p}_{V}|,
\end{split}
\end{equation}
where  $q^2$ has the kinematical range 
\begin{equation}
m_l^2 \leq q^2 \leq (M_{D} - M_V)^2 .
\end{equation}
In comparison with the pseudoscalar case, we have now a considerably more complex situation due to the amplitudes $H^{V}_{V,+}$, $H^{V}_{V,-}$, $H^{V}_{V,0}$ and $H^{V}_{V,t}$, which involve various hadronic form factors. Unfortunately, to the best of our knowledge, the most recent lattice calculation of the $D \to \rho,K^*$ form factors dates back to 1995 \cite{Bowler:1994zr}. As significant improvements have recently been made in lattice QCD, a calculation of the $D \to V$ form factors exploiting the current state-of-the-art methods would be very desirable for testing LFU in the charm sector. 

In our work, we complement the lattice QCD (LQCD) calculation in Ref.\ \cite{Bowler:1994zr} with the results from Ref.\ \cite{Wu:2006rd}, using light-cone sum rules (LCSR). The latter calculation is done in the framework of Heavy Quark Effective Field Theory (HQEFT). It should be mentioned that HQEFT, as the name suggests, relies on the assumption of a heavy quark, which in the case of $D$ mesons has to be treated carefully \cite{Fael:2019umf}. The definitions and different parametrizations of the form factors are given in Appendix \ref{AppFFs}. The resulting SM predictions for the branching fractions and the corresponding experimental results are 
summarized in Table \ref{Tab_SM_BFsVecSummary}. For the SM predictions, we use the values for $|V_{cd}|$ and $|V_{cs}|$ 
in Eqs.~(\ref{Vcd-SM}) and (\ref{Vcs-SM}), respectively.

\begin{table}[h!]
\renewcommand{\arraystretch}{1.2}
\begin{center}
\begin{tabular}{ cccc } 
 \hline
 Decay & LQCD & LCSR & Experiment  \\ 
 \hline
 \hline
   $\mathcal{B}(D^+ \to \rho^0 e^+ \nu_{e} )$ & $(2.23 \pm 0.70) \times 10^{-3}$ &$(2.25 \pm 0.28) \times 10^{-3}$& $(2.18^{+0.17}_{-0.25})\times 10^{-3}$ \\ 
    $ \mathcal{B}(D^+ \to \overline{K}^*(892)^0 e^+ \nu_{e} )$ & $(6.26 \pm 1.84) \times 10^{-2}$ &$(5.02 \pm 0.56) \times 10^{-2}$& $(5.40 \pm 0.10)\times 10^{-2}$\\ 
    \hline
  $\mathcal{B}(D^+ \to \rho^0 \mu^+ \nu_{\mu} )$ & $(2.13 \pm 0.64) \times 10^{-3}$ &$(2.14 \pm 0.27) \times 10^{-3}$& $(2.4 \pm 0.4)\times 10^{-3}$ \\ 
    $ \mathcal{B}(D^+ \to \overline{K}^*(892)^0 \mu^+ \nu_{\mu} )$ & $(5.95 \pm 1.67) \times 10^{-2}$ &$(4.75 \pm 0.53) \times 10^{-2}$& $(5.27 \pm 0.15)\times 10^{-2}$ \\ 
  \hline
\end{tabular}
\end{center}
\caption{SM predictions for $D \to V \bar{l} \nu_l$ branching ratios using LQCD and LCSR form-factor information from Refs.\ \cite{Bowler:1994zr} and \cite{Wu:2006rd}, respectively; the experimental values are from Ref.\ \cite{Tanabashi:2018oca}.}\label{Tab_SM_BFsVecSummary}
\end{table}

\subsubsection{Constraints on Pseudoscalar Coefficients}\label{sec:semileppseuNP}

\begin{figure}[t!]
\centering
\begin{subfigure}{.45\linewidth}
  \centering
  \includegraphics[width=.8\linewidth]{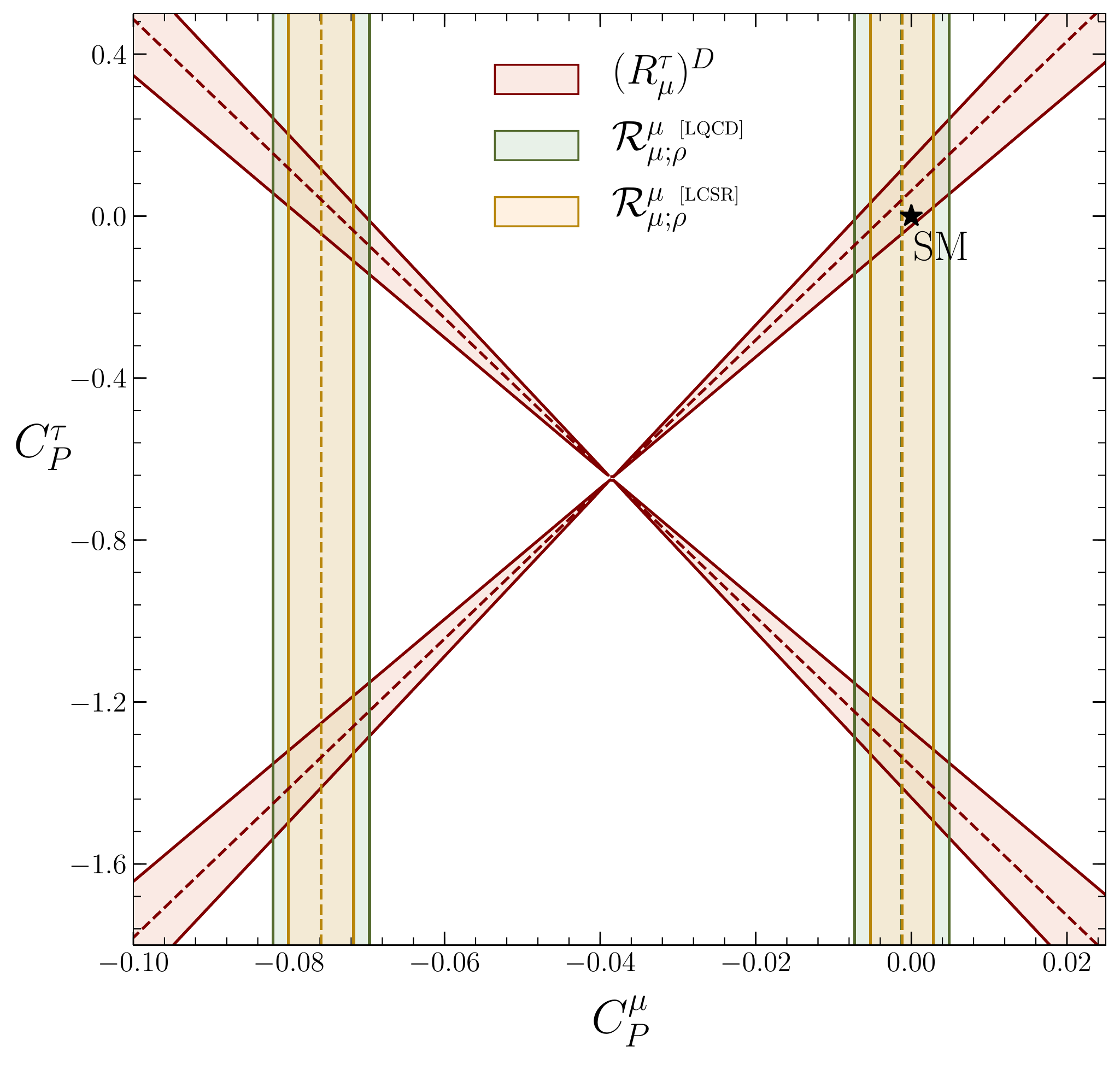}
\end{subfigure}
\begin{subfigure}{.45\linewidth}
  \centering
  \includegraphics[width=.8\linewidth]{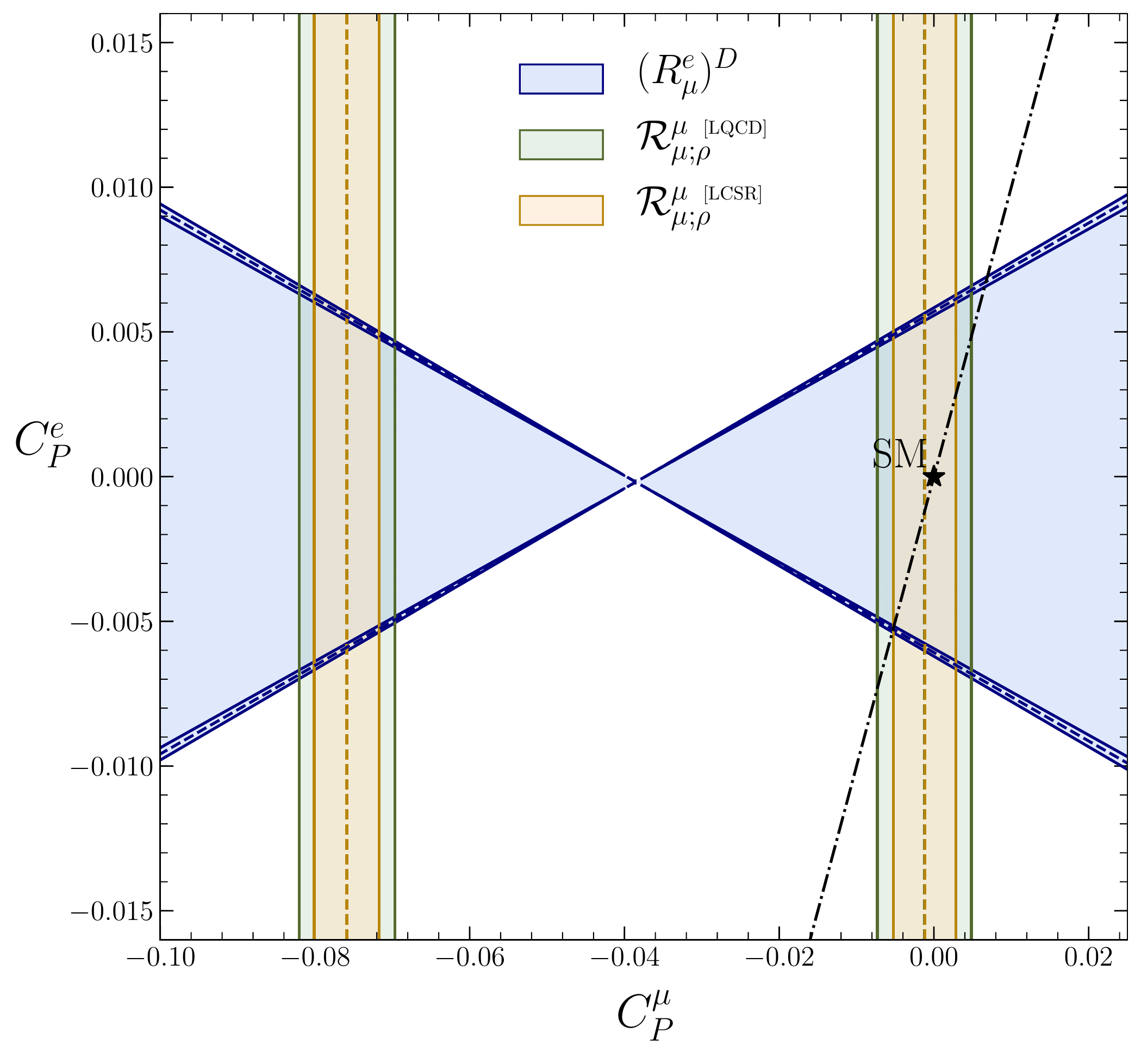}
\end{subfigure}
\caption{Allowed regions in the $C^{\mu}_P$--$C^{\tau}_P$ (left) and $C^{\mu}_P$--$C^{e}_P$ (right) planes using the ratios 
$(R^{\tau}_{\mu})^D$, $\mathcal{R}^{\mu}_{\mu ; \rho}$ and $(R^{e}_{\mu})^{D}$, $\mathcal{R}^{\mu}_{\mu ; \rho}$, 
respectively. The dashed-dotted line corresponds to $C^e_P=C^{\mu}_P$. }
\label{Fig_Cmu_Ctau_Cel_PseuNP_lepsemilep_D}
\end{figure}

\begin{figure}[h!]
\centering
\begin{subfigure}{.45\linewidth}
  \centering
  \includegraphics[width=.8\linewidth]{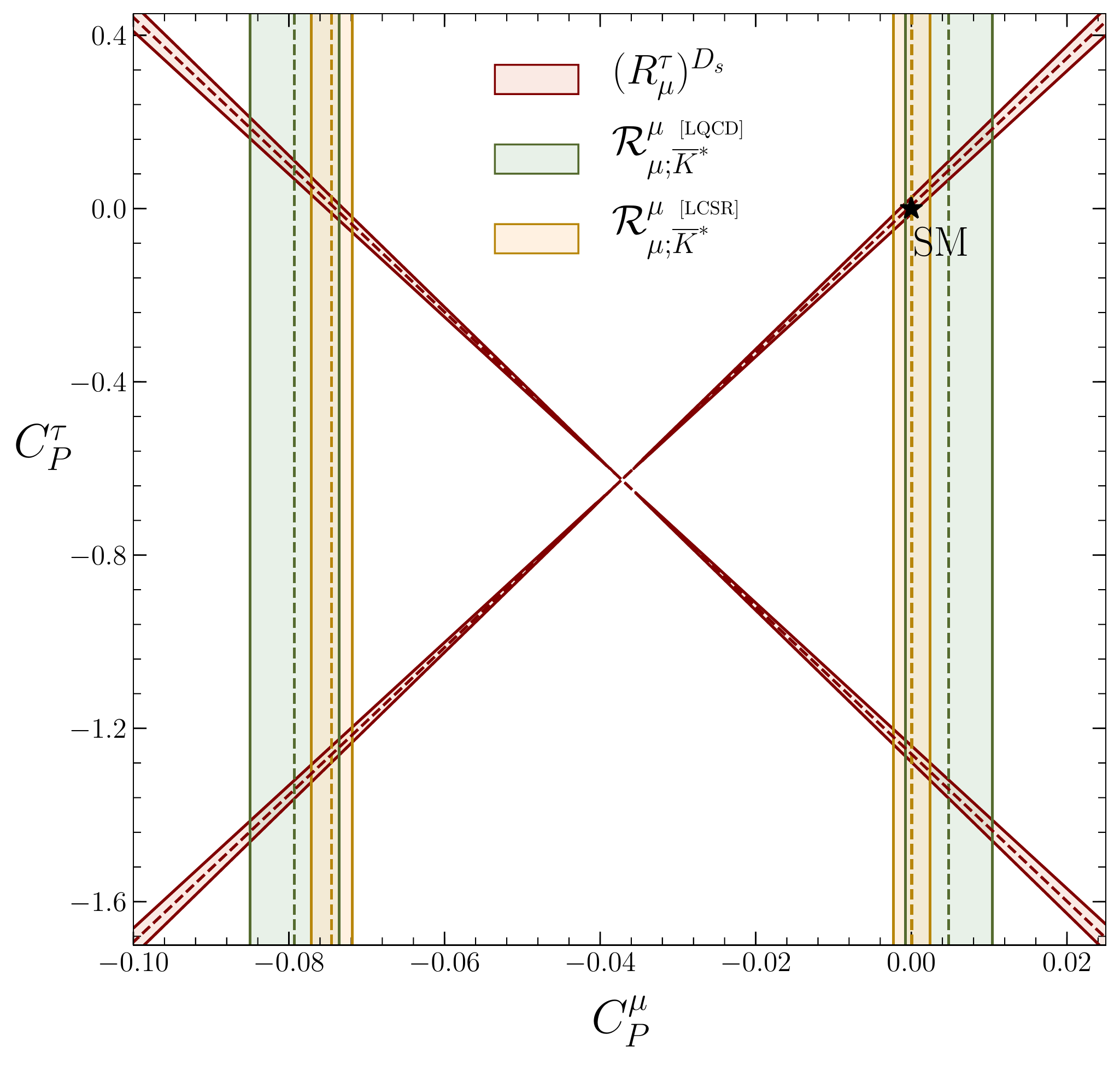}
\end{subfigure}
\begin{subfigure}{.45\linewidth}
  \centering
  \includegraphics[width=.8\linewidth]{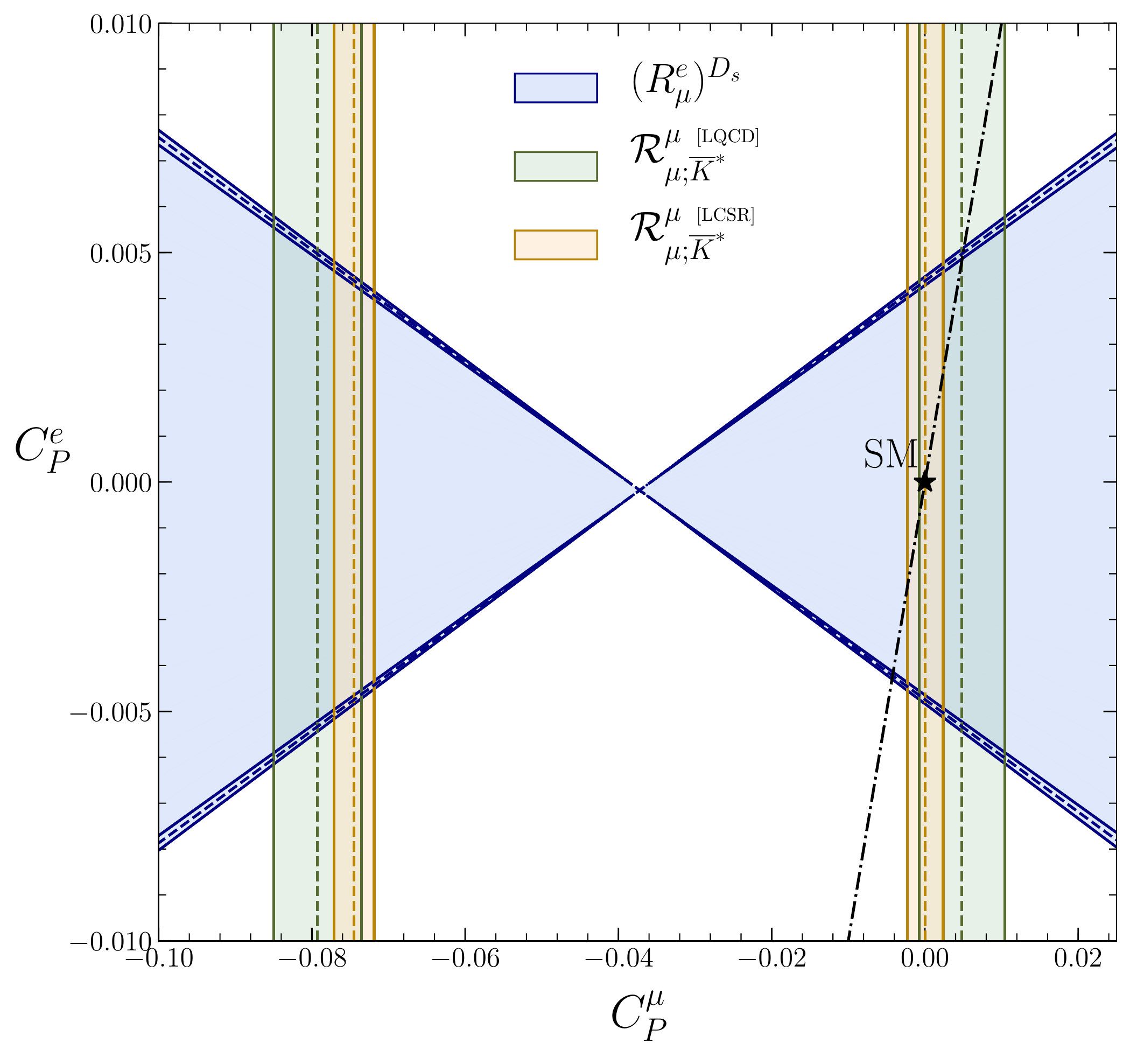}
\end{subfigure}
\caption{Allowed regions in the $C^{\mu}_P$--$C^{\tau}_P$ (left) and $C^{\mu}_P$--$C^{e}_P$ (right) planes using the ratios $(R^{\tau}_{\mu})^{D_s}$, $\mathcal{R}^{\mu}_{\mu ; \overline{K}^*}$  and $(R^{e}_{\mu})^{D_s}$, $\mathcal{R}^{\mu}_{\mu ; \overline{K}^*}$, respectively. The dashed-dotted line corresponds to $C^e_P=C^{\mu}_P$. }
\label{Fig_Cmu_Ctau_Cel_PseuNP_lepsemilep_Ds}
\end{figure}

If we allow for pseudoscalar NP interactions, we obtain the following expression for the differential branching fraction for semileptonic $D$ decays with vector mesons in the final state: 
\begin{equation}
\begin{split}
&\frac{d \mathcal{B}(D \rightarrow V \bar{l} \nu_l)}{dq^2} = \frac{G_F^2 \tau_{D} |V_{cq}|^2}{24 \pi^3 M^2_{D}} \Bigg\{ \Bigg[ \frac{1}{4} \Big( 1 + \frac{m_l^2}{2q^2} \Big) \Big[ (H^{V}_{V,+})^2 + (H^{V}_{V,-})^2 + (H^{V}_{V,0})^2 \Big] 
\\
&+ \frac{3}{8}\frac{m_l^2}{q^2} (H^{V}_{V,t})^2 \Bigg] + \frac{3}{8} |C_P^l |^2 (H^{V}_{S})^2 + \frac{3}{4} \mathcal{R}e(C_P^{l*} ) \frac{m_l}{\sqrt{q^2}} H^{V}_{S}H^{V}_{V,t} \Bigg\} \frac{(q^2 - m_l^2)^2}{q^2}|\vec{p}_{V}|,
\end{split}
\end{equation}
where the terms in the large square brackets represent the SM part. In order to constrain the pseudoscalar NP coefficients $C^l_P$, we
introduce the ratios
\begin{equation}\label{Eq_Rmumu_Vec_PseuNP}
\begin{split}
\mathcal{R}^{\mu}_{\mu;\rho} &= \frac{\mathcal{B}(D^+ \to \mu^+ \nu_{\mu})}{ \mathcal{B}(D^+ \to \rho^0 \mu^+ \nu_{\mu} )}, \hspace{10mm} \mathcal{R}^{\mu}_{\mu;\overline{K}^{*}} = \frac{\mathcal{B}(D_s^+ \to \mu^+ \nu_{\mu})}{ \mathcal{B}(D^+ \to \overline{K}^*(892)^0 \mu^+ \nu_{\mu} }.
\end{split}
\end{equation}
Since each of these observables depends on $C^{\mu}_P$, they can be used to complement the constraints for this coefficient 
following from the leptonic $D$ decays. Note that for the $D^+ \to \rho^0 \mu^+ \nu_{\mu}$ channel, a factor of 1/2 has to be taken into account due to the wave function of the $\rho^0$ meson.
Using the measured branching fractions in Table \ref{Tab_SM_BFsVecSummary}, we obtain the following experimental values:
\begin{equation}
\begin{split}\label{Eq_Rmumu_Vec_exp}
\mathcal{R}^{\mu}_{\mu;\rho} &= (1.56 \pm 0.27) \times 10^{-1} , \hspace{10mm} \mathcal{R}^{\mu}_{\mu;\overline{K}^{*}} =(1.04 \pm 0.05) \times 10^{-1} .\\
\end{split}
\end{equation}
These constraints can be converted correspondingly into allowed ranges for $C^{\mu}_P$, employing $c \to d$ and
$c\to s$ transitions. The results are listed in Table \ref{Tab_CmuP_Values_LQCD_LCSR}, utilizing information on the form factors 
both from LQCD and from LCSR calculations.
\begin{table}[]
\renewcommand{\arraystretch}{1.2}
\begin{center}
\begin{tabular}{ ccc } 
 \hline
 Coefficient & LQCD & LCSR    \\ 
 \hline
 \hline
  \multirow{2}{*}{$C^{\mu}_{P}\big|_{cd} $}   & $(-7.58 \pm 0.62)\times 10^{-2}$ &$(-7.59 \pm 0.42)\times 10^{-2}$\\
   & $(-1.23 \pm 6.08)\times 10^{-3}$ &$(-1.19 \pm 4.04)\times 10^{-3}$\\
   \hline
     \multirow{2}{*}{$C^{\mu}_{P}\big|_{cs} $}& $(-7.93 \pm 0.57)\times 10^{-2}$ &$(-7.45 \pm 0.26)\times 10^{-2}$\\
    & $(4.82 \pm 5.59)\times 10^{-3} $ &$(0.05 \pm 2.34)\times 10^{-3}$\\
    \hline
    \end{tabular}
\end{center}
\caption{Allowed values for the coefficients $C^{\mu}_{P}\big|_{cd} $ and $C^{\mu}_{P}\big|_{cs}$, obtained through the ratios $\mathcal{R}^{\mu}_{\mu;\rho}$ and $\mathcal{R}^{\mu}_{\mu;\overline{K}^{*}}$, respectively, using LQCD \cite{Bowler:1994zr} and LCSR \cite{Wu:2006rd} form-factor information. }\label{Tab_CmuP_Values_LQCD_LCSR}
\end{table}

We may use the obtained ranges for $C^{\mu}_P$ to further constrain the allowed regions shown in Figs.\ \ref{Fig_Lep_Constr_PseuNP_D} and \ref{Fig_Lep_Constr_PseuNP_Ds}. This yields the vertical bands  in Figs.\ \ref{Fig_Cmu_Ctau_Cel_PseuNP_lepsemilep_D} and  \ref{Fig_Cmu_Ctau_Cel_PseuNP_lepsemilep_Ds}. The constraints obtained using LQCD and LCSR form-factor information are indicated by the green and yellow bands, respectively. The dashed-dotted lines in Figs.\ \ref{Fig_Cmu_Ctau_Cel_PseuNP_lepsemilep_D} (right) and  \ref{Fig_Cmu_Ctau_Cel_PseuNP_lepsemilep_Ds} (right) indicate the correlations arising from $C^e_P = C^{\mu}_P$. In both cases, we find no large discrepancies with the SM predictions where the pseudoscalar NP coefficients vanish.

\subsubsection{Constraints on Vector Coefficients}\label{Sec_vec_Contr_vectors}

\begin{figure}[h!]
\centering
\begin{subfigure}{.45\linewidth}
  \centering
  \includegraphics[width=.8\linewidth]{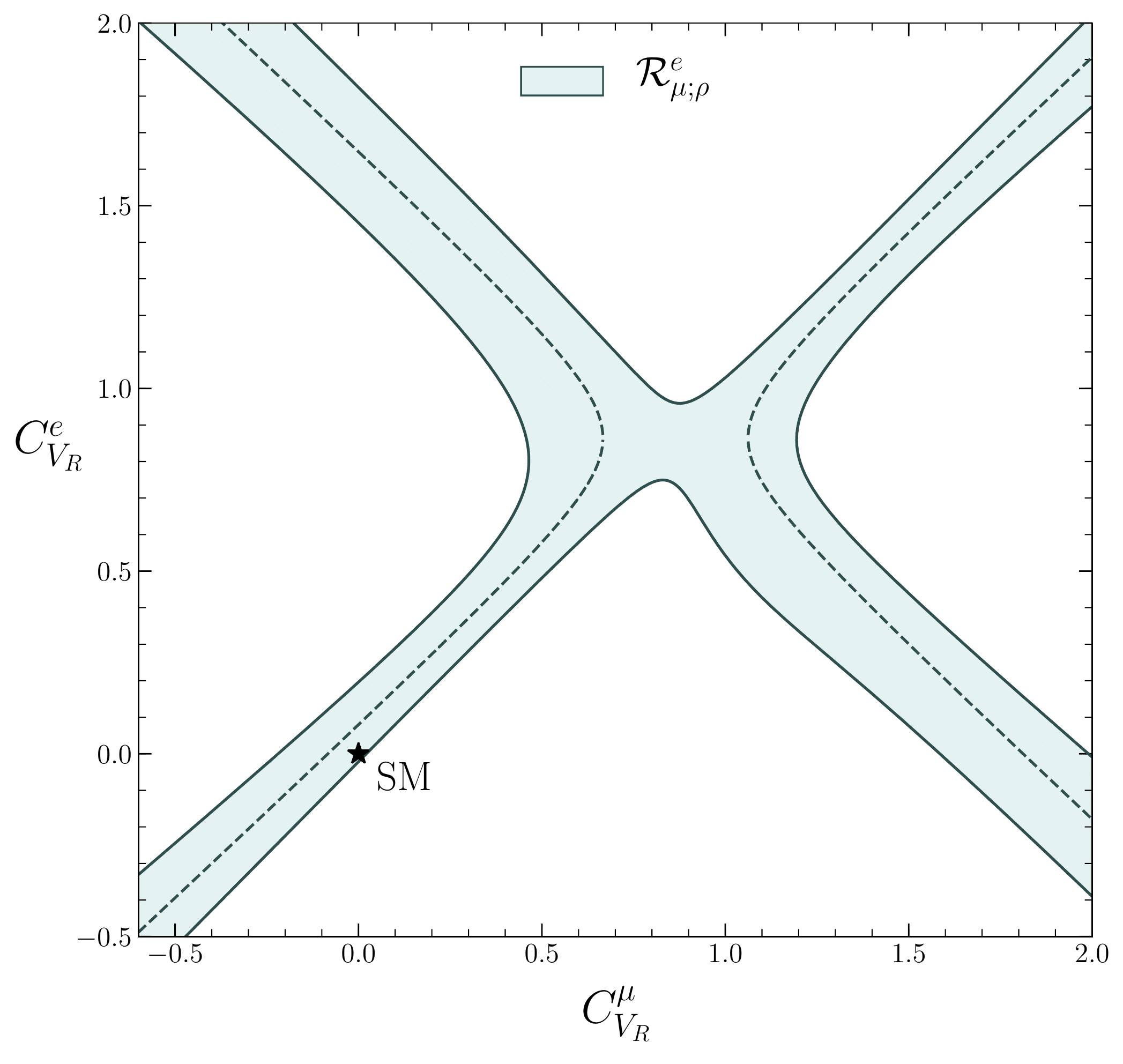}
\end{subfigure}
\begin{subfigure}{.45\linewidth}
  \centering
  \includegraphics[width=.8\linewidth]{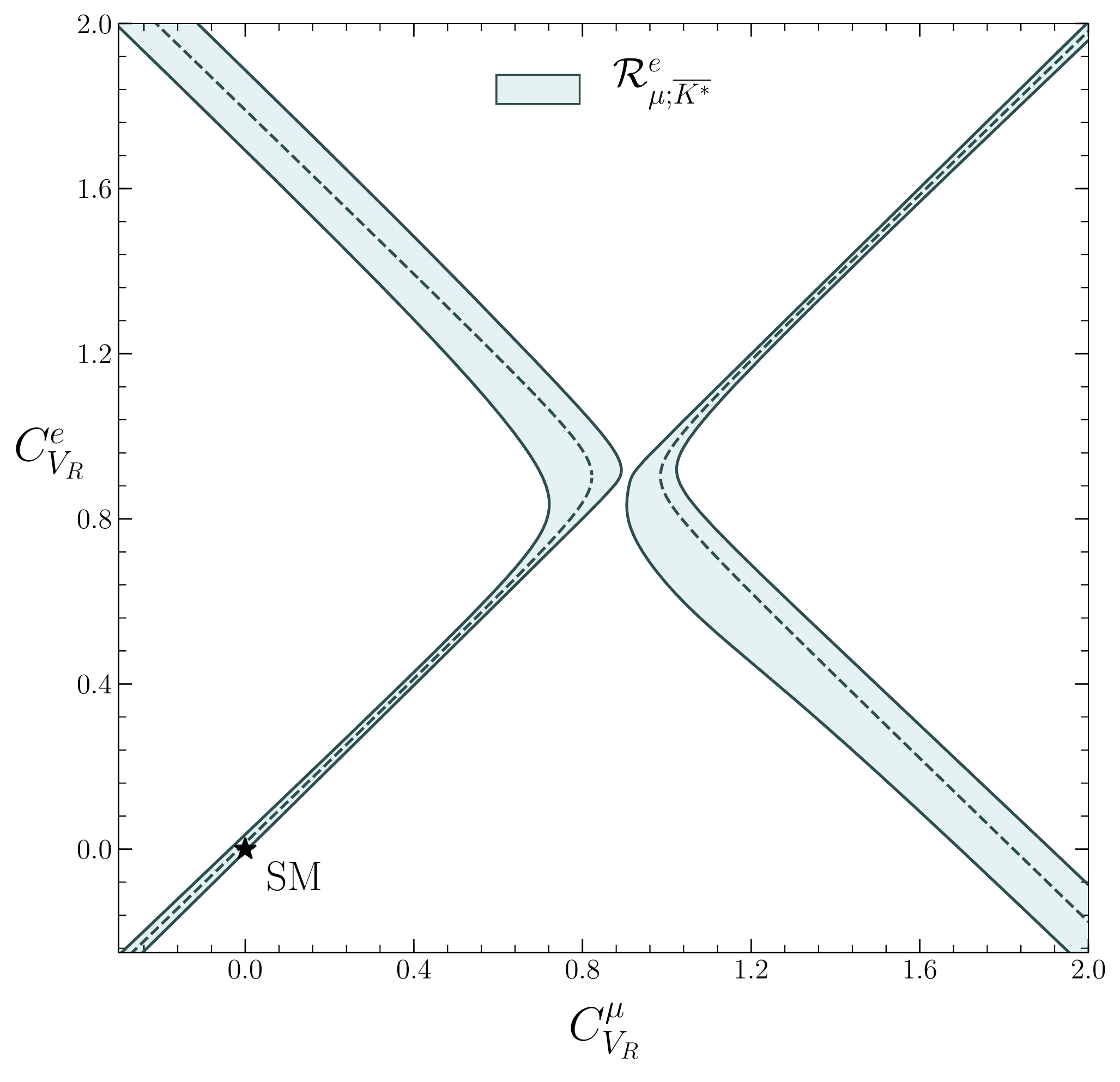}
\end{subfigure}
\caption{Allowed regions in the $C^{\mu}_{V_R}$--$C^{e}_{V_R}$ plane using the ratios $\mathcal{R}^{e}_{\mu ; \rho}$ (left) and $\mathcal{R}^{e}_{\mu ; \overline{K}^*}$ (right).}
\label{Fig_Ce_Cu_RH_Vector_contours}
\end{figure}

Besides pseudoscalar NP interactions, semileptonic $D$ decays with a vector meson in the final state are also sensitive probes
of LH and RH vector contributions from physics beyond the SM. The differential branching fraction in the presence of LH vector NP interactions can be written as
\begin{equation}\label{D_to_V_lv_BF_NP_LHvec}
\begin{split}
\frac{d \mathcal{B}(D \rightarrow V \bar{l} \nu_l)}{dq^2} =\frac{d \mathcal{B}(D \rightarrow V \bar{l} \nu_l)}{dq^2} \Bigg|_{\text{SM}} \big| 1 + C_{V_L}|^2 .
\end{split}
\end{equation}
In analogy to decays with pseudoscalar mesons in the final states, the ratio between a leptonic and semileptonic decay does not yield further constraints since the additional LH vector contributions cancel. The differential branching fraction in the presence of RH vector NP interactions is slightly different from the pseudoscalar case \cite{Sakaki:2013bfa}:
\begin{equation}
\begin{split}
&\frac{d \mathcal{B}(D \rightarrow V \bar{l} \nu_l)}{dq^2} = \frac{G_F^2 \tau_{D} |V_{cq}|^2}{24 \pi^3 M^2_D} \Bigg\{ \Bigg[ \frac{1}{4} \Big( 1 + \frac{m_l^2}{2q^2} \Big) \Big[ (H^{V}_{V,+})^2 + (H^{V}_{V,-})^2 + (H^{V}_{V,0})^2 \Big] \\
&+ \frac{3}{8}\frac{m_l^2}{q^2} (H^{V}_{V,t})^2 \Bigg](1 + |C^l_{V_R}|^2)  - 2 \mathcal{R}e(C_{V_R}^{l*} ) \Bigg[ \frac{1}{4} \Big( 1 + \frac{m_l^2}{2q^2} \Big) \Big[ (H^{V}_{V,0})^2 + 2H^{V}_{V,+} H^{V}_{V,-} \Big]\\
& + \frac{3}{8}\frac{m_l^2}{q^2} (H^{V}_{V,t})^2 \Bigg]  \Bigg\} \frac{(q^2 - m_l^2)^2}{q^2}|\vec{p}_{V}|.
\end{split}
\end{equation}
However, the structure of the formulae does prohibit the extractions of further constraints through the ratio between a leptonic and a semileptonic decay with the same lepton flavour. Therefore, in analogy to Eqs.\ (\ref{Eq_Remu_Pseu_VecNP_Pi}) and (\ref{Eq_Remu_Pseu_VecNP_K}), we define the following ratios between two semileptonic decays with different flavours of leptons:
\begin{equation}\label{Eq_R_emu_rho_Kstar}
\mathcal{R}^{e}_{\mu;\rho} \equiv \frac{\mathcal{B}(D^+ \to \rho^0 e^+ \nu_{e} )}{\mathcal{B}(D^+ \to \rho^0 \mu^+ \nu_{\mu} )} , \hspace{10mm} \mathcal{R}^{e}_{\mu;\overline{K}^*} \equiv \frac{\mathcal{B}(D^+ \to \overline{K}^*(892)^0 e^+ \nu_{e} )}{\mathcal{B}(D^+ \to \overline{K}^*(892)^0 \mu^+ \nu_{\mu} )}. 
\end{equation}
From the measured branching fractions in Table \ref{Tab_SM_BFsVecSummary}, we obtain the following experimental constraints:
\begin{equation}
\mathcal{R}^{e}_{\mu;\rho} =(9.08 \pm 1.75) \times 10^{-1} , \hspace{10mm} \mathcal{R}^{e}_{\mu;\overline{K}^*} = 1.02 \pm 0.03. 
\end{equation}
By comparing Eqs.\ (\ref{D_to_P_lv_BF_NP_vec}) and (\ref{D_to_V_lv_BF_NP_LHvec}) with each other, 
it is clear that the contours in the $C^{\mu}_{V_L}$--$C^{e}_{V_L}$ planes will be of the same shape; the expressions differ only up to a scale-factor. As the form-factor information for $D \to P$ decays is significantly more precise than for $D \to V$ decays, and the experimental precision is more or less the same, the constraints coming from $D \to V$ decays do not improve the constraints in Figs.\ \ref{Fig_SemilepPseu_Constr_VecNP_Pi} and \ref{Fig_SemilepPseu_Constr_VecNP_K}. For RH vector interactions, however, the slightly different structure is worth investigating. For $\mathcal{R}^{e}_{\mu;\rho}, $ this leads to the contour shown in Fig.\ \ref{Fig_Ce_Cu_RH_Vector_contours} (left), and for $\mathcal{R}^{e}_{\mu;\overline{K}^*} $, this leads to the contour shown in Fig.\ \ref{Fig_Ce_Cu_RH_Vector_contours} (right). In both plots, the SM prediction is included in the 1$\sigma$ contours.

\subsubsection{Constraints on Tensor Coefficients}

\begin{figure}[h!]
\centering
\begin{subfigure}{.45\linewidth}
  \centering
  \includegraphics[width=.8\linewidth]{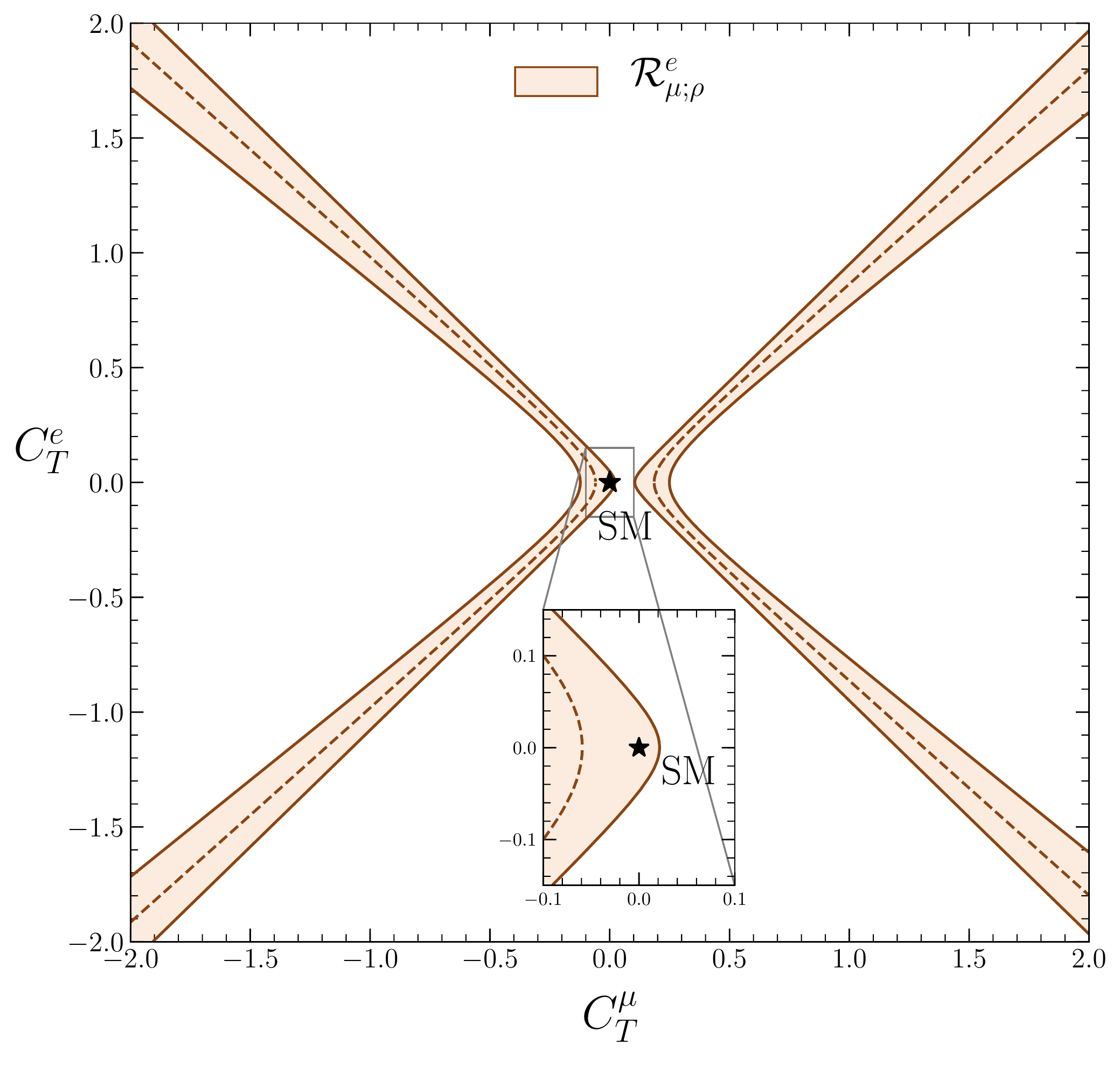}
\end{subfigure}
\begin{subfigure}{.45\linewidth}
  \centering
  \includegraphics[width=.8\linewidth]{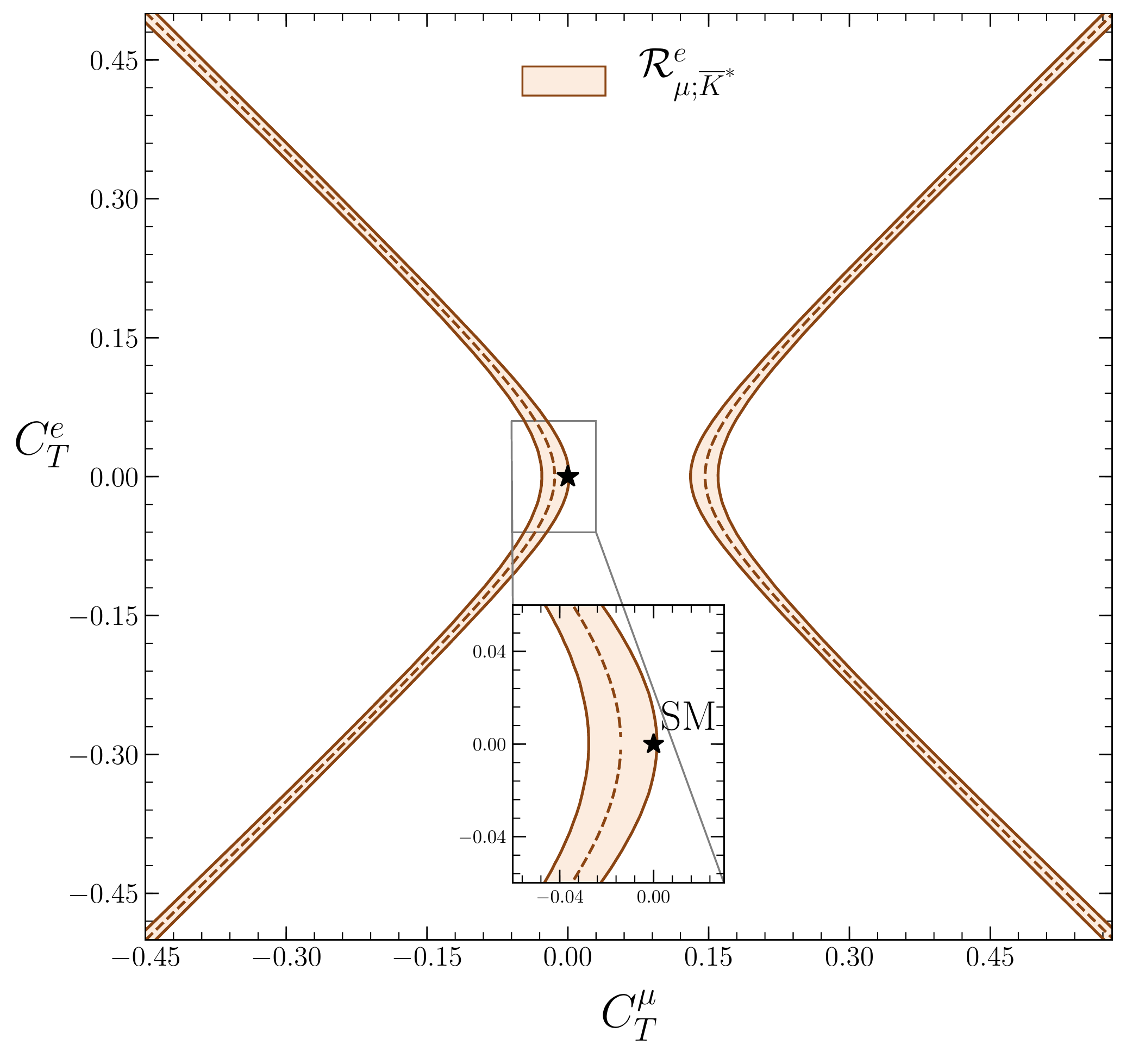}
\end{subfigure}
\caption{Allowed regions in the $C^{\mu}_{T}$--$C^{e}_{T}$ plane using the ratios $\mathcal{R}^{\mu}_{\mu ; \rho}$ (left) and $\mathcal{R}^{\mu}_{\mu ; \overline{K}^*}$ (right).}
\label{Fig_Ce_Cu_RH_Tensor_contours}
\end{figure}

Finally, let us have a closer look at NP tensor contributions, which are the final category of potential NP contributions to
semileptonic $D$ decays with a vector meson in the final state. Allowing for such an effect, we obtain the following differential
branching ratio:
\begin{equation}
\begin{split}
&\frac{d \mathcal{B}(D \rightarrow V \bar{l} \nu_l)}{dq^2} = \frac{G_F^2 \tau_{D} |V_{cq}|^2}{24 \pi^3 M^2_D} \Bigg\{ \Bigg[ \frac{1}{4} \Big( 1 + \frac{m_l^2}{2q^2} \Big) \Big[ (H^{V}_{V,+})^2 + (H^{V}_{V,-})^2 + (H^{V}_{V,0})^2 \Big]  \\
&+ \frac{3}{8}\frac{m_l^2}{q^2} (H^{V}_{V,t})^2 \Bigg] + 2 |C^l_T|^2 \Big( 1 + \frac{2 m_l^2}{q^2}\Big) \Big[ (H^{V}_{T,+})^2 + (H^{V}_{T,-})^2 + (H^{V}_{T,0})^2 \Big] \\
& - 3 \mathcal{R}e(C^{l*}_T) \frac{m_l}{\sqrt{q^2}} \big( H^{V}_{T,0} H^{V}_{V,0} + H^{V}_{T,+}H^{V}_{V,-} - H^{V}_{T,-}H^{V}_{V,+} \big) \Bigg\} \frac{(q^2 - m_l^2)^2}{q^2}|\vec{p}_{V}|.
\end{split}
\end{equation}
However, to the best of our knowledge, there is no explicit lattice calculation of the tensor form factors for $D \to V$ decays available. In Ref.\ \cite{Wu:2006rd}, the tensor form factors are related to the vector and scalar form factors $A_1$, $A_2$, $A_3$, $V$ in the framework of HQEFT. A lattice determination of the tensor form factors in $D \to V$ transitions would be very desirable. However, as currently no lattice information is available for these quantities, we will use the HQEFT relations to obtain constraints, albeit for illustrative purpose. The relations for the form factors are given as follows \cite{Wu:2006rd}:
\begin{align}
T_1(q^2) &= \frac{M_D^2 - m_V^2 + q^2}{2 M_D} \frac{V(q^2)}{M_D + m_V} + \frac{M_D + m_V}{2 M_D} A_1(q^2) ,\\
T_2(q^2) &= \frac{2}{M_D^2 - m_V^2} \Bigg[ \frac{(M_D - y)(M_D + m_V)}{2} A_1(q^2) + \frac{M_D (y^2 - m_V^2)}{M_D + m_V} V(q^2) \Bigg] ,\\
T_3(q^2) &= -\frac{M_D + m_V}{2 M_D} A_1(q^2) + \frac{M_D - m_V}{2 M_D} \big[ A_2(q^2) - A_3(q^2)\big] + \frac{M_D^2 + 3 m_V^2 - q^2}{2 M_D (M_D + m_V)} V(q^2) .
\end{align}
where $y = (M_D^2 + m_V^2 - q^2)/(2 M_D)$ is the energy of the vector meson. We use the ratios defined in Eq.\ (\ref{Eq_R_emu_rho_Kstar}) and allow for tensor contributions instead of vector contributions. Using the HQEFT relations and the experimental information, we find the allowed regions in the $C^{\mu}_T$--$C^e_T$ plane  shown in Fig.\ \ref{Fig_Ce_Cu_RH_Tensor_contours}. We observe that the SM points fall just within in the 1\,$\sigma$ regions in both cases.

\section{\boldmath Determination of $|V_{cd}|$ and $|V_{cs}|$}\label{Ch_CKM_elements}\label{sec:VcdVcsNP}

The determination of $|V_{cd}|$ or $|V_{cs}|$ is done by exploiting data from charged-current interactions involving
$c \to d \bar{l} \nu_l$ or $c \to s \bar{l} \nu_l$ quark-level transitions, respectively. Usually, these determinations assume just the
corresponding SM expressions. However, there may be NP contributions present in these processes, as is the main focus of our
analysis. In the case of (semi)-leptonic $D_{(s)}$ decays, their SM expressions could be used to determine $|V_{cd(s)}|$ from the
data and non-perturbative information on the hadronic parameters. In Section \ref{Sec_Vcd_Vcs_Unitarity}, we have determined 
$|V_{cd(s)}|$ independently of $D_{(s)}$ decays, utilizing the Wolfenstein parametrization of the CKM matrix. Here, we have assumed the SM in the corresponding kaon and $B$ decay processes. 

Physics beyond the SM may also affect the values of the CKM matrix elements. Consequently, we have probed NP through observables that do not involve such parameters. Combining the theoretical expressions for these observables with experimental information on the branching fractions allows us to constrain NP coefficients, independently of the CKM matrix elements. 
Finally, having these constraints available, we may use them to extract $|V_{cd(s)}|$ from $D_{(s)}$ decays, even in the presence of possible NP effects in the corresponding transition amplitudes.

\subsection{The Strategy}

In our analysis, we follow a strategy first proposed for $B$ decays in Ref.\ \cite{Banelli:2018fnx}. It has not yet been 
applied to the charm sector. Using the NP constraints obtained from the observables $\mathcal{R}^{\mu}_{\mu;\rho}$ and $\mathcal{R}^{\mu}_{\mu;\overline{K}^*}$, we will finally also determine $|V_{cd}|$ or $|V_{cs}|$ in the presence of pseudoscalar NP contributions. To distinguish the coefficients corresponding either to a $c \to d$ transition or a $c \to s$ transition, we denote the latter with $\tilde{C}^{\mu}_P$. 
There are three steps in this approach, after which we obtain the CKM matrix elements:
\begin{itemize}
	\item{We start with the expressions for $\mathcal{R}^{\mu}_{\mu;\rho}$ and $\mathcal{R}^{\mu}_{\mu;\overline{K}^*}$ in Eq.\ (\ref{Eq_Rmumu_Vec_PseuNP}), which depend only on one coefficient, either $C^{\mu}_{P}$ or $\tilde{C}^{\mu}_{P}$. Consequently, we can solve for $C^{\mu}_{P}$ or $\tilde{C}^{\mu}_{P}$.}
	\item{This results in the functions $C^{\mu}_{P}(\mathcal{R}^{\mu}_{\mu;\rho})$ and $\tilde{C}^{\mu}_{P}(\mathcal{R}^{\mu}_{\mu;\overline{K}^*})$. For both functions, there are two independent solutions for the coefficients, corresponding to the two bands 
	shown in Figs.\ \ref{Fig_Cmu_Ctau_Cel_PseuNP_lepsemilep_D} and \ref{Fig_Cmu_Ctau_Cel_PseuNP_lepsemilep_Ds}. }
	\item{We then evaluate any of the individual branching fractions for each of these ratios and solve for the corresponding CKM matrix element. Comparing the resulting expression to individual measurements of the branching fractions allows us to obtain the value of $|V_{cd}|$ or $|V_{cs}|$.}
\end{itemize}

\subsection{\boldmath $|V_{cd}|$ and $|V_{cs}|$}

We apply the described strategy to leptonic and semileptonic $D_{(s)}$ decays to determine the CKM matrix 
elements $|V_{cd}|$ and $|V_{cs}|$. To this end, we recall the following ratios:
\begin{equation}
\begin{split}
\mathcal{R}^{\mu}_{\mu;\rho} &= \frac{\mathcal{B}(D^+ \to \mu^+ \nu_{\mu})}{ \mathcal{B}(D^+ \to \rho^0 \mu^+ \nu_{\mu} )}, \hspace{10mm} \mathcal{R}^{\mu}_{\mu;\overline{K}^{*}} = \frac{\mathcal{B}(D_s^+ \to \mu^+ \nu_{\mu})}{ \mathcal{B}(D^+ \to \overline{K}^*(892)^0 \mu^+ \nu_{\mu} },\\
\end{split}
\end{equation}
which were given in Section \ref{Sec_vec_Contr_vectors}, along with the theoretical expressions for the branching fractions including pseudoscalar NP contributions. Using these expressions, we write the ratios as follows:
\begin{equation}\label{Ratiolepsemilepvec} 
\mathcal{R}^{l}_{l;V} =  \frac{   \tilde{\alpha}^{l}  \Big|1 + \beta^{l} C^{l}_{P} \Big|^2}
{ \mathcal{I}_1  + \mathcal{I}_2 |C_P^{l} |^2 + \mathcal{I}_3 \mathcal{R}e(C_P^{l*}) },
\end{equation}
where $V = \rho, \overline{K}^*$ and $\tilde{\alpha}^l = \alpha^l f_{D}^2 $. To keep the notation clear, the subscript $s$ is omitted; in the case $\mathcal{R}^{\mu}_{\mu;\overline{K}^{*}}$ the notation $D \to D_s$ and $C^l_P \to \tilde{C}^l_P$ is implied. The integrals $\mathcal{I}_1$, $\mathcal{I}_2$, $\mathcal{I}_3$ take 
the forms
\begin{equation}
\begin{split}
\mathcal{I}_1 &= I_0  \int_{m_l^2}^{(M_{D} - m_{V})^2} d q^2  \Bigg[  \frac{1}{4} \Big( 1 + \frac{m_l^2}{2q^2} \Big) \Big[ (H^{+}_{V})^2 + (H_{V}^{-})^2 + (H_{V}^{0})^2 \Big] + \frac{3}{8}\frac{m_l^2}{q^2} (H^{t}_{V})^2 \Bigg]\frac{(q^2 - m_{l}^2)^2}{q^2}|\vec{p}_{V}|,\\
\mathcal{I}_2 &=I_0  \int_{m_l^2}^{(M_{D} - m_{V})^2}  d q^2 \frac{3}{8}  (H^{0}_{P})^2  \frac{(q^2 - m_{l}^2)^2}{q^2}|\vec{p}_{V}|,\\
\mathcal{I}_3 &= I_0 \int_{m_l^2}^{(M_{D} - m_{V})^2}  d q^2  \frac{3}{4}   \frac{m_l}{\sqrt{q^2}} H^{0}_{P}H^{t}_{V}\frac{(q^2 - m_{l}^2)^2}{q^2}|\vec{p}_{V}| ,
\end{split}
\end{equation}
where $I_0 = 1/(3 \pi^2 M^2_D)$. Note that for $ \mathcal{R}^{\mu}_{\mu;\overline{K}^{*}}$, an additional factor of $\tau_{D_s}/\tau_D$ should be included, as the decaying D mesons in the leptonic and semileptonic decays are different. We assume the coefficients to be real and solve for $C^{\mu}_{V_R}$, $\tilde{C}^{\mu}_{V_R}$. The resulting expressions are the functions $C^{\mu}_{P}(\mathcal{R}^{\mu}_{\mu;\rho})$ and $\tilde{C}^{\mu}_{P}(\mathcal{R}^{\mu}_{\mu;\overline{K}^*})$. Subsequently, we insert $C^{\mu}_{P}(\mathcal{R}^{\mu}_{\mu;\rho})$ and $\tilde{C}^{\mu}_{P}(\mathcal{R}^{\mu}_{\mu;\overline{K}^*})$ in the expressions for the (semi)-leptonic branching fractions and solve for the corresponding CKM element. For the leptonic branching fractions, this yields:
\begin{equation}\label{eq:vckmlepbffunc}
\begin{split}
|V_{\text{cq}}| &= \Bigg( \frac{ 8 \pi \mathcal{B}(D_{(s)}^+ \rightarrow \mu^+ \nu_{\mu})}{G_F^2 \tau_{D_{(s)}} \alpha^{\mu}   \big|1 + \beta^{\mu} C^{\mu}_P(\mathcal{R}^{\mu}_{\mu;V})  \big|^2} \Bigg)^{1/2}.
    \end{split}
\end{equation}
As there are two solutions for $C^{\mu}_{P}$ and $\tilde{C}^{\mu}_{P}$, corresponding to the either of the two vertical bands in 
Figs.\ \ref{Fig_Cmu_Ctau_Cel_PseuNP_lepsemilep_D} and  \ref{Fig_Cmu_Ctau_Cel_PseuNP_lepsemilep_Ds}, we obtain two independent solutions for the each of the CKM matrix elements. Using the form factors calculated in LQCD \cite{Bowler:1994zr}, these results are equivalent at the current level of precision. In both cases, we obtain 
\begin{equation}\label{eq:vcdlqcdnp}
|V_{cd}| = 0.227 \pm 0.037, 
\end{equation}
\begin{equation}\label{eq:vcslqcdnp}
|V_{cs}| = 0.880 \pm 0.115 .
\end{equation}
Using the LCSR \cite{Wu:2006rd} form-factor information, we obtain the following results:
\begin{equation}\label{eq:vcdlcsrnp}
|V_{cd}| = 0.227 \pm 0.027 \hspace{.5cm} \lor \hspace{.5cm} |V_{cd}| = 0.227 \pm 0.025 ,
\end{equation}
\begin{equation}\label{eq:vcslcsrnp}
|V_{cs}| = 0.992 \pm 0.080 \hspace{.5cm} \lor \hspace{.5cm} |V_{cs}| = 0.993 \pm 0.066 .
\end{equation}
In Table \ref{Tab_VcdVcsvalues}, we list the values for the CKM elements obtained in Section \ref{Sec_Vcd_Vcs_Unitarity}, denoted as $V_{\text{UT}}$, the values determined in the presence of pseudoscalar NP, denoted as $V^{\text{NP}}_{\text{LQCD}}$ and $V^{\text{NP}}_{\text{LCSR}}$, and the PDG values, denoted as $V_{\text{PDG}}$. For both $|V_{cd}|$ and $|V_{cs}|$, the obtained values allowing for pseudoscalar NP agree with the SM and PDG values at the 1$\sigma$ level. This was to be expected, as we did not find large deviations from the SM when investigating NP contributions. Although our determined values have considerably larger uncertainties, the aim here is to show how to properly account for NP in the determination of CKM elements. Interestingly, there are discrepancies between our UT value and the PDG value assuming the SM of 1.5\,$\sigma$ and 1.4$\,\sigma$ for $|V_{cd}|$ 
and $|V_{cs}|$, respectively.

\begin{table}[t]
\renewcommand{\arraystretch}{1.3}
\begin{center}
\begin{tabular}{ |ccccc| } 
\hline
&$V_{\text{UT}}$ &$V^{\text{NP}}_{\text{LQCD}}$& $V^{\text{NP}}_{\text{LCSR}}$   & $V_{\text{PDG}}$  \\ 
 \hline
 \multirow{2}{*}{$|V_{cd}|$}  &  \multirow{2}{*}{$0.2242 \pm 0.0005$}&$0.227 \pm 0.037$&$0.227 \pm 0.027$   & \multirow{2}{*}{$0.218 \pm 0.004$}\\
    & &$0.227 \pm 0.037$ &$0.227 \pm 0.025$&\\
    \hline
 \multirow{2}{*}{$|V_{cs}|$} &\multirow{2}{*}{$0.9736 \pm 0.0001$}&$0.880 \pm 0.115$  &$0.992 \pm 0.080$& \multirow{2}{*}{$0.997 \pm 0.017$}\\
    &&$0.880 \pm 0.115$& $0.993 \pm 0.066$  &\\
   \hline
   \hline
   &$V^{\text{NP}}_{\text{LQCD}}- V_{\text{UT}}$& $V^{\text{NP}}_{\text{LCSR}}- V_{\text{UT}}$ &$V^{\text{NP}}_{\text{LQCD}}- V_{\text{PDG}}$ &  $V^{\text{NP}}_{\text{LCSR}}- V_{\text{PDG}}$\\ 
   \hline
  \multirow{2}{*}{$\Delta_{|V_{cd}|}$}   &$0.07$&$0.09$ &   $0.24$& $0.31$\\
    &$0.08$ &$0.10$&$0.24$&$0.34$\\
\hline
  \multirow{2}{*}{$\Delta_{|V_{cs}|}$}   &$0.82$&$0.23$ &   $1.01$& $0.06$\\
    &$0.81$ &$0.29$&$1.01$&$0.07$\\
\hline
    \end{tabular}
\end{center}
\caption{The CKM matrix elements $|V_{cd}|$ and $|V_{cs}|$ obtained in the present analysis and their PDG values \cite{Tanabashi:2018oca}. $\Delta_{|V_{cd}|}$ and $\Delta_{|V_{cs}|}$ denote the differences between the indicated values in standard deviations. }\label{Tab_VcdVcsvalues}
\end{table}


%
%
%
\section{\boldmath Predictions for Branching Fractions}\label{sec:predic}
\begin{figure}[]
\centering
\begin{subfigure}{.45\linewidth}
  \centering
  \includegraphics[width=.8\linewidth]{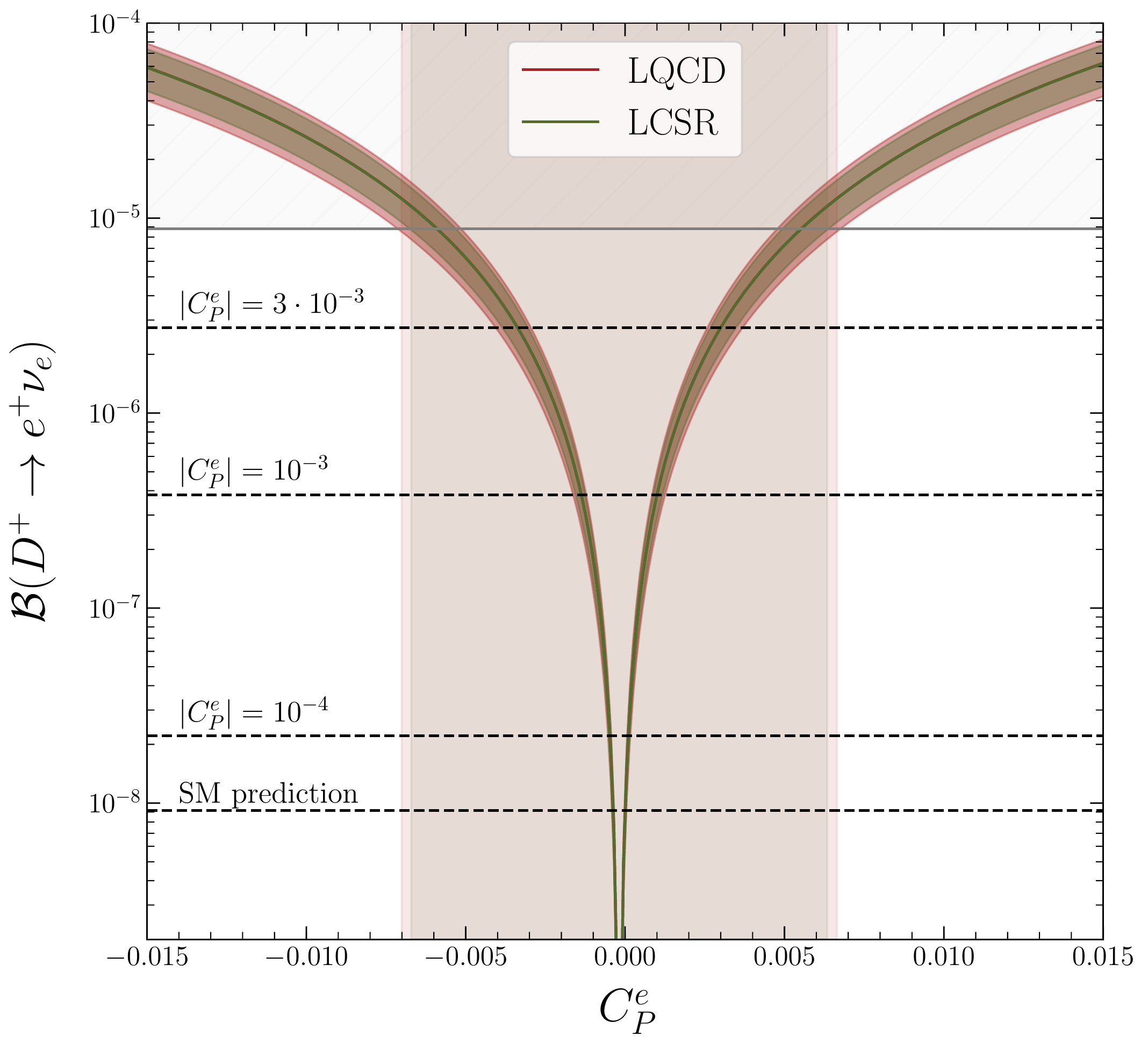}
\end{subfigure}
\begin{subfigure}{.45\linewidth}
  \centering
  \includegraphics[width=.8\linewidth]{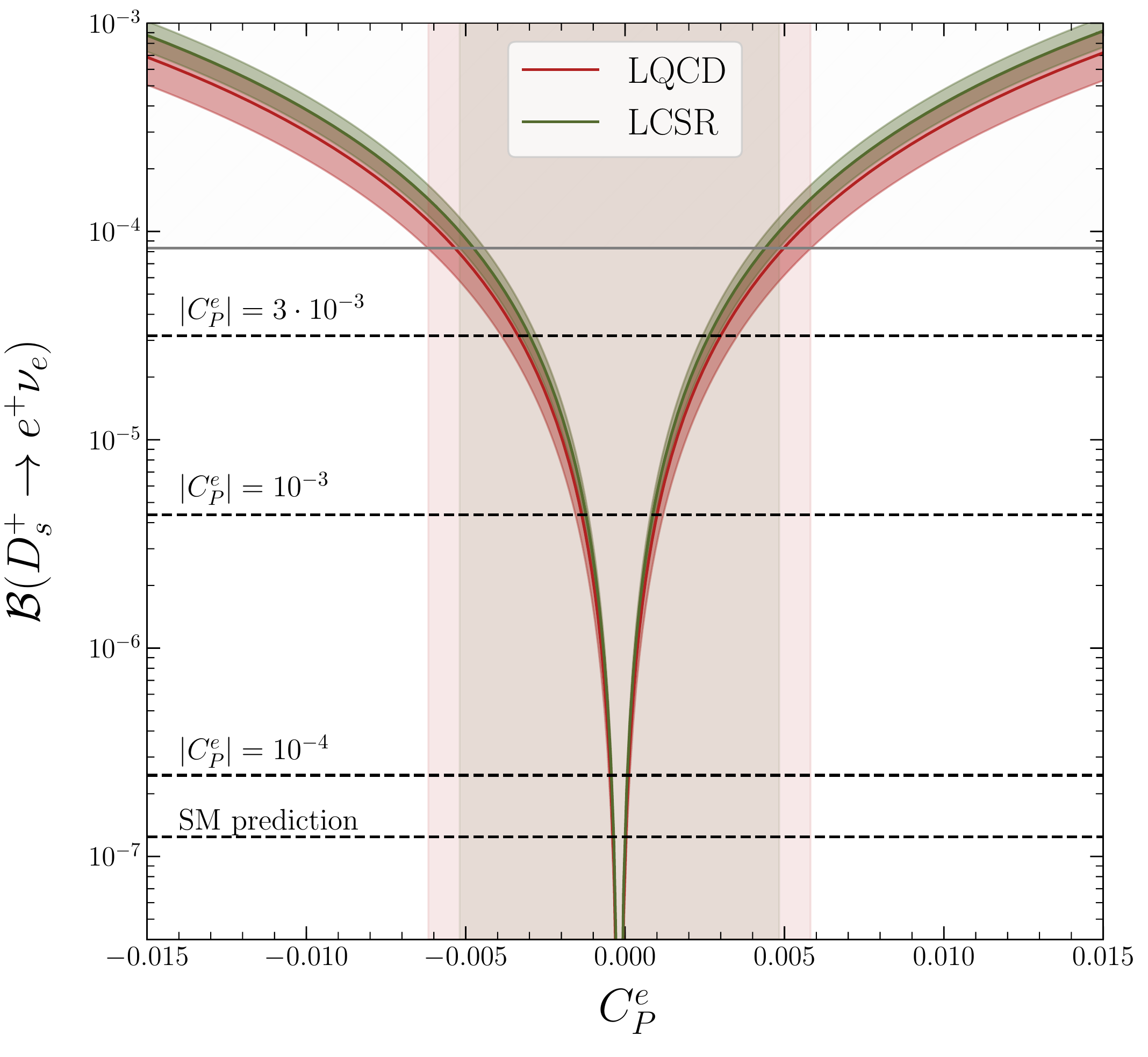}
\end{subfigure}
\caption{$\mathcal{B}(D_{(s)}^+ \to e^+ \nu_e)$ as functions of $C^e_P$. The grey regions indicate experimentally excluded regions, 
while the red and green ones show the allowed values for $C^e_P$. }
\label{Fig_BFel_DDs_Ce_P_Func}
\end{figure}

Now that we have constraints on the pseudoscalar NP coefficients and determinations of the CKM elements at our disposal, we may combine them to make predictions for the following leptonic branching fractions that have yet to be measured: 
\begin{equation}
\mathcal{B}(D^+ \to e^+ \nu_e), \hspace{1cm} \mathcal{B}(D_s^+ \to e^+ \nu_e) .
\end{equation}
In the SM, these decays are strongly helicity suppressed, as indicated by the smallness of the corresponding SM predictions in Table \ref{SM_BFsLep_summary}. However, this suppression may be lifted by new pseudoscalar interactions. This is shown in Fig.\ \ref{Fig_BFel_DDs_Ce_P_Func}. Here, the branching fractions $\mathcal{B}(D^+ \to e^+ \nu_e)$ and $\mathcal{B}(D_s^+ \to e^+ \nu_e)$ are plotted as a function of the coefficient $C^e_P$. The labels ``LQCD" and ``LCSR" refer to the form-factor information used to determine the relevant CKM elements in the presence of pseudoscalar NP, given in Eqs.\ (\ref{eq:vcdlqcdnp}--\ref{eq:vcslcsrnp}). The grey regions indicate the experimental upper bounds on the branching fractions. In order to illustrate the sensitivity of the branching fractions to the absolute value of $C^e_P$, the SM prediction and three predictions for different values of $|C^e_P|$ are shown in each plot. We see that even a small value for $|C^e_P|$ has a potentially large effect on the branching fraction, which is a direct consequence of the pseudoscalar NP contributions lifting the helicity suppression in these decays.

\begin{figure}[]
\centering
  \includegraphics[width=.8\linewidth]{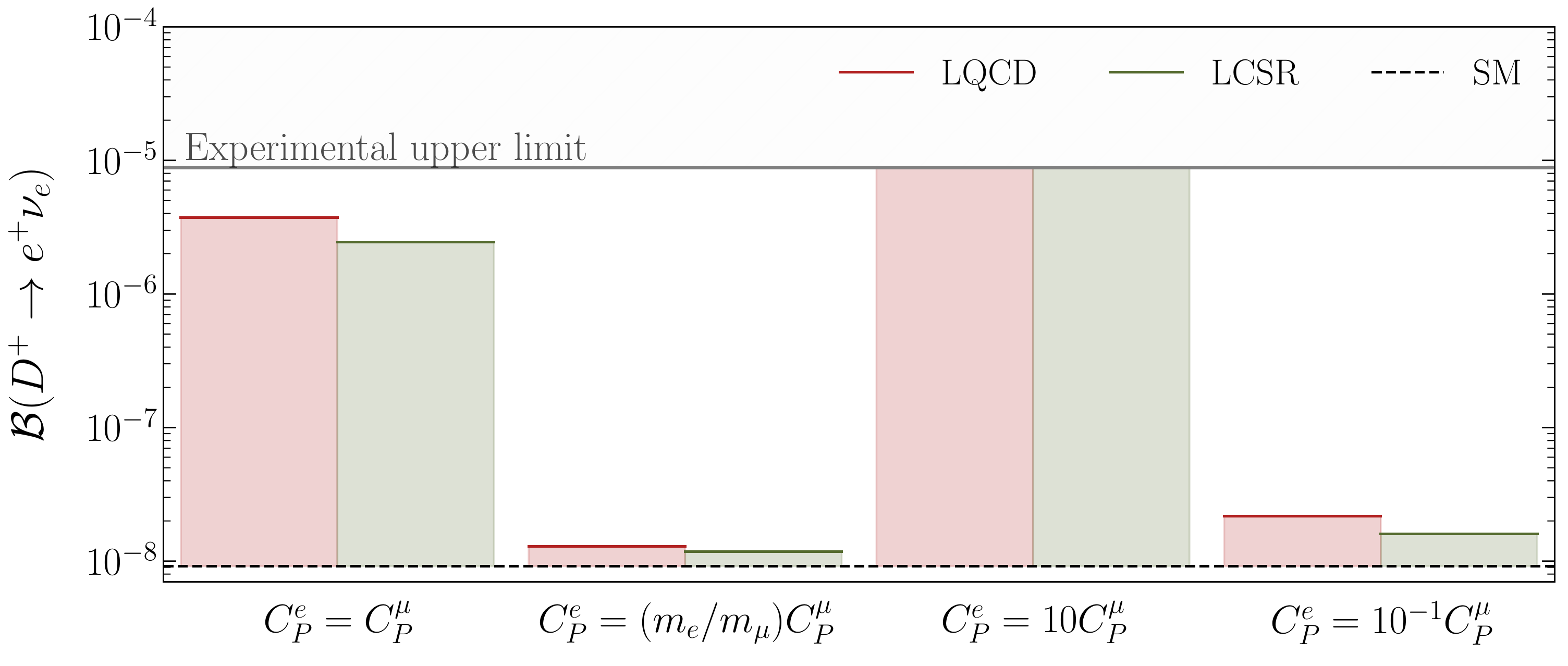}
\caption{Predictions of $\mathcal{B}(D^+ \to e^+ \nu_e)$ for the scenarios discussed in the text. The red and green bands indicate the LQCD and LCSR form-factor information used. The grey line is the experimental upper limit, while the dashed line is the SM prediction.}
\label{fig:propga:D}
\end{figure}

\begin{figure}[]
\centering
  \includegraphics[width=.8\linewidth]{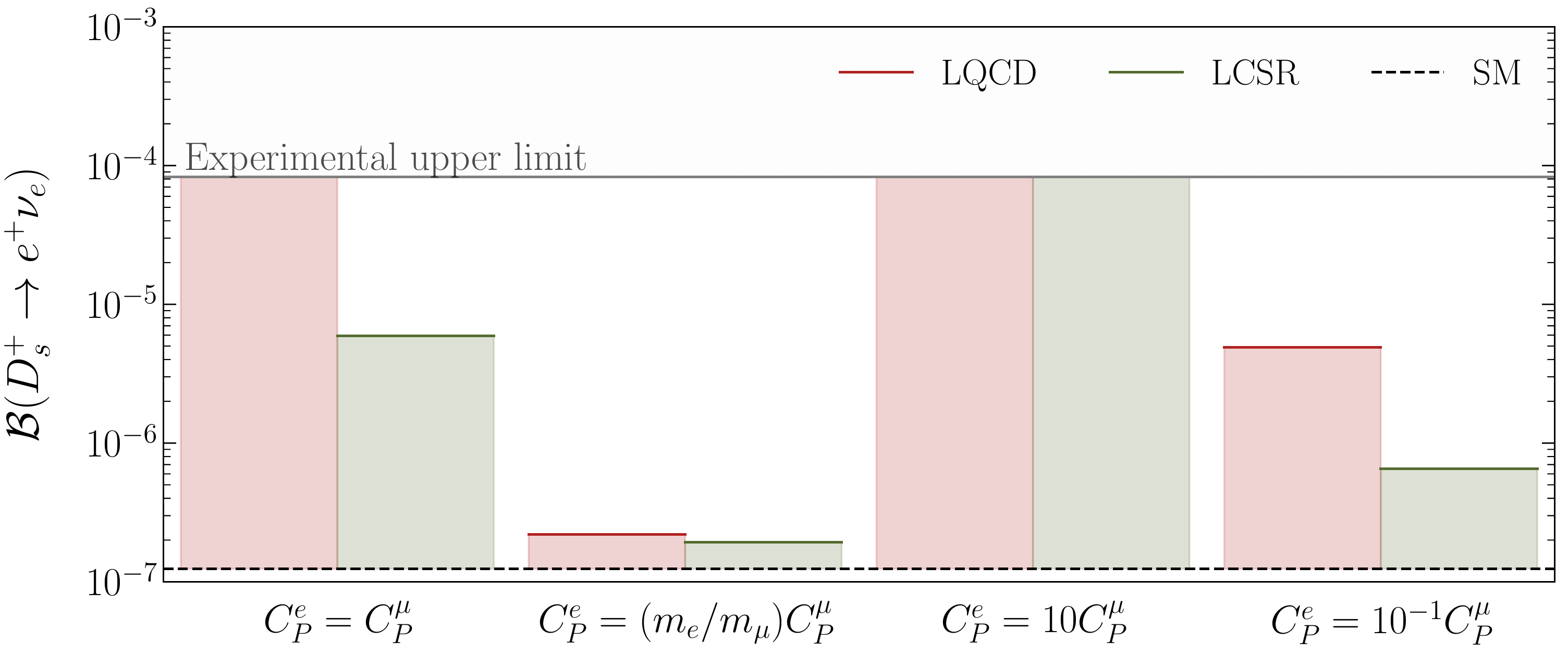}
\caption{Predictions of $\mathcal{B}(D_s^+ \to e^+ \nu_e)$ for the scenarios discussed in the text. The red and green bands indicate the LQCD and LCSR form-factor information used. The grey line is the experimental upper limit, while the dashed line is the SM prediction.}
\label{fig:propga:Ds}
\end{figure}

At the same time, the obtained constraints on the NP coefficients allow us to predict the branching fractions discussed here in the context of different scenarios related to LFU violation. Let us consider the following scenarios: 
\begin{itemize}
	\item $C^e_P = C^{\mu}_P$. This correlation is indicated by the dashed-dotted lines in Figs.\ \ref{Fig_Cmu_Ctau_Cel_PseuNP_lepsemilep_D} (right) and \ref{Fig_Cmu_Ctau_Cel_PseuNP_lepsemilep_Ds} (right).  
	\item $C^e_P = (m_e/m_{\mu})C^{\mu}_P$. This correlation arises, for instance, in the context of a type II Two-Higgs-Doublet model. 
	\item $C^e_P \gg C^{\mu}_P$. We probe the impact of $C^e_P = 10 C^{\mu}_P$ on the relevant branching fractions.
	\item $C^e_P \ll C^{\mu}_P$. Similar to the previous scenario, but focusing on the effect of setting $C^e_P = 10^{-1} C^{\mu}_P$ for the corresponding branching fractions.
\end{itemize}
We use the obtained constraints on $C^{\mu}_{P}$, given in Table \ref{Tab_CmuP_Values_LQCD_LCSR}, and relate them to $C^e_P$ in each scenario accordingly. Subsequently, we use the corresponding values for the CKM elements, given in Table \ref{Tab_VcdVcsvalues}, and apply the results to Eq.\ (\ref{Eq_Lep_BF_PseuNP}) to obtain a theoretical prediction for the branching fractions $\mathcal{B}(D^+ \to e^+ \nu_e)$ and $\mathcal{B}(D_s^+ \to e^+ \nu_e)$ in each of the scenarios. In Figs.\ \ref{fig:propga:D} and \ref{fig:propga:Ds}, we compare our predictions to the corresponding SM predictions, where $C^e_P = C^{\mu}_P = 0$. For the SM prediction, the $|V_{cd(s)}|$ values obtained from unitarity were used. The determination of the relevant CKM elements using LQCD or LCSR form-factor information is indicated by the red and green colours of the bars, respectively. Furthermore, the grey regions indicate the values that are currently already excluded by experiment. 

It is interesting to see that in the case $C^e_P = C^{\mu}_P$ in Fig.\ \ref{fig:propga:D}, our predictions for $\mathcal{B}(D^+ \to e^+ \nu_e)$ are very close to the experimental upper limit. For $\mathcal{B}(D_s^+ \to e^+ \nu_e)$ illustrated in Fig.\ \ref{fig:propga:Ds}, our predictions using either LQCD or LCSR based form-factor calculations differ substantially, underlining the importance of precise lattice information on the semileptonic form factors. An observation of spectacularly enhanced $D_{(s)}^+ \to e^+ \nu_e$ modes close to the current experimental limits would be an unambiguous signal of physics beyond the SM.

%
%
%
\section{Conclusions}\label{sec:concl}
We have presented a comprehensive analysis of (semi)-leptonic $D_{(s)}$-meson decays to constrain possible effects of physics 
from beyond the SM arising from new (pseudo)-scalar, vector and tensor operators, allowing also for violations of LFU.
The central role is played by various ratios of
decay rates that are independent of the CKM matrix elements $|V_{cd}|$ and $|V_{cs}|$. In the case of leptonic decays, the decay constants cancel there and in the case of combinations of leptonic and semileptonic decays, ratios of non-perturbative hadronic parameters arise that are usually more precise than the individual parameters. 

We obtain a picture in agreement with the SM, including a few deviations at the $1\,\sigma$ level. Following our strategy, we may also determine  $|V_{cd}|$ and $|V_{cs}|$ from the (semi)-leptonic decays in the presence of NP contributions. The corresponding results 
are fully consistent with values arising from the unitarity of the CKM matrix and the Wolfenstein parameterization that do not involve any experimental input from the charm system. We have identified various form factors with interesting potential for future improvement through lattice QCD calculations. 

The leptonic $D_{(s)}^+\to e^+\nu_e$ decays are hugely helicity suppressed in the SM. However, this suppression may be lifted through new pseudoscalar interactions. Using the constraints following from the interplay between leptonic $D_{(s)}$ decays with
muon or tau leptons in the final state and $D\to V\bar l\nu_l$ decays, we illustrate that the branching ratio for $D_{(s)}^+\to e^+\nu_e$
may be hugely enhanced and could enter the regime close to the current experimental upper bounds. A future observation of these
modes would be an unambiguous signal of NP effects. 

We look forward to obtaining stronger constraints on possible violations of LFU in the charged-current interactions in 
$D_{(s)}$ decays in the high-precision era of quark-flavour physics which is ahead of us.


\section*{Acknowledgements}
This research has been supported by the Netherlands Foundation for Fundamental Research of Matter (FOM) programme 156, ``Higgs as Probe and Portal'', and by the National Organisation for Scientific Research (NWO). We would like to thank 
Gilberto Tetlalmatzi-Xolocotzi for valuable discussions.


\newpage
\appendix

\section{Form Factors}\label{AppFFs}

\subsection{\boldmath $D \to P$}

The form factors for $D \to P$ decays in the helicity basis are defined as follows \cite{Sakaki:2013bfa}:

\begin{subequations}
\noindent \begin{minipage}{0.45\linewidth}
\centering

\begin{align}
&H^{P}_{V, s} = \frac{M_{D_{(s)}}^2-m_P^2}{\sqrt{q^2}} f_0(q^2), \\
&H^P_{V,0} = \sqrt{\frac{\lambda_P(q^2)}{q^2}}f_+(q^2), \\
&H^P_{V,\pm} = 0,
\end{align}
\end{minipage}
\hfill
\begin{minipage}{0.45\linewidth}
\centering
\begin{align}
&H^{P}_{S, s} = \frac{M_{D_{(s)}}^2 - m_P^2}{m_c-m_q} f_0(q^2), \label{Eq_Hs_pseu_hel_ampl} \\
&H^P_{P,s} = 0, \\
&H^{P}_{T} = -\frac{\sqrt{\lambda_P(q^2)}}{M_{D_{(s)}} + m_P} f_T(q^2),
\end{align}
\end{minipage}
\end{subequations}

\vline

\noindent where $\lambda_{P}(q^2) = [(M_{D_{(s)}}-m_P)^2 - q^2][(M_{D_{(s)}} + m_P)^2-q^2]$. For our calculations, we adopt the $z$-series parametrization from Ref.\ \cite{Lubicz:2017syv}. For the $D \to \pi$ case, the scalar and vector form factors are given by
 \begin{align}
 f^{D \to \pi}_+(q^2) = \frac{f^{D \to \pi}(0)+c^{D \to \pi}_+ (z-z_0) (1+ \frac{1}{2}(z+z_0))}{1- P_V q^2},\\
 f^{D \to \pi}_0(q^2) = \frac{f^{D \to \pi}(0)+c^{D \to \pi}_0 (z-z_0) (1+ \frac{1}{2}(z+z_0))}{1- P_S q^2},
 \end{align}
where $z_0 = z(0,t_0^{\pi})$. In the case of $D \rightarrow K$ transitions, the scalar and vector form factors are parametrized as
 \begin{align}
 f^{D \to K}_+(q^2) = \frac{f^{D \to K}(0)+c^{D \to K}_+ (z-z_0) (1+ \frac{1}{2}(z+z_0))}{1- q^2/M^2_{D_s^*}},\\
 f^{D \to K}_0(q^2) = f^{D \to K}(0)+c^{D \to K}_0 (z-z_0) \Big(1+ \frac{1}{2}(z+z_0)\Big),
 \end{align}
where $z_0 = z(0,t_0^{K})$. The fit parameters are listed in Table \ref{fitparametersZseries}.

\begin{table}[h!]
\renewcommand{\arraystretch}{1.2}
\begin{center}
\begin{tabular}{ c c c c c c } 
 \hline
 Decay &$f(0)$ & $c_+$ & $P_V$ (GeV)$^{-2}$ &$c_0$ & $P_S$ (GeV)$^{-2}$\\ 
 \hline
\hline
 $D\rightarrow \pi$ & 0.6117 (354) & -1.985 (347) & 0.1314 (127) &-1.188 (256) & 0.0342 (122)\\ 
 $D\rightarrow K$ & 0.7647 (308) & -0.066 (333)  &-& -2.084 (283) & -  \\ 
 \hline
\end{tabular}
\end{center}
\caption{Fit parameters for $f_0$, $f_+$ in the $z$-series expansion \cite{Lubicz:2017syv}.}\label{fitparametersZseries}
\end{table}

\noindent For the tensor form factor, we use the lattice calculation of Ref.\ \cite{Lubicz:2018rfs}. Here, $f_T$ is parametrized in the following way:
 \begin{align}
 f^{D \to \pi}_T(q^2) = \frac{f^{D \to \pi}_T(0)+c^{D \to \pi}_T (z-z_0) (1+ \frac{1}{2}(z+z_0))}{1- P^{D \to \pi}_T q^2},\\
 f^{D \to K}_T(q^2) = \frac{f^{D \to K}_T(0)+c^{D \to K}_T (z-z_0) (1+ \frac{1}{2}(z+z_0))}{1- P^{D \to K}_T q^2}.
 \end{align}
The fit parameters are given in Table \ref{Tab_Fit_Params_z_series_tensor}.
 
\begin{table}[h!]
\renewcommand{\arraystretch}{1.2}
\begin{center}
\begin{tabular}{ c c c c } 
 \hline
 Decay &$f_T(0)$ & $c_T$ & $P_T$ (GeV)$^{-2}$ \\ 
 \hline
 \hline
 $D\rightarrow \pi$ & 0.5063 (786) & -1.10 (1.03) & 0.1461 (681)\\
 $D\rightarrow K$ & 0.6871 (542) & -2.86 (1.46)  &0.0854 (671) \\ 
 \hline
\end{tabular}
\end{center}
\caption{Fit parameters for $f_T$ in the $z$-series expansion \cite{Lubicz:2018rfs}.}\label{Tab_Fit_Params_z_series_tensor}
\end{table}

\subsection{\boldmath $D \to V$}

The non-zero form factors for $D \to V$ decays in the helicity basis are defined as follows \cite{Sakaki:2013bfa}:

\begin{subequations}
\begin{align}
&H^{V}_{V,\pm}  = (M_D + m_{V}) A_1(q^2) \mp \frac{\sqrt{\lambda_{V}(q^2)}}{M_D + m_{V}} V(q^2),\\
&H^{V}_{V,0}  = \frac{M_D + m_{V}}{2M_{V} \sqrt{q^2}} \big[ -(M^2_D - M_{V}^2 - q^2) A_1(q^2) +  \frac{\lambda_{V}(q^2)}{(M_D + m_{V})^2} A_2(q^2)\big],\\
&H^{V}_{V,t}  = \sqrt{\frac{\lambda_{V}(q^2)}{q^2}} A_0(q^2),\\
&H^{V}_{P} = - \frac{\sqrt{\lambda_{V}(q^2)}}{m_c + m_q} A_0(q^2),\\
&H^{V}_{T,\pm} = \frac{1}{\sqrt{q^2}} \Big[\pm (M^2_D - m_{V}^2) T_2(q^2) + \sqrt{\lambda_{V}(q^2)} T_1(q^2)    \Big],\\
&H^{V}_{T,0} = \frac{1}{2 m_{V}} \big[ -(M^2_D + 3 m^2_{V} - q^2) T_2(q^2) + \frac{\lambda_{V}(q^2)}{M_D^2 - m_{V}^2} T_3(q^2) \big],
\end{align}
\end{subequations}

\noindent where $\lambda_{V}(q^2) = [(M_D-m_V)^2 - q^2][(M_D + m_V)^2-q^2]$. We use the lattice determination of the $D \to \rho$ and $D \to K^*$ semileptonic form factors from Ref.\ \cite{Bowler:1994zr}. The $q^2$ dependence is obtained through a single-pole parametrization:
\begin{equation}
        V(q^2) = \frac{V(0)}{1 - q^2/m^2_{1^-}} , \hspace{5mm} A_0(q^2) = \frac{A_0(0)}{1 - q^2/m^2_{0^-}} , \hspace{5mm} A_i(q^2) = \frac{A_i(0)}{1 - q^2/m^2_{1^+}} , 
\end{equation}
where $i=1,2,3$ and $m_{J^P}$ denotes the mass of the meson with spin $J$ and parity $P$ corresponding to the relevant quark transition, i.e., $c\bar{d}$ for $D \to \rho$ and $c\bar{s}$ for $D \to K^*$. The values for the parameters are given in Table \ref{Tab_Fit_Params_UKQCD}. The masses are listed in units of the inverse lattice spacing, which is given by 
$a^{-1} = (2.73 \pm 0.05)$\,GeV. The form factor $A_3$ is related to the form factors $A_1$ and $A_2$ in the following way:
\begin{equation}
A_3(q^2) = \frac{M_D + m_V}{2 m_V} A_1(q^2) - \frac{M_D - m_V}{2 m_V} A_2(q^2).
\end{equation}

\begin{table}[h!]
\renewcommand{\arraystretch}{1.5}
\begin{center}
\begin{tabular}{c c c c  } 
 \hline
 Decay & Form Factor & $F(0)$ & $m_{J^P}  [a^{-1}]$   \\ 
 \hline
 \hline
 $D \rightarrow \rho$ & $A_0$   & $0.70^{+0.05}_{-0.12}$ & $m^{c\bar{d}}_{0^-} = 0.60^{+0.07}_{-0.05}$ \\ 
 & $A_1$   & $0.63^{+0.06}_{-0.09}$ &   $m^{c\bar{d}}_{1^+} = 1.1^{+0.3}_{-0.2} $  \\ 
 & $A_2$   & $0.51^{+0.10}_{-0.15}$ &   $m^{c\bar{d}}_{1^+} = 0.44^{+0.09}_{-0.05} $  \\ 
 &     $V$   & $0.95^{+0.29}_{-0.14}$ &   $m^{c\bar{d}}_{1^-} = 0.91^{+0.36}_{-0.18}$  \\ 
 \hline
 $D \rightarrow K^*$ & $A_0$   & $0.75^{+0.05}_{-0.11}$ & $m^{c\bar{s}}_{0^-} = 0.59^{+0.06}_{-0.05}$ \\ 
 & $A_1$   & $0.70^{+0.07}_{-0.10}$ &   $m^{c\bar{s}}_{1^+} = 1.1^{+0.3}_{-0.2} $  \\ 
 & $A_2$   & $0.66^{+0.10}_{-0.15}$ &   $m^{c\bar{s}}_{1^+} = 0.46^{+0.16}_{-0.07} $  \\ 
 &     $V$   & $1.01^{+0.30}_{-0.13}$ &   $m^{c\bar{s}}_{1^-} = 0.85^{+0.24}_{-0.15}$  \\ 
 \hline
\end{tabular}
\end{center}
\caption{LQCD fit parameters for the $D \to \rho$ and $D \to K^*$ form factors, taken from Ref.\ \cite{Bowler:1994zr}.}
\label{Tab_Fit_Params_UKQCD}
\end{table}

\noindent We complement the lattice calculation with the LCSR calculation of Ref.\ \cite{Wu:2006rd}. This work dates back to 2006, but there exists a more recent determination of the $D \to \rho$ form factors \cite{Fu:2018yin}. In this analysis, 
the massless lepton limit is taken, neglecting the form factor $A_0$. In our study, we investigate the difference between different lepton flavours, therefore requiring information on $A_0$. We therefore utilize the form-factor information of Ref.\ \cite{Wu:2006rd}, where
the following double-pole parametrization was adopted:
\begin{equation}
F^i(q^2) = \frac{F^i(0)}{1 - a_{F^i} q^2/M_D^2 + b_{F^i} (q^2/M_{D_{(s)}}^2)^2} ,
\end{equation}
with $F^i$ denoting any of the form factors $A_1$, $A_2$, $A_3$ or $V$. Here, a different convention is used: instead of the set $\{A_0, A_1, A_2\}$, the hadronic matrix elements are parametrized by the set $\{A_1, A_2, A_3\}$. The relation between the different conventions is given by the following relation \cite{Wang:2002zba}:
\begin{equation}
A_0(q^2) = \frac{1}{2 m_V} \Big[(M_D + m_V) A_1(q^2) - (M_D - m_V) A_2(q^2) - \frac{q^2}{M_D + m_V} A_3(q^2) \Big].
\end{equation}
The results from LCSR for the fit parameters for $D \to \rho$ and $D \to K^*$ transitions are listed in 
Table \ref{Tab_Fit_Params_Wu_LCSR}.

\begin{table}[h!]
\renewcommand{\arraystretch}{1.5}
\begin{center}
\begin{tabular}{c c c c c } 
 \hline
 Decay & Form Factor & $F(0)$ & $a_F$ & $b_F$   \\ 
 \hline
  \hline
 $D \rightarrow \rho$ & $A_1$   & $0.590^{+0.031}_{-0.029}$ & $0.44^{-0.04}_{+0.05}$&  $0.20^{-0.03}_{+0.10}$\\ 
 & $A_2$   & $0.528^{+0.036}_{-0.031}$ &   $ 0.91^{+0.07}_{-0.13} $ &				$-1.01^{+0.22}_{-0.41}$ \\ 
 & $A_3$   & $-0.528^{+0.031}_{-0.026}$ &   $ 0.91^{-0.13}_{+0.07} $ &				$-1.01^{-0.41}_{+0.22}$\\ 
 &     $V$   & $0.735^{+0.032}_{-0.025}$ &   $0.48^{-0.11}_{+0.21}$ & 				$2.25^{-0.21}_{+0.61}$\\ 
 \hline
 $D \rightarrow K^*$ & $A_1$   & $0.601^{+0.030}_{-0.029}$ & $ 0.51^{-0.02}_{+0.02}$& $ 0.04^{+0.01}_{-0.01}$\\ 
 & $A_2$   & $0.541^{+0.038}_{-0.033}$ &   $ 0.91^{+0.05}_{-0.10} $ & 				$ -0.68^{+0.12}_{-0.21}$\\ 
 & $A_3$   & $-0.541^{+0.033}_{-0.038}$ &   $ 0.91^{-0.10}_{+0.05} $&  				$ -0.68^{-0.21}_{+0.12}$\\ 
 &     $V$   & $0.796^{+0.032}_{-0.027}$ &   $ 0.60^{-0.07}_{+0.13}$&  				$ 1.53^{-0.13}_{+0.30}$\\ 
 \hline
\end{tabular}
\end{center}
\caption{LCSR fit parameters for the $D \to \rho$ and $D \to K^*$ form factors, taken from Ref.\ \cite{Wu:2006rd}. }
\label{Tab_Fit_Params_Wu_LCSR}
\end{table}


\end{document}